\documentclass[12pt]{article}
\usepackage[margin=1.2in]{geometry}
\usepackage[onehalfspacing]{setspace}

\usepackage{amsfonts, amsmath, amssymb}
\usepackage[utf8]{inputenc}

\usepackage[authoryear]{natbib}
\usepackage{graphicx}
\usepackage{tikz}

\usepackage{float} 
\usepackage{booktabs, makecell} 
\usepackage{longtable,multirow,multicol} 
\usepackage[labelformat=simple]{subcaption} 

\usepackage{placeins} 

\usepackage{caption} 
\usepackage{array} 
\usepackage{afterpage}
\usepackage{bbm}

\makeatletter
\renewcommand\paragraph{\@startsection{paragraph}{4}{\z@}%
  {1.25ex \@plus .5ex \@minus .2ex}%
  {-1em}%
  {\normalfont\normalsize\bfseries}}
\makeatother

\usepackage{xcolor}
\definecolor{bl}{rgb}{0,0.1,0.5}
\definecolor{darkblue}{rgb}{0,0,0.65}
\definecolor{darkgreen}{rgb}{0.1,0.35,0.1} 
\definecolor{darkorange}{rgb}{1.0, 0.55, 0.0}
\definecolor{olive}{rgb}{0.5, 0.5, 0.0}
\definecolor{ao(english)}{rgb}{0.0, 0.6, 0.0}

\usepackage{xr-hyper}
\makeatletter
\newcommand*{\addFileDependency}[1]{
  \typeout{(#1)}
  \@addtofilelist{#1}
  \IfFileExists{#1}{}{\typeout{No file #1.}}
}
\makeatother

\makeatletter
\@ifundefined{ifusebackref}{\newif\ifusebackref\usebackreftrue}{}
\makeatother
\ifusebackref
\usepackage[unicode=true,
  bookmarks=true,bookmarksnumbered=true,
  bookmarksopen=true, bookmarksopenlevel=2,   
  backref=page,                       
  colorlinks=true,allcolors=darkblue, 
  breaklinks=true,
  pdfborder={0 0 1},pdfborderstyle=, 
  ]{hyperref}
\else
\usepackage[unicode=true,
  bookmarks=true,bookmarksnumbered=true,
  bookmarksopen=true, bookmarksopenlevel=2,   
  backref=false,
  colorlinks=true,allcolors=darkblue, 
  breaklinks=true,
  pdfborder={0 0 1},pdfborderstyle=, 
  ]{hyperref}
\providecommand{\backcite}[2]{}
\fi

\usepackage{xspace} 

\ifusebackref
\renewcommand*{\backref}[1]{}
\renewcommand*{\backrefalt}[4]{{\ifcase #1%
    \or(Cited on page~#2)%
    \else(Cited on pages #2)%
  \fi%
  }}
\fi

\usepackage[textwidth=1.25in,colorinlistoftodos]{todonotes} 
\ifx11 
    \newcommand{\chiaram}[1]{\todo[color=pink!30,caption={}]{\textbf{CDB:} #1}}
    \newcommand{\chiarai}[1]{\todo[inline,color=pink!30,caption={}]{\textbf{CDB:} #1}}
    \newcommand{\andream}[1]{\todo[color=green!30,caption={}]{\textbf{AM:} #1}}
    \newcommand{\andreai}[1]{\todo[inline,color=green!30,caption={}]{\textbf{AM:} #1}}
\else
    \newcommand{\chiaram}[1]{}
    \newcommand{\chiarai}[1]{}
    \newcommand{\andream}[1]{}
    \newcommand{\andreai}[1]{}
\fi

\usepackage{siunitx}
\newcommand{\hindex}{\emph{Hindex}\xspace}
\newcommand{\linreest}{Linear -- Estimated}

\newcommand{\sym}[1]{#1} 

\newcommand{\targetedcaption}{Data (dotted lines) and simulated (solid lines) age profiles of average assets (in \pounds1000), fraction of people in disability, and employment rate by four health quantiles.}
\newcommand{\nontargetcaption}{Data and simulated age profiles of health in disability, probability of being in disability if in disability the previous period (Persistence) or if not (Inflow), and average assets (in \pounds1000) by health.}

\hypersetup{
  pdftitle={The economic effects of nonlinear health dynamics: estimates from a dynamic life-cycle model},
  pdfauthor={Chiara Dal Bianco and Andrea Moro}
}

\begin{document}

\setlength{\abovedisplayskip}{6pt} 
\setlength{\belowdisplayskip}{6pt}
\title{The economic effects of nonlinear health dynamics: estimates from a dynamic life-cycle model\thanks{We thank Marco Bertoni, Guglielmo Weber, and several participants in conferences and seminars for helpful suggestions. We also thank Stéphane Bonhomme for his advice on the estimation of the nonlinear health process.
}}

\author{Chiara Dal Bianco\setcounter{footnote}{2}\footnote{Department of Economics and Management, University of Padova, Italy. ORCID: 0000-0002-5449-0645. \emph{E-mail}: \texttt{chiara.dalbianco@unipd.it}} \and Andrea Moro\footnote{Department of Economics, Vanderbilt University, Nashville, TN, USA. ORCID: 0000-0001-5570-8151 \emph{E-mail}: \texttt{andrea@andreamoro.net}}. }
\date{\today}
\maketitle
\begin{abstract}
\noindent

We study how nonlinear, state-dependent health dynamics shape economic behavior, inequality, and the evaluation of disability insurance at older ages.
Using English panel data, we construct a continuous health index and estimate its dynamics with a flexible quantile-based method that allows persistence to vary
across health states.
We find that adverse health realizations are both larger and more persistent among individuals in poor health.
Embedding the estimated process into a life-cycle model, we show that these state-dependent nonlinearities generate substantial losses in assets and welfare for economically vulnerable individuals--those with poor health and low wealth. Misspecifying health dynamics as state-independent attenuates these losses and leads to distorted savings behavior, with effects concentrated among vulnerable individuals.
Finally, we find that the welfare losses of removing disability insurance are highly heterogeneous across health types, and are overstated by a state-independent health process.

\end{abstract}
\vfill


	\vfill
\thispagestyle{empty}

\section{Introduction}

Health deteriorates with age, but the pace and pattern of decline vary substantially across individuals. Understanding this heterogeneity is essential for evaluating the lifetime economic consequences of health risk and the value of social insurance programs. Yet quantifying these effects requires confronting two fundamental measurement challenges: constructing a health measure that aggregates information from multiple indicators while mitigating reporting bias, and modeling its evolution in a way that allows persistence and shock responses to vary across health states.

This paper addresses both challenges. We construct a continuous health index that combines subjective health reports with objective health indicators and estimate its dynamics using a flexible quantile-based panel methodology that allows persistence to vary nonlinearly across the health distribution. We embed the estimated health process into a life-cycle model of consumption, saving, labor supply, and disability insurance (DI) participation to quantify how health dynamics shape economic outcomes, inequality, and welfare over the later life cycle. \label{sec:intro-di}

Our central finding is that state-dependent nonlinearities in health dynamics have first-order implications for economic outcomes and their distribution. Adverse health realizations are both larger in magnitude and more persistent when experienced by individuals in poor health, generating substantial cumulative losses in assets and welfare. These effects are concentrated among economically vulnerable individuals--those entering older ages with poor health and low wealth--for whom a single adverse health shock can lead to long-lived deteriorations in both health and economic well-being. Accounting for these state-dependent patterns is therefore essential for evaluating both the distributional burden of health risk and the insurance value of disability programs.

\paragraph{Health measurement and dynamics.}
Our health measure combines subjective and objective health information. We construct a continuous index as the component of self-reported health explained by objective indicators \citep{blundell2016dynamic, hosseini2021evolution}--diagnoses, functional limitations, and physiological measures. This strategy mitigates reporting bias and measurement error while yielding a smooth distribution well-suited to estimating nonlinear (state-dependent) dynamics.

To model health evolution, we adopt the quantile-based panel approach of \citet{ABB2017}, originally developed for earnings dynamics. This framework nests standard linear autoregressive specifications but allows both the magnitude of realizations and their persistence to vary flexibly with lagged health and the rank of the innovation. Recent work by \citet{hosseini2021evolution, hosseini2024important} introduces nonlinearity through a log transformation of health, a useful and parsimonious way to capture state-dependent dynamics. The quantile-based approach we adopt offers complementary flexibility by allowing heterogeneous persistence patterns across the entire health distribution without imposing functional form restrictions. This is particularly valuable for capturing asymmetric responses to positive and negative shocks and for studying outcomes across different segments of the health distribution.

The data exhibit pronounced state dependence in health dynamics. Persistence varies substantially across the health distribution, being considerably higher in low-health states and much weaker in good health. As a result, adverse shocks that move individuals into poor health are associated with trajectories that are strongly shaped by past health histories, while health in better states displays faster mean reversion. These features generate asymmetric adjustment patterns that are not captured by specifications with state-invariant persistence.

\label{sec:dynamic-fixedeffects}In addition, the health process incorporates time-invariant individual heterogeneity to account for persistent differences in health endowments. The importance of such permanent heterogeneity for life-cycle outcomes has been emphasized in the literature, most notably by \citet{de2022lifetime}, and is a common feature of recent models of health dynamics \citep{hosseini2021evolution, hosseini2024important}. While De Nardi and coauthors model this heterogeneity through unobserved latent health types, we follow the approach of \citet{hosseini2021evolution, hosseini2024important} and allow for individual fixed effects in the estimation of the health process. This specification enables us to distinguish between ex-ante heterogeneity in baseline health trajectories and the state-dependent propagation of realized health shocks, a distinction that is central to understanding inequality and welfare dynamics over the life cycle.

To assess the quantitative importance of these patterns, we also consider a process with age-dependent but state-invariant persistence (we label it the \emph{linear} process). While this specification reproduces aggregate life-cycle profiles comparably well, incorporating state dependence in health dynamics proves important for understanding heterogeneity across individuals. In particular, it materially affects predictions for economically vulnerable groups, especially with respect to asset accumulation and welfare following adverse health realizations.

\paragraph{Life-cycle model and estimation.}
To assess the economic importance of these patterns, we incorporate the estimated health process into a  life-cycle model in which individuals make consumption, saving, and labor supply decisions under uncertainty over health, earnings, survival, and DI acceptance. Health affects the time available for work and leisure, earnings capacity, survival probabilities, and eligibility for DI.

We estimate the model parameters using a simulated method of moments, targeting age profiles of assets, labor force participation by health status, and DI receipt using data from the English Longitudinal Study of Ageing (ELSA). We focus on low-educated men aged 50 and above, who are most exposed to disability risk and most reliant on public insurance. To reduce heterogeneity and improve precision, we condition on partnership status and restrict attention to men living with a partner, the most numerous group in the data. The model fits targeted moments well and replicates salient non-targeted patterns, including asset accumulation by health status and transitions into and out of DI.

\paragraph{Economic and welfare effects.}
Using the estimated model, we conduct four complementary exercises to isolate how health dynamics shape economic outcomes.

First, we compare aggregate outcomes across alternative health process specifications: our baseline nonlinear model, a specification with state-invariant persistence,
and a specification fully re-estimated under the alternative dynamics to fit the same data moments. This comparison reveals that health dynamics have particularly large effects on wealth accumulation. Among economically vulnerable individuals (those with below-median health and wealth at age 50), differences in health process specifications lead to asset differences of up to about 7\% at age 70, while effects on earnings and labor supply are more modest. These patterns reflect the distinction between stock and flow adjustments: asset accumulation integrates expectations about lifetime health risk and is highly sensitive to persistence patterns, while labor supply responds more strongly to current health conditions. The different model dynamics also affect inequality patterns, particularly for wealth accumulation.

Second, we quantify the total burden of realized bad health by comparing baseline outcomes to a counterfactual where all individuals experience persistently good health throughout the life cycle (99th percentile), following a methodology employed in \citet{de2022lifetime}. Median welfare losses are substantial and heterogeneous across unobserved types, declining monotonically with earnings capacity. This heterogeneity in costs across types proves important for understanding who benefits most from health-related social insurance.

Third, we study impulse responses to one-time adverse health shocks. For individuals starting in poor health and low wealth, experiencing a shock placing them at the 10th percentile of the health distribution (versus a median realization) reduces assets at age 70 by substantial amounts relative to a median realization and generates larger and more persistent increases in asset inequality. These distributional consequences are most pronounced for vulnerable groups, highlighting the importance of state-dependent persistence for understanding inequality dynamics.

\label{sec:intro-channels}
Fourth, we decompose health effects by channel--mortality, time costs, earnings capacity, and DI eligibility--to clarify which mechanisms drive aggregate and distributional outcomes. Mortality and time costs emerge as the dominant channel accounting for the largest share of welfare losses. This finding complements recent work by \citet{hosseini2024important} on the US, where DI plays a more prominent role in earnings inequality dynamics. The different relative magnitudes likely reflect institutional differences: UK disability programs provide flat-rate benefits with less generous coverage compared to earnings-related US DI, shifting the balance toward direct time costs as the primary economic burden of poor health.

\paragraph{Disability insurance.}
We conclude by evaluating the insurance value of DI. Removing DI generates welfare losses concentrated among poor health types, that is, those with a lower permanent health level and limited capacity to self-insure through assets. These losses are slightly larger under the alternative health dynamics with state-invariant persistence, consistent with differences in the ex ante distribution of health risk and average wealth accumulation across model specifications. When DI removal is implemented in a revenue-neutral way through reductions in labor and pension taxes, average welfare effects turn positive because the tax cuts are financed over a relatively low-earnings population in our baseline sample. These gains are highly uneven: individuals in better underlying health benefit substantially, while those in poor underlying health benefit little or may experience welfare losses. These distributional patterns underscore the redistributive role of DI, providing insurance disproportionately valuable to individuals facing adverse permanent health conditions.

\paragraph{Contribution and main insights.}
\label{sec:contributionx}

To summarize, our paper shows that allowing for nonlinear, state-dependent health dynamics yields three main insights for life-cycle behavior, inequality, and the evaluation of social insurance. First, misspecifying health dynamics primarily distorts savings behavior
understating the severity and persistence of adverse health realizations.
This explains why aggregate wealth levels and relative wealth inequality are sensitive to health dynamics even when earnings and employment responses are modest.

Second,
nonlinear dynamics amplify inequality following adverse health realizations, generating large and persistent wealth losses when bad shocks occur. Linear dynamics, by contrast, compress average asset holdings and mechanically increase aggregate relative inequality, despite attenuating individual shock responses.

Third, the welfare value of DI is concentrated among individuals who belong to poor permanent health types and depends importantly on the persistence of adverse health histories. State-dependent health processes overstate the value of DI.

The remainder of the paper proceeds as follows. In the next section, we place this paper in the context of the existing literature. Section \ref{sec:health} describes our health measure construction and documents key empirical patterns in health dynamics. Section \ref{sec:dynamic} presents the quantile-based estimation framework and reports estimates from both flexible and restricted specifications. Section \ref{sec:model} describes the life-cycle model, calibration, and estimation. Section \ref{sec:results} quantifies the economic and welfare effects of health dynamics through a series of counterfactual exercises. Section \ref{sec:conclusion} concludes.

\section{Related literature}\label{sec:intro-litreview}
A large literature has documented the important role of health in shaping economic outcomes over the life cycle, highlighting the complex interactions between health, labor supply, earnings, and savings. Structural models have shown that health risk plays a central role in wealth accumulation and decumulation \citep{de2010elderly, de2016savings, ameriks2020, nakajima2020home, de2022lifetime}, as well as in labor supply and earnings dynamics \citep{low2015disability, french2011effects, capatina2015life, hosseini2024important, keane2024health}.

Much of this literature relies on discrete measures of health--typically binary or ordered indicators based on self-assessed health--reflecting their widespread availability in survey data. An alternative approach constructs continuous health measures that aggregate information from multiple indicators. Examples include principal-component–based indices \citep{dalbianco2022} and deficit-accumulation (frailty) indices \citep{hosseini2021evolution, hosseini2024important}. Existing evidence suggests that, with sufficiently rich information, different continuous measures display similar predictive power for key outcomes such as mortality, labor supply, and disability insurance receipt \citep{hosseini2021evolution}.
We adopt a continuous health index that combines subjective self-reported health with detailed objective health indicators, following the approach of \citet{blundell2016dynamic, blundell2021impact}. Our choice is motivated by the suitability of a smooth health measure for studying nonlinear and state-dependent health dynamics.

A growing body of work recognizes that health dynamics are inherently nonlinear and exhibit strong state dependence. In models with discrete health states, this has motivated extensions of the canonical first-order Markov specification. For example, \citet{de2022lifetime} enriches a binary health process by allowing transition probabilities to depend on fixed heterogeneity and the duration spent in a given health state, showing that such history dependence is necessary to replicate observed patterns of persistence and the welfare costs of poor health. In the context of continuous health measures, \citet{hosseini2021evolution, hosseini2024important} propose using a frailty index and show that its dynamic behavior differs markedly from that implied by discrete self-reported health measures. They model frailty in logs,  offering a parsimonious way to allow health dynamics to vary with health levels and improves the ability of the model to match observed outcomes.

Our quantile-based specification for health dynamics allows both persistence and the distribution of shocks to vary flexibly across health states, without imposing parametric functional forms. This approach, originally developed by \citet{ABB2017} for earnings dynamics, is well suited to capturing asymmetric responses to health shocks and heterogeneity across the health distribution.
Our results complement existing approaches that enrich discrete-state models or impose parametric nonlinearities, by demonstrating that accurately modeling state-dependent health risk is quantitatively important.

Our work is related to \citet{de2022lifetime}, who quantify the costs of bad health over the life cycle using discrete health states, permanent heterogeneity, and duration dependence. While their framework highlights the importance of persistent health differences and survival risk, we take a complementary approach using a continuous health measure and a flexible nonlinear process. This allows us to unpack health dynamics across the health distribution, and to study how misspecifying persistence affects the distribution of welfare and income losses.

An important benchmark for our accounting and decomposition exercises is \citet{capatina2015life}, who quantify the relative importance of productivity, medical expenditures, time endowments, and survival probabilities by selectively shutting down health channels in a life-cycle model. Our decomposition follows a similar logic, while abstracting from medical expenditures and allowing for flexible nonlinear health dynamics. As in her analysis, time-related health costs emerge as a central mechanism, while our framework additionally allows us to evaluate the welfare contribution of mortality risk in a manner consistent with survival differences across health states.

Our paper is also related to \citet{hosseini2024important}, who show that DI accounts for a substantial share of lifetime earnings inequality in the US. We study related mechanisms in a different institutional context and show that institutional features shape the relative importance of health channels, with non-pecuniary time costs and survival risk playing a prominent role in welfare outcomes.

Finally, unlike studies that model the full working life \citep{capatina2015life, hosseini2024important, de2022lifetime}, we focus on older individuals aged 50 and above. This choice reflects both data availability and the objectives of the paper. Rich objective health information is most readily available in surveys of older populations, and later working ages are particularly relevant for studying DI. Longitudinal aging surveys such as ELSA therefore provide a natural setting for our analysis.

\section{Health measurement and its dynamics}\label{sec:health}
\subsection{Health measurement}\label{sec:hmeasure}



In this paper, we adopt a continuous health measure that aggregates information from multiple health indicators. Specifically, we extract the component of subjective health explained by a rich set of objective health conditions. This method was initially developed to address measurement and reporting issues in subjective health indicators and has subsequently been used to construct continuous health indices in empirical and structural analyses \citep{Bound1991, blundell2016dynamic, blundell2021impact}.



We assume we observe a subjective measure of health, denoted by $h^s_{it}$, which summarizes individuals’ self-assessments of their overall health status. Following a large literature, we interpret this measure as a noisy proxy for an underlying latent health variable, $\psi_{it}$, and write:
\begin{align}
h^s_{it} = \psi_{it} + \mu_{it}, \label{eq:latenth}
\end{align}
where $\mu_{it}$ captures measurement error and heterogeneity in reporting behavior across individuals and over time.

While subjective health measures are informative and widely used, household surveys increasingly collect a rich set of more objective health indicators describing physical, mental, and functional limitations. These indicators provide complementary information on individuals' health status and can be used to approximate the latent health component $\psi_{it}$.
Specifically, we assume that latent health can be expressed as a linear combination of objective health conditions:
\begin{align}
\psi_{it} = Z'_{it}\alpha + \xi_{it},
\end{align}
where $Z_{it}$ is a vector of objective health indicators, $\alpha$ is a vector of weights, and $\xi_{it}$ captures health dimensions that are not directly observed in the data.
Substituting this expression into equation~\eqref{eq:latenth} yields:
\begin{align}
h^s_{it} = Z'_{it}\alpha + \xi_{it} + \mu_{it}.
\label{eq:Hhat1}
\end{align}

Parameter identification relies on objective indicators not being affected by the same reporting behavior as $h^s_{it}$ and on measurement error in $Z_{it}$ being uncorrelated with $\mu_{it}$ \citep{blundell2021impact}. Under these conditions, the component of subjective health predicted by objective indicators provides a meaningful summary of latent health.

In our implementation, $h^s_{it}$ is the first principal component of three self-reported ELSA indicators: self-rated general health, work-related limitations, and activity-related limitations. The objective vector $Z_{it}$ includes eyesight, hearing, mobility, ADL/IADL limitations, depression, diagnosed cardiovascular/respiratory conditions, other chronic diseases, eye problems, incontinence, BMI, and grip strength. Variable definitions and descriptive statistics are reported in External Appendix \ref{sec:app-health}.

\begin{figure}[t]
\centering
\caption{Moments of health shocks by age and previous health deciles. Data (top panel) and Simulations (bottom panel).}
\includegraphics[width=1\textwidth]{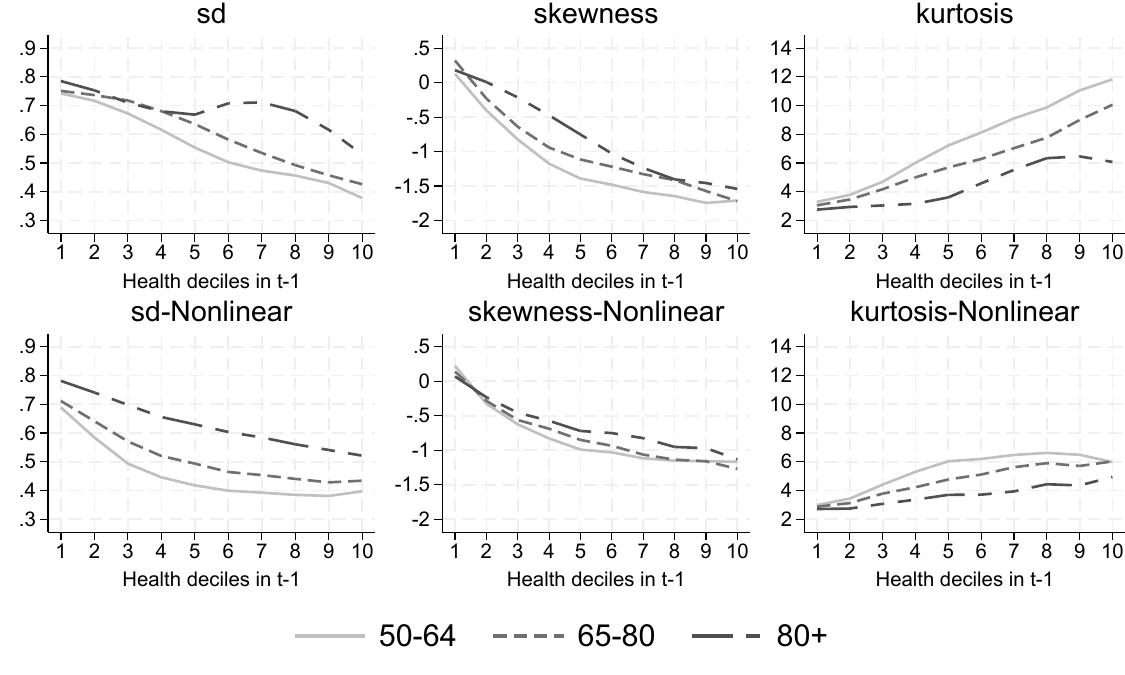}
\label{fig-1nonlinear}
\end{figure}

We define the health index as the predicted component of subjective health:
\begin{align}
\hat{h}_{it} = Z'_{it}\hat{\alpha},
\end{align}
and use this continuous index as the health state variable throughout the paper (denoted \hindex).
To estimate health dynamics, we use residual health $h_{it}$, defined as the component of \hindex that remains after removing systematic variation due to demographics (a third-order polynomial in age, year of birth, education, and partnership status). This removes predictable components and isolates individual fluctuations around life-cycle profiles. All empirical moments of health changes in Section~\ref{subsec:hdynamic} (Figures~\ref{fig-1nonlinear} and \ref{fig-cond2}, top panels) are computed using these residuals.

By construction, \hindex\ is fully continuous and free of mass points, which simplifies the estimation of the health process considered later in the paper. 

\begin{figure}[!t]
\centering
    \caption{Health index distribution}\label{fig:hdensity}
    \includegraphics[width=0.7\textwidth]{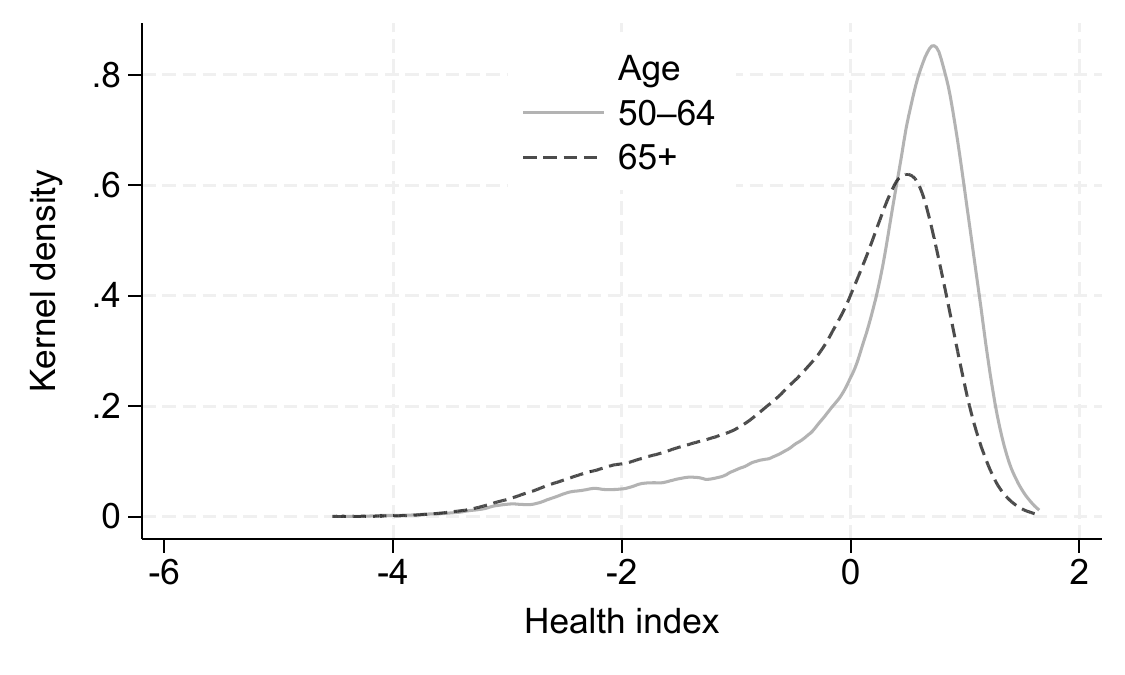}

\caption*{\footnotesize\normalfont \textit{Note:} Cross-sectional distribution of the health index in the data (Equation $h_{it}=Z'_{it}\hat{\alpha}$), conditional on age being less than 65 (solid line), or over age 65 (dashed line). ELSA, waves 1-7.
}
\end{figure}

Figure~\ref{fig:hdensity} shows \hindex follows a left-skewed distribution, with lower health becoming more likely at older ages, closely resembling the frailty distribution (Figure \ref{fig-frailty}).
\hindex is strongly correlated with frailty (correlation $-0.93$) and with SRH (Figures~\ref{fig-corrFrailty} and \ref{fig-corrSRH} in External Appendix \ref{sec:app-health}).

In External Appendix~\ref{sec:app-health}, we compare the predictive performance of \hindex, frailty, and SRH for DI receipt, labor supply, formal/informal care, and mortality in $t+1$. Both \hindex and frailty outperform SRH. The \hindex performs slightly better for mortality, DI receipt, and labor supply, while frailty is marginally better for care receipt.
\label{sec:hmeasure-properties}

\paragraph{Relation to the literature} \label{sec:hmeasure-relation}
The health measure adopted in this paper is closely related to a broader class of health indices that aggregate information from multiple health indicators. Recent work has shown that, once a sufficiently rich set of objective health conditions is used, alternative continuous health measures tend to exhibit very similar empirical properties. In particular, \citet{hosseini2021evolution} compare instrumented subjective health measures, such as those used in \citet{blundell2016dynamic, blundell2021impact} and closely related to our \hindex, with PCA-based health indices and frailty measures constructed from objective deficits. They find that these measures display comparable predictive power for a wide range of outcomes, including mortality, medical expenditures, DI receipt, and labor supply, and that all of them substantially outperform raw self-reported health. This evidence suggests that differences across health measures are largely driven by the underlying information contained in the health indicators, rather than by the specific aggregation method or weighting scheme. In this sense, the measure adopted here can be viewed as one of several empirically equivalent ways of summarizing multidimensional health into a single index.

\subsection{Health dynamics} \label{subsec:hdynamic}

We begin by documenting a set of empirical facts on the dynamics of health changes over the life cycle. Health shocks are defined as $\Delta h_t = h_t - h_{t-1}$.\footnote{Recall that observations are biennial, so one unit of time corresponds to two years.} Figure \ref{fig-1nonlinear} reports three moments of the distribution of health shocks--variance, skewness, and kurtosis--conditional on age and on the initial level of health, measured by deciles of $h_{t-1}$.

\begin{figure}[t]
\centering
\caption{Variances and covariances of health shocks, conditional on health in $t-1$ and age. Data (top panel) and Simulations (bottom panel).}
\label{fig-cond2}
\begin{subfigure}{0.32\textwidth}
    \centering
    \includegraphics[width=.95\textwidth]{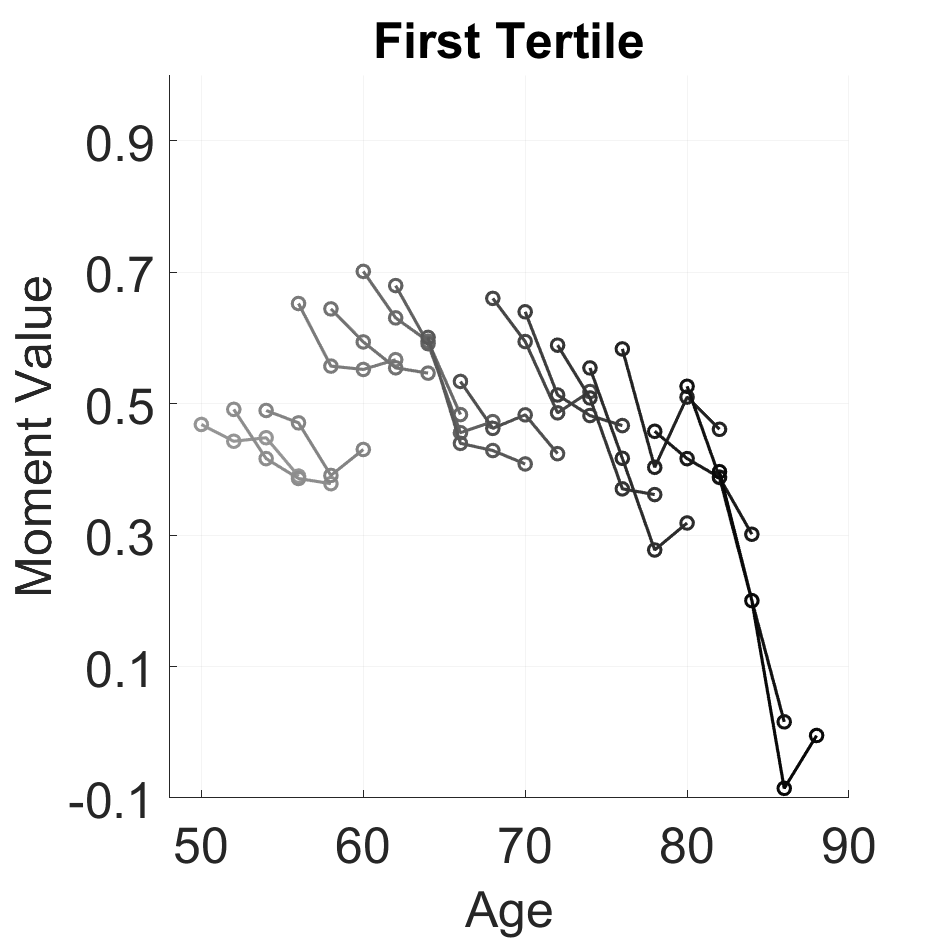}
    \label{fig-cond2a}
\end{subfigure}
\hfill
\begin{subfigure}{0.32\textwidth}
    \centering
    \includegraphics[width=.95\textwidth]{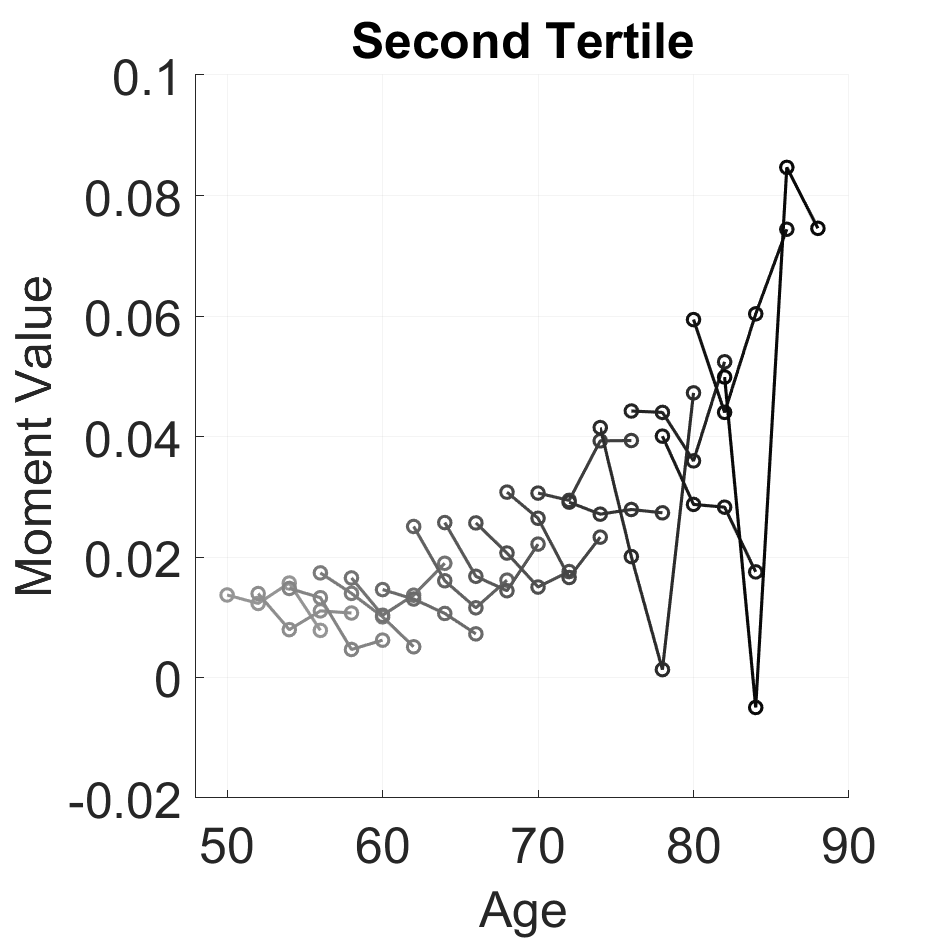}
    \label{fig-cond2b}
\end{subfigure}
\hfill
\begin{subfigure}{0.32\textwidth}
    \centering
    \includegraphics[width=.95\textwidth]{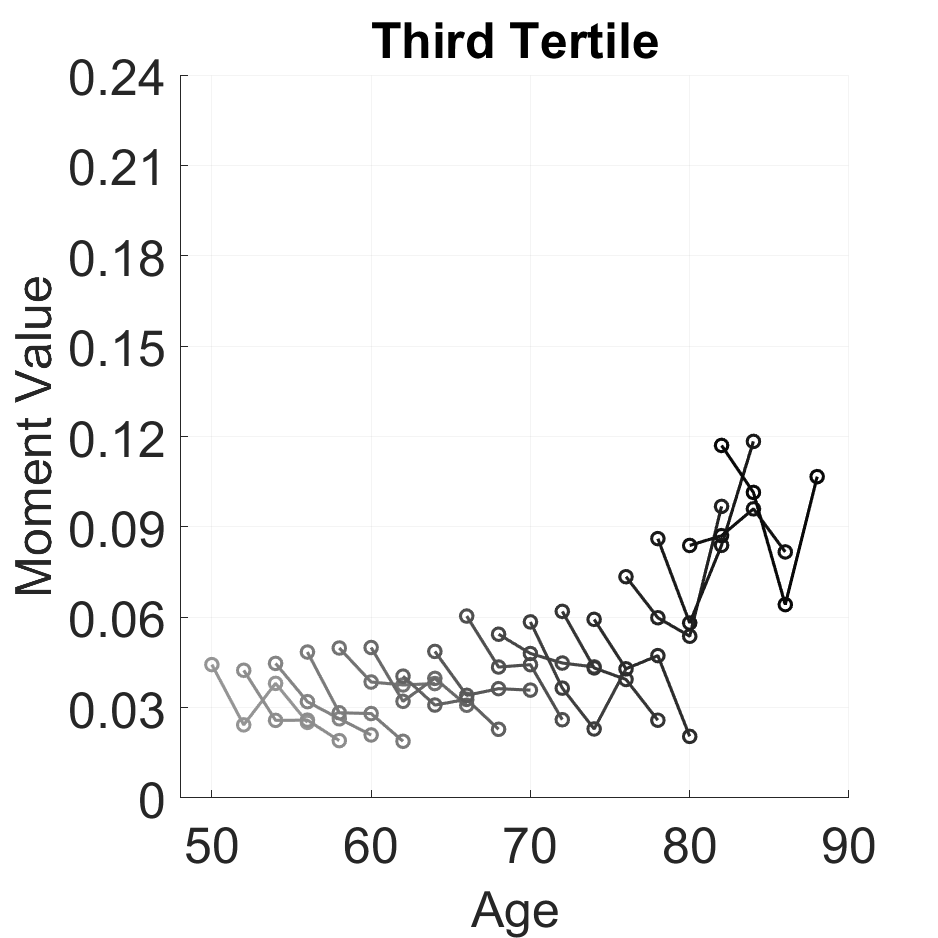}
    \label{fig-cond2c}
\end{subfigure}
\begin{subfigure}{0.32\textwidth}
    \centering
    \includegraphics[width=.95\textwidth]{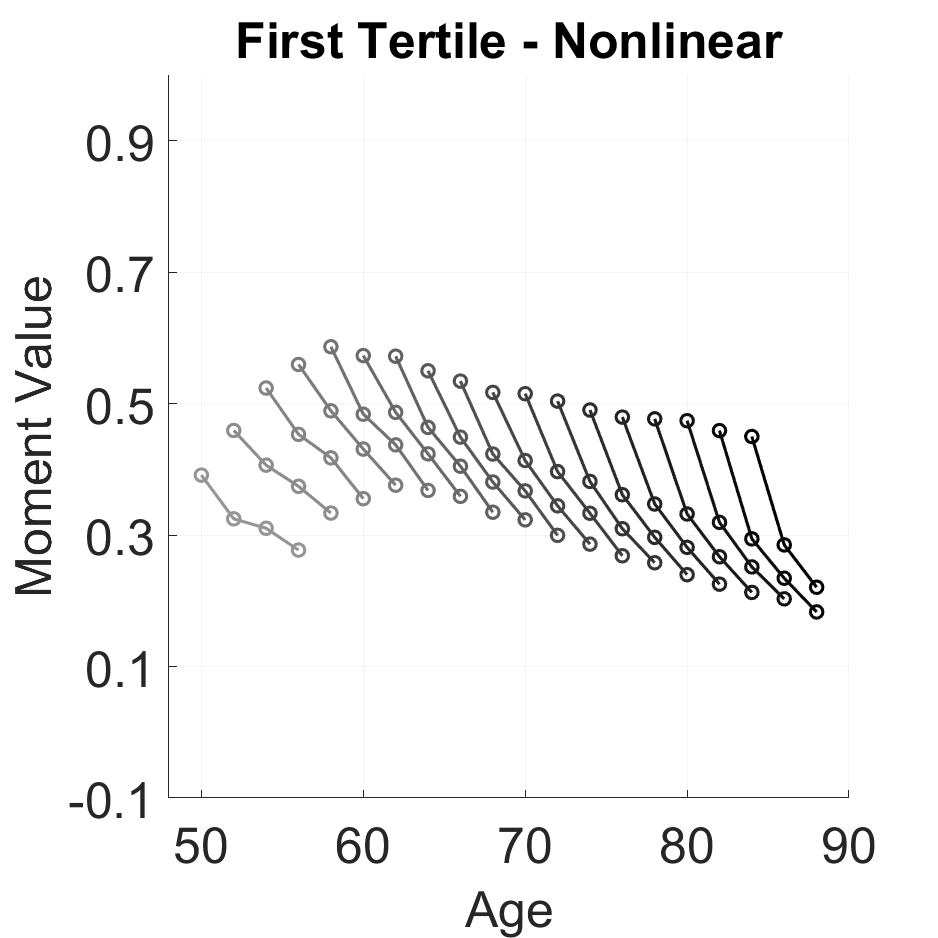}
    \label{fig-cond2aS}
\end{subfigure}
\hfill
\begin{subfigure}{0.32\textwidth}
    \centering
    \includegraphics[width=.95\textwidth]{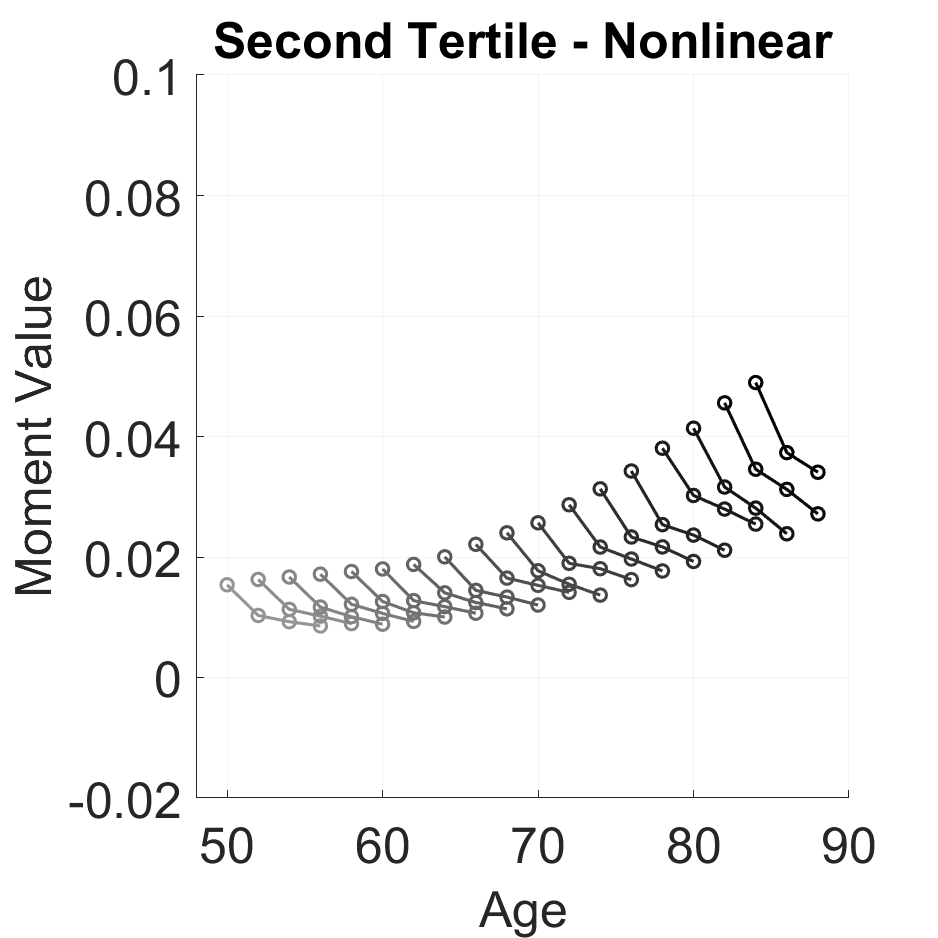}
    \label{fig-cond2bS}
\end{subfigure}
\hfill
\begin{subfigure}{0.32\textwidth}
    \centering
    \includegraphics[width=.95\textwidth]{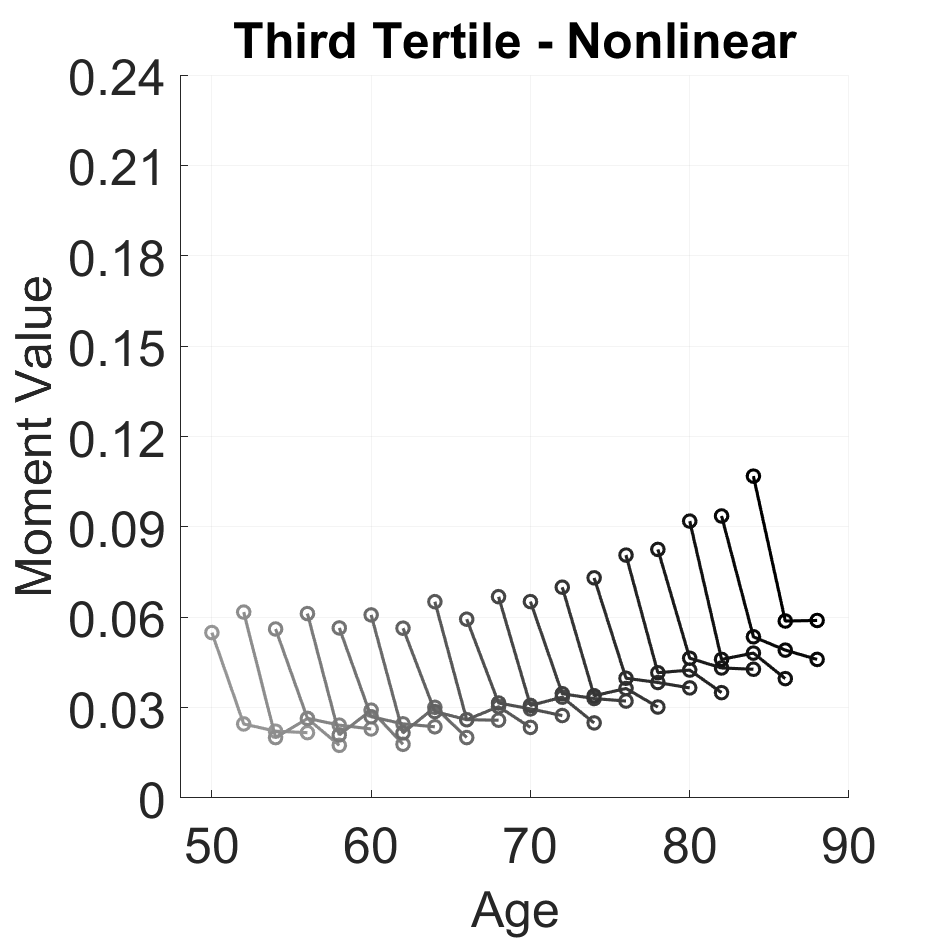}
    \label{fig-cond2cS}
\end{subfigure}
\end{figure}
For individuals in poor initial health, the distribution of shocks is approximately symmetric and close to normal, with a standard deviation of about 0.7. As the initial level of health improves, the dispersion of shocks declines markedly, the distribution becomes increasingly left-skewed, and excess kurtosis rises. These patterns indicate that individuals in good health are exposed to smaller but more asymmetric shocks, with a higher probability of rare but large deteriorations. Age also plays an important role: for older individuals, health shocks are more dispersed, less negatively skewed, and display lower excess kurtosis. Reassuringly, very similar patterns emerge when health shocks are computed using the frailty index rather than the \hindex (see Figure \ref{fig-1nonlinearF} in External Appendix \ref{sec:app-health}).

Beyond the marginal distribution of shocks, we study second-order moments at different lags, allowing them to vary with age and with health status in the previous period. Figure \ref{fig-cond2} reports variances and covariances of health shocks conditional on age and on tertiles of health in $t-1$ (unconditional moments are shown in Figure \ref{fig-uncond2} in External Appendix \ref{sec:app-health}). The conditional moments display substantial heterogeneity across the health distribution. In particular, covariances increase monotonically with age when conditioning on the second and third tertiles of initial health, while they exhibit a hump-shaped profile for individuals in the lowest tertile. This evidence points to systematic differences in the persistence and accumulation of health shocks depending on both age and prior health status.

A further key feature of health dynamics emphasized in the literature is the high degree of persistence. In the case of SRH, persistence manifests itself in strong state dependence: individuals reporting poor health at time $t$ are very likely to report poor health again at $t+1$. While persistence in SRH can be directly assessed using transition matrices that allow transition probabilities to vary with initial health,\footnote{Using SRH, \citet{de2022lifetime} document strong lag dependence in both poor and good health realizations.} for continuous health measures persistence is typically summarized by an autoregressive coefficient. Using this approach, \citet{hosseini2021evolution} estimate a persistence parameter of about 0.99 for log(frailty) using US-PSID data, \citet{dalbianco2022} find a value of 0.97 for England (ELSA), and \citet{blundell2016dynamic} report estimates ranging from 0.90 to 1.06 for England (ELSA) and from 0.89 to 0.97 for the US (using Health and Retirement Study data).

Borrowing from the earnings dynamics literature, we also investigate whether persistence varies across the health distribution, we find that persistence is higher for individuals in worse health and lower for those in better health (see Figure \ref{fig-surface}, left panel). This pattern reinforces the evidence from the moments of health shocks, suggesting that a linear and homoskedastic specification is unlikely to capture the true dynamics of health over the life cycle.

Taken together, these findings indicate that health follows a highly nonlinear process characterized by strong persistence, state dependence, and pronounced age effects. These features motivate our choice of the nonlinear health process as the benchmark specification, against which linear alternatives are evaluated in the structural analysis.

\subsection{Modeling health dynamics}\label{sec:dynamic}
Motivated by strong persistence, age dependence, and nonlinearities, we model residual health as:
\begin{align}
h_{it} = \eta_{it} + \zeta_i + \varepsilon_{it},
\qquad i=1,\ldots,N,\quad t=1,\ldots,T,
\label{eq:health}
\end{align}
where $\eta_{it}$ is a persistent component following a first-order Markov process, $\zeta_i$ is time-invariant unobserved heterogeneity, and $\varepsilon_{it}$ is a transitory shock with zero mean, independent of $\eta_{it}$, $\zeta_i$, and past realizations.

\paragraph{Quantile-based specification of health dynamics.}
We estimate the dynamics of $\eta_{it}$ using the quantile-based panel approach of \citet{ABB2017}. Let $Q_x(\tau \mid \cdot)$ denote the conditional $\tau$-th quantile of $x$. The evolution of the persistent component is written as:
\begin{equation}
\eta_{it} = Q_{\eta}(u_{it} \mid \eta_{i,t-1}, t),
\qquad u_{it} \sim \text{Uniform}(0,1), \quad t>1.
\label{quantilefunc}
\end{equation}
This formulation allows the effect of a health realization to depend both on the individual's previous health status $\eta_{i,t-1}$ and on the rank of the innovation $u_{it}$.\footnote{The uniform distribution assumption is without loss of generality in this framework: since $u_{it}$ represents the rank of the shock within its conditional distribution, any distributional shape is accommodated through the conditional quantile function $Q_{\eta}(\cdot)$.}

Within this framework, persistence after a realization of rank $\tau$ is:
\begin{equation}
\rho(\tau \mid \eta_{i,t-1}, t)
= \frac{\partial Q_{\eta}(\tau \mid \eta_{i,t-1}, t)}{\partial \eta_{i,t-1}}.
\label{persistencef}
\end{equation}
Persistence is therefore allowed to vary across the health distribution and with age. A special case of (\ref{quantilefunc}) is the linear AR(1) process $\eta_{it} = \rho \eta_{i,t-1} + \nu_{it}$, where persistence is constant and equal to $\rho$.

We model the remaining components of health--the time-invariant fixed effect $\zeta_i$, the initial persistent state $\eta_{i1}$, and the transitory component $\varepsilon_{it}$--using the same quantile-based framework, allowing their conditional distributions to vary with age and (when relevant) with time-invariant heterogeneity.


\paragraph{Comparison to alternative approaches.}\label{sec:dynamic-logar1}An alternative approach to modeling nonlinear health dynamics is to specify the process in logs, as in \citet{hosseini2021evolution, hosseini2024important}. In that framework, health is modeled as a frailty index with a mass point at zero, while positive frailty evolves as the sum of a persistent AR(1) component and a transitory shock in logs. Estimation therefore delivers a single persistence parameter governing the dynamics of log frailty. This log-linear specification provides a parsimonious and elegant way to introduce nonlinear dynamics in levels through a small number of parameters.
While a log specification mechanically implies nonlinear dynamics in levels, this nonlinearity is tightly constrained: state dependence arises indirectly through the transformation. By contrast, our quantile-based approach directly estimates how persistence varies with health, allowing the data to determine whether and where persistence is stronger or weaker.

\begin{figure}[t!]
\centering
 \caption{Changes in health after health shocks with same likelihood}\label{fig:health-count}
\begin{subfigure}{0.48\textwidth}
    \centering
    \caption*{Nonlinear}
    \includegraphics[width=\textwidth]{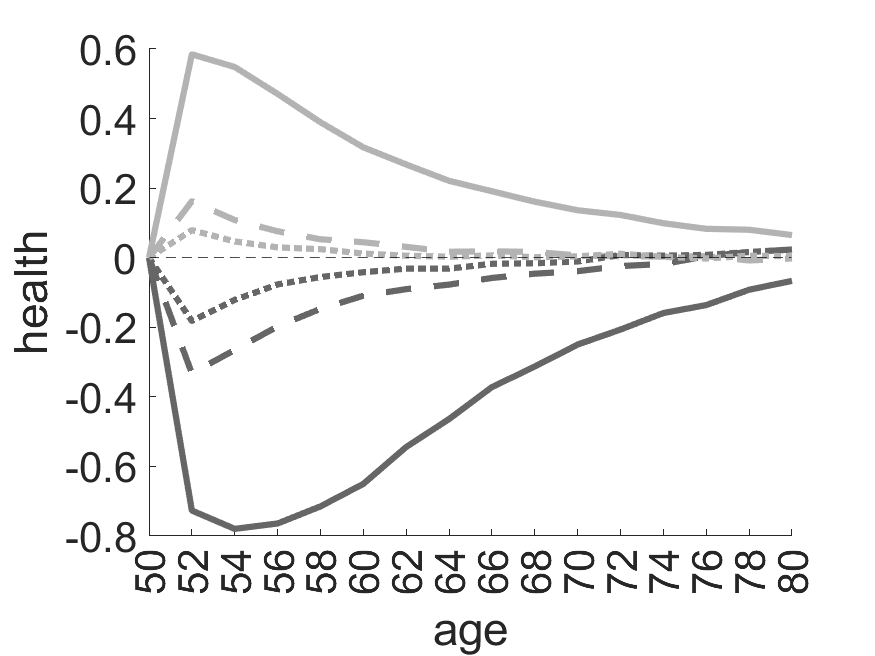}
    \label{fig-nonlinear}
\end{subfigure}
\hfill
\begin{subfigure}{0.48\textwidth}
    \centering
    \caption*{Linear}
    \includegraphics[width=\textwidth]{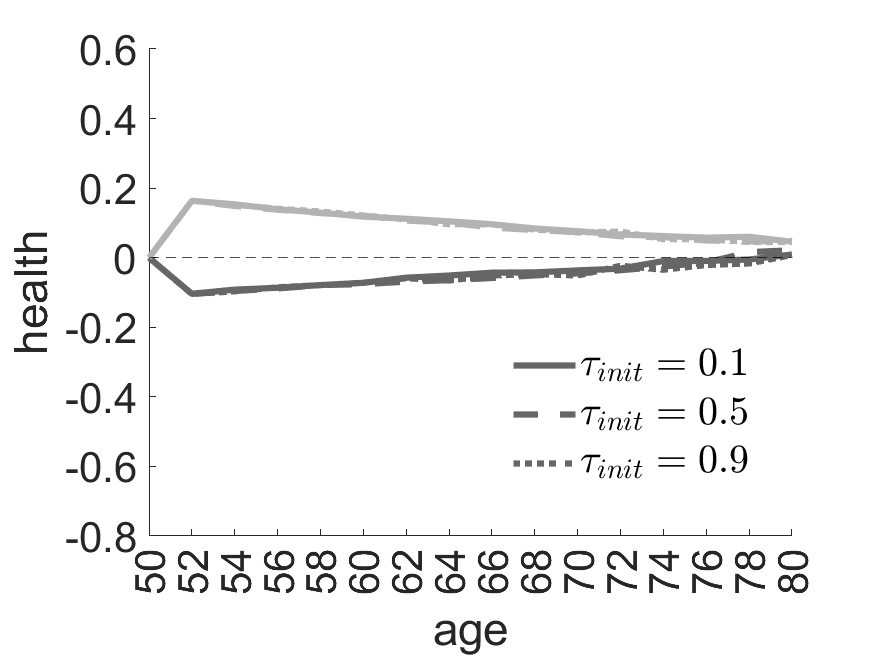}
    \label{fig-linear}
\end{subfigure}
 \caption*{\footnotesize\normalfont \textit{Note:} Age profiles of the difference in health between individuals subject to a permanent component of health shock $\tau_{shock}$ ($\tau_{shock}=0.9$ black lines and $\tau_{shock}=0.1$ grey lines) and individuals subject to  $\tau_{shock}=0.5$, starting from different initial levels of the permanent component ($\tau_{init}$).}
\end{figure}

From this perspective, our contribution is not to introduce nonlinearity per se, but to unpack it. The quantile-based framework makes explicit the state dependence that is implicit in log-linear specifications and allows us to assess its empirical relevance, including asymmetric responses to positive and negative shocks documented in Figure~\ref{fig:health-count}. Moreover, this flexibility makes it straightforward to shut down state dependence and recover linear benchmarks, facilitating transparent comparisons across alternative health processes.
A formal comparison with a log-AR(1) specification would require imposing strong and nontrivial restrictions on the quantile functions to replicate the log structure, and is therefore outside the scope of our estimation framework.

\paragraph{Estimation and numerical implementation.}
We parametrize the conditional quantiles of each health component using products of low-order Hermite polynomials $\psi_\ell$%
\footnote{In practice, the Hermite bases may differ across equations.
For $Q_\eta$, we use a tensor-product basis with order $3$ in lagged $\eta_{i,t-1}$ and order $1$ in age.
For $Q_{\eta_1}$, we use order $1$ in $\zeta_i$ and order $2$ in $age_{i1}$.
For $Q_\varepsilon$ and $Q_\zeta$, we use univariate Hermite bases in age of order $2$ and $1$, respectively.}
\begin{align}
Q_{\eta}(\tau \mid \eta_{i,t-1}, age_{it})
&= \sum_{\ell=0}^{L} a_\ell^{\eta}(\tau)\, \psi_\ell(\eta_{i,t-1}, age_{it}),
\label{eq-Qeta} \\
Q_{\zeta}(\tau \mid age_{i1})
&= \sum_{\ell=0}^{L} a_\ell^{\zeta}(\tau)\, \psi_\ell(age_{i1}),
\label{eq-Qzeta} \\
Q_{\eta_1}(\tau \mid \zeta_i, age_{i1})
&= \sum_{\ell=0}^{L} a_\ell^{\eta_1}(\tau)\, \psi_\ell(\zeta_i, age_{i1}),
\label{eq-Qeta1} \\
Q_{\varepsilon}(\tau \mid age_{it})
&= \sum_{\ell=0}^{L} a_\ell^{\varepsilon}(\tau)\, \psi_\ell(age_{it}).
\label{eq-Qeps}
\end{align}

Estimation proceeds using a stochastic EM–type algorithm for nonlinear panel models with latent states, following \citet{ABB2017} and \citet{arellano2016nonlinear}. The persistent component $\eta_{it}$ and the individual fixed effect $\zeta_i$ are treated as latent variables in estimation, while the transitory component $\varepsilon_{it}$ is modeled through its conditional quantile function and integrated out. At each iteration, the E-step relies on simulation-based draws from the joint posterior distribution of $(\eta_{it},\zeta_i)$, implemented via a Metropolis--Hastings sampler. The M-step updates the conditional quantile functions by solving a sequence of convex quantile regressions, rather than maximizing a parametric likelihood.

Conditional quantiles are approximated using low-order Hermite polynomial bases in lagged health and age. The coefficients of these bases are modeled as piecewise linear splines over a finite grid of quantiles $\tau_1<\dots<\tau_P$ with $P=11$ grid points, as in \citet{ABB2017}. To ensure stability and well-behaved extrapolation beyond the central grid, the intercept terms of the quantile functions are extended in the lower and upper tails using parametric exponential distributions, while slope coefficients are held constant outside the grid.

Given the potential presence of local modes in the objective function, we assess robustness to initialization by experimenting with multiple starting values, varying both the initial variance components and the tail extrapolation parameters. Final estimates correspond to the solution delivering the highest average fit across iterations. Identification of the model requires at least five consecutive observations per individual; we therefore estimate the process on a balanced panel covering six periods.

Because the coefficients of the Hermite polynomial basis and the tail extrapolation parameters depend on normalization choices and have no direct structural interpretation, we do not report them individually. Model implications are instead evaluated through implied persistence surfaces, conditional moments, and simulated impulse responses, which are the economically relevant objects of interest.

\paragraph{Linear versus nonlinear dynamics.}
To assess the role of nonlinearities, we compare the flexible specification above (the \emph{nonlinear model}) to a restricted state-invariant \emph{linear model} in which persistence depends only on age:%
\footnote{\label{fn:linear-terminology}We adopt the ``linear/nonlinear" terminology following \citet{ABB2017} and \citet{de2020nonlinear} in the context of earnings dynamics. In this classification, ``linear" denotes state-invariant persistence (i.e., $\rho(\tau | \eta_{i,t-1}, t) = \rho_t$), while ``nonlinear" indicates state-dependent persistence (i.e., $\rho$ follows eq. \ref{persistencef}), regardless of functional transformations applied to the variable itself.}
\[
\rho(\tau \mid \eta_{i,t-1}, t) = \rho_t.
\]
In the linear model, persistence is time-varying but independent of both the previous health realization and the magnitude of the shock.%
\footnote{The estimated persistence in the \emph{linear} specification is 0.92 showing little variation with age.}

\begin{figure}[t]
\caption{Health distribution: data and model simulations}
    \label{fig:alldensities}
    \centering
    \includegraphics[width=0.8\textwidth]{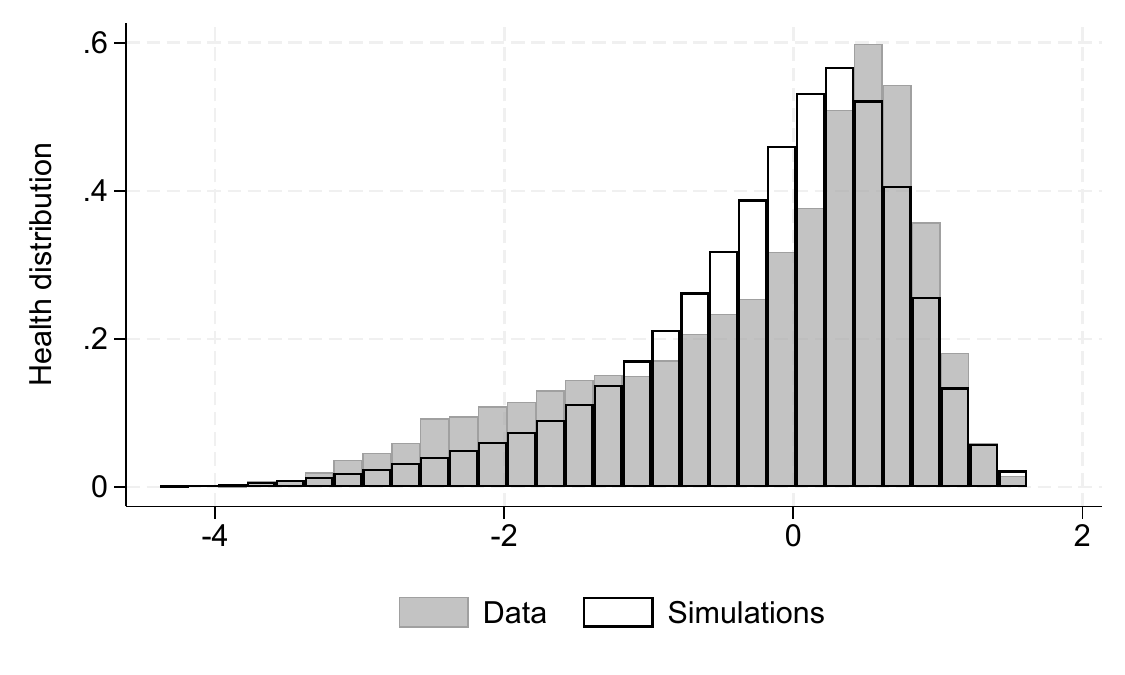}
\caption*{\footnotesize\normalfont \emph{Note:} Distributions of \hindex (Data), and nonlinear model simulated data (Simulations)}
\end{figure}%

\paragraph{Model fit.}
Figure~\ref{fig:alldensities} shows that the nonlinear model closely matches the cross-sectional distribution of \hindex. Figures~\ref{fig-1nonlinear} and~\ref{fig-cond2} further show that the nonlinear specification provides a good fit to the dynamics of health shocks: simulations closely reproduce the conditional moments observed in the data, including the decline in dispersion across previous health deciles, the asymmetric shape of shocks, and the age profiles of conditional variances and covariances across health tertiles.

For comparison, Appendix Figures~\ref{fig-1linear} and~\ref{fig-cond2L} report the corresponding moments implied by the linear specification. While the linear model delivers a similar fit to cross-sectional distributions (not shown), it performs worse in the dynamics. In particular, it understates the decline in dispersion with health in $t-1$, generates excessive skewness at low health levels, and produces lower kurtosis throughout the distribution. The largest discrepancies arise in second-order dynamics: the linear specification fails to reproduce the strong heteroskedasticity and age-dependent comovement of shocks observed in the data, especially for individuals in the lowest health tertile.

Overall, these results indicate that nonlinear health dynamics are essential to match not only the marginal distribution of health shocks, but also their conditional and joint behavior over the life cycle.

\begin{figure}[t]
\centering
 \caption{Persistence, nonlinear model}\label{fig:persistence-2}
\begin{subfigure}{0.48\textwidth}
    \centering
    \caption*{Data}
    \includegraphics[width=\textwidth]{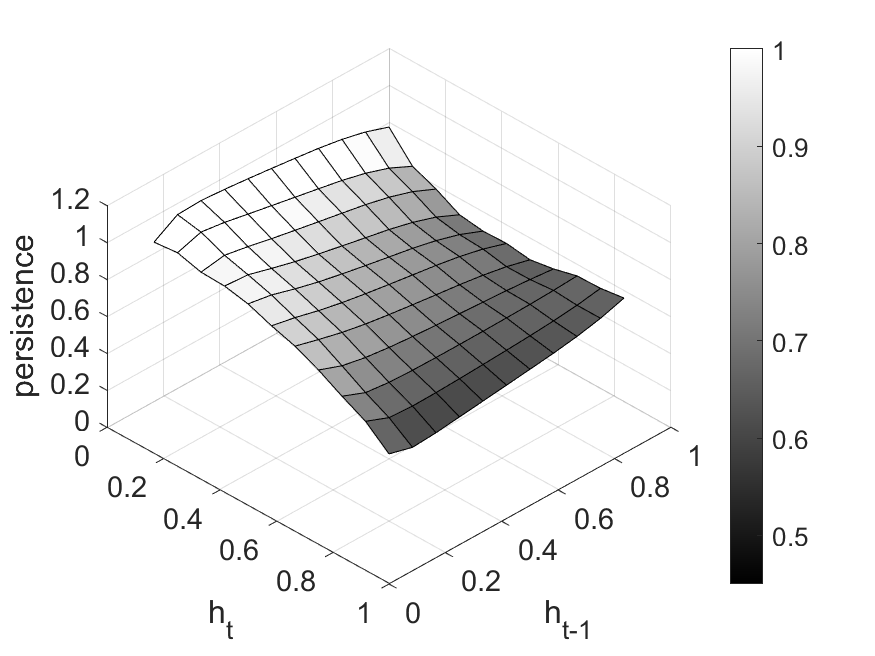}
    \label{fig-surData}
\end{subfigure}
\hfill
\begin{subfigure}{0.48\textwidth}
    \centering
    \caption*{Simulation}
    \includegraphics[width=\textwidth]{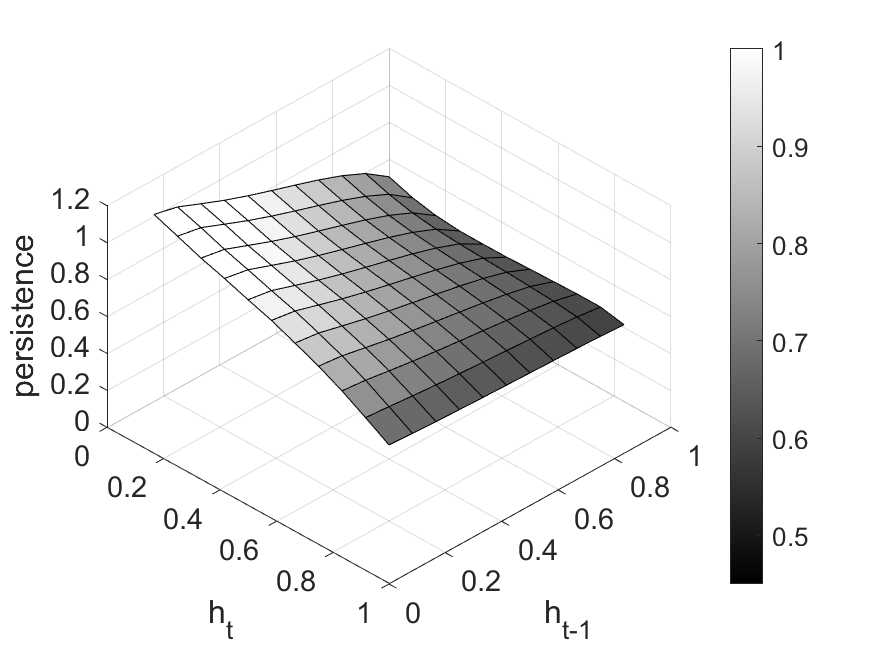}
    \label{fig-surSim}
\end{subfigure}
\label{fig-surface}
\caption*{\footnotesize\normalfont \textit{Note:} Quantile autoregressions of \hindex. Data (left panel) and Simulations (right panel). Residuals $h_{it}$ of \hindex. Estimates of the average derivative of the conditional quantile function of $h_{it}$ given $h_{i,t-1}$ with respect to $h_{i,t-1}$. 

Quantile functions are specified as third--order Hermite polynomials. 
A similar pattern emerges using a piecewise--linear specification quantile regression.}
\end{figure}

\paragraph{Persistence and impulse responses.}
Figure~\ref{fig:persistence-2} reports average persistence as a function of current and lagged health deciles, computed from quantile autoregressions from the data, and from data simulated after estimating the nonlinear model. The nonlinear model reproduces the pattern in the data, with persistence ranging from  0.5 to 1.1. Persistence is strongly state dependent in poor health: when persistence is high, current realizations are largely driven by past health rather than by contemporaneous innovations. In good health, the weight on past health is lower and varies less across current realizations.

To illustrate the implications of state-dependent persistence, we simulate health histories starting from the 10th, 50th, and 90th percentiles of the persistent health component at age 50. At age 52, individuals are exposed to a bad, median, or good health shock, each occurring with the same likelihood under both the linear and nonlinear processes. These shocks place individuals at the 10th, 50th, and 90th percentiles of the health distribution conditional on their initial persistent component.

Differences arise solely from the realization at age 52, which moves the persistent component to different post-shock states.
Figure~\ref{fig:health-count} reports impulse responses relative to the median realization.

Two margins distinguish nonlinear from linear dynamics.
First, shocks of the same likelihood differ in \emph{magnitude}. In the nonlinear model, adverse realizations are larger---they move individuals farther below the median path on impact, especially when starting from poor health. In the linear specification, shocks of the same likelihood generate smaller initial deviations.

Second, persistence is \emph{state dependent}. When an adverse realization places individuals in low-health states where persistence is high, the subsequent path remains anchored near the post-shock level and converges slowly. This interaction between larger initial deteriorations and higher persistence is absent under the linear model, which imposes a flat persistence surface.

Together, these impulse responses show that assessing health dynamics requires considering both the size of shocks and how persistence varies across states. This motivates embedding the estimated health process in the life-cycle model, where the economic relevance of health shocks depends on how often individuals enter, and how long they remain, in vulnerable low-health states.
\label{sec:dynamic-end}
\section{The life-cycle model}\label{sec:model}

We model the life-cycle behavior of low-educated men living in a couple from age 50 onward. Individuals choose consumption and saving, labor supply, and DI under health, earnings, and survival risk. Time is discrete in two-year periods. Agents enter the model at age 50 and face uncertainty over health, labor income, DI eligibility, and survival until a terminal age of 90.

Health affects preferences (the disutility of work and leisure), earnings, survival probabilities, and the likelihood of receiving DI. Through these channels, health dynamics generate heterogeneity in labor supply, income, and wealth accumulation over the later life cycle.

In each period, individuals choose consumption, asset accumulation, and whether to work. Before 65, the state pension age, they may also apply for DI. Survival to the next period is uncertain and depends on age and health. If death occurs, individuals derive utility from bequest.

We solve the model recursively and use it to compute simulated life-cycle histories to study the implications of alternative health processes and counterfactual policies for behavior and welfare.

\paragraph{Preferences.}

Individuals derive utility from consumption and leisure while alive. Period utility is given by a CRRA specification over a Cobb--Douglas aggregator:
\begin{equation}
U(c_t,l_t)=\frac{1}{1-\nu}\left(c_t^\gamma l_t^{1-\gamma}\right)^{1-\nu} + \bar{b},
\label{eq:utility}
\end{equation}
where $\nu$ denotes relative risk aversion and $\gamma$ governs the weight on consumption. Leisure $l_t$ is determined by labor supply and health-related time costs.
Following \citet{hall2007value} and \citet{de2022lifetime}, we include a positive constant $\bar{b}$ to ensure that continuation utility from being alive always exceeds the value of death, independently of consumption or health.
Upon death, individuals derive utility from bequest:
\begin{equation}
b(a_t)=\phi_B \frac{(a_t + K)^{(1-\nu)\gamma}}{1-\nu},
\label{eq:bequest}
\end{equation}
where $\phi_B$ governs the strength of bequest motives and $K>0$ ensures finite utility at zero bequests.

\paragraph{Time allocation and labor supply.}

Before age 70, individuals may choose to work,  $w_t \in \{0,1\}$.
Each period, agents are endowed with one unit of time. Working and poor health reduce available leisure according to:
\begin{equation}
l_t = 1 - \phi_w(t) w_t - \phi_h(h_t),
\label{eq:time}
\end{equation}
where $\phi_w(t)$ captures the age-dependent disutility of work and $\phi_h(h_t)$ health-related disutility.
Through this channel, poor health increases the utility cost of working by reducing available leisure, over and above its effects on earnings and survival.

The disutility of work varies flexibly with age:
\begin{equation}
\phi_w(t)=\phi_{w0} + \phi_{w1}\left(\frac{t}{\max(t)}\right)^{\phi_{w2}},
\end{equation}
while the health-related time cost is specified as:
\begin{equation}
\phi_h(h_t)=\phi_h \frac{\bar{h}-h_t}{\bar{h}-\underline{h}},
\end{equation}
so that individuals in maximum health $\bar{h}$ face no time cost, and individuals in minimum health face cost $\phi_h$.

\paragraph{Earnings and income.}
\label{sec:model-earnings}

When working, individuals earn labor income $e_t$:
\begin{equation}
\log e_t = f_i + \omega_e(h_t,t) + \vartheta_t + \upsilon_t,
\label{eq:wage}
\end{equation}
where $\omega_e(h_t,t)$ captures the effect of age and health on productivity, $f_i$ is an individual fixed effect, and $\vartheta_t$ and $\upsilon_t$ are persistent and transitory shocks.
The persistent component follows a random walk,
\begin{equation}
\vartheta_t = \vartheta_{t-1} + \nu^e_t, \qquad \nu^e_t \sim \mathcal{N}(0,\sigma^2_{\nu^e}),
\end{equation}
while $\upsilon_t \sim \mathcal{N}(0,\sigma^2_{\upsilon^e})$. Individuals observe current health and productivity when making decisions. After age 68, labor supply and earnings risk are absent.

Earnings process parameters are estimated outside the structural model, accounting for endogenous participation following \citet{low2015disability}. Details are provided in External Appendix \ref{sec:app-wage}.

\paragraph{Health in the model.}

Health is measured by the continuous residual \hindex constructed in Section~\ref{sec:hmeasure}. Individuals observe current health $h_t$, which affects preferences, earnings, survival, and DI eligibility.

Health evolves exogenously according to the stochastic process estimated in Section~\ref{sec:dynamic}. The baseline specification is nonlinear; a linear specification is used for comparison. To embed health into the discrete-state life-cycle problem, we discretize the estimated process and construct age-specific transition matrices from simulated health histories.\footnote{Cohort, partnership status, and education are fixed to match the representative individual in the life-cycle model (see Section \ref{sec:sample-selection}). Estimating the health process on this restricted subsample--low-educated males, having a partner, and born in 1948-1952--is infeasible due to sample size.} Details are in External Appendix~\ref{sec:discretization}.

\paragraph{Disability insurance.}
\label{sec:model-di}

Before the state pension age, individuals may apply for DI. DI provides a flat benefit and insures against earnings losses due to poor health.\footnote{DI rules are based on the UK Incapacity Benefit program in force between 1995 and 2008.}
Between ages 50 and 64, eligible individuals decide whether to apply. Acceptance is random, with the probability decreasing with health $\psi_d(h_t)$. Accepted applicants receive DI and cannot work while enrolled. Once on DI, individuals may continue receiving benefits without reapplying, capturing persistence in DI receipt. They may also choose to exit DI and start working.

When applying, individuals evaluate expected utility by comparing the acceptance-weighted value of DI to alternative labor supply options, including working if the application is rejected. See Appendix \ref{sec:app-bellman} for details.

\paragraph{Budget constraint and transfers.}

Assets evolve according to:
\begin{equation}
a_{t+1} = (1+r)a_t + y_t - tax_t + tr_t - c_t,
\label{eq:budget}
\end{equation}
where income $y_t$ depends on labor supply, DI status, and age. Pension wealth accumulates through contributions while working and is annuitized at retirement.
A consumption floor $\underline{c}$ is imposed through means-tested transfers, ensuring that consumption does not fall below $\underline{c}$ in any period.
The model abstracts from out-of-pocket medical expenditures, consistent with the UK institutional context, where healthcare is publicly provided.\footnote{See Figure \ref{fig:medicalexp} in the External Appendix for evidence on the limited role of medical expenditures in the UK.}

\paragraph{Survival, pensions, and timing.}

Survival from $t$ to $t+1$ occurs with probability $\pi^{t+1}=\pi(h_t,t)$, which depends on age and health. Individuals face a terminal age of 90. Survival probabilities are constructed using ELSA data linked to death records and matched to life tables; details are in External Appendix~\ref{sec:app-mortality}.

Importantly, the health process is estimated using an iterative procedure that ensures consistency between health dynamics and survival, thereby accounting for selective mortality. In particular, the simulated health distribution among survivors matches the age profile observed in the data once mortality is applied. Details of this selection correction are provided in External Appendix~\ref{sec:app-mortality_sel}.\label{ref:selection-mortality}

From the state pension age onward, individuals receive pension income and no longer face earnings risk.

The state vector at the beginning of period $t$ is
\[
X_t = \{a_t, h_t, \vartheta_t, p_t, di_{t-1}\}.
\]
    Given $X_t$, individuals choose consumption, saving, labor supply, and DI decisions to maximize expected discounted (at rate $\beta$) utility. The full recursive problem is presented in Appendix~\ref{sec:app-bellman}.

\subsection{Calibration and estimation}\label{sec:model-estimation}

\paragraph{Sample and targeted moments.}
\label{sec:sample-selection}
We estimate the life-cycle model using data from ELSA, focusing on low-educated male respondents living with a partner. We restrict attention to men who left education at the compulsory school-leaving age. This group is the most likely to claim DI and is the primary target of disability programs in the UK, which provide relatively limited insurance and disproportionately cover individuals with weaker labor market opportunities. \label{sec:dynamic-controls} We condition on partnership status to study a homogeneous and highly represented group in the data and to abstract from endogenous household formation.\footnote{Appendix Table \ref{tab:descriptives} shows that among male respondents in ELSA wave 1--7, 82\% have a partner and 45\% are low-educated. DI is 13.8\% among low-educated and it is 4\% among high-educated individuals.}

Unless otherwise stated, empirical moments are adjusted for cohort effects following the procedure described in External Appendix~\ref{sec:app-momest}. Profiles by age (and health) are estimated using the full sample, including individuals with different education levels and partnership status, in order to reduce sampling noise. These profiles are then evaluated for low-educated men living with a partner born between 1948 and 1952, which is the group targeted in the structural estimation to construct initial conditions and to simulate life-cycle outcomes.

We estimate the model by matching (i) average assets by age, (ii) labor force participation by age and health status (using thresholds at the 20th, 30th, and 50th percentiles of the health distribution)\footnote{The thresholds used to discretize health are reported in Table~\ref{tab:healthperc} in the External Appendix.}, and (iii) the fraction receiving DI by age.

\paragraph{Externally set parameters.}
A subset of parameters is fixed to values commonly used in the literature or directly observed in the data. The yearly consumption floor is set to $\underline{c}=\pounds 1{,}660$, corresponding to 10\% of average male earnings in the sample, as in \citet{capatina2015life}. Relative risk aversion is fixed at $\nu=3$. The annual discount factor is set to $\sqrt{\beta}=0.95$, in the range of estimates from \citet{gourinchas2002consumption} and \citet{cagetti2003wealth}.

Institutional parameters are calibrated using UK data. The pension contribution rate is set to $c_p=6\%$, the average contribution rate to defined-contribution pension schemes. Pension wealth is annuitized at a constant rate $r_p=3.94\%$, computed as the median actuarially fair annuity rate among individuals aged 55--65 in the data. Taxation follows the UK 2003/04 tax schedule and is described in External Appendix~\ref{sec:tax}.

Finally, we calibrate the constant term $\bar b$ in period utility to match a target value of a statistical life (VSL). Specifically, $\bar b$ is chosen so that the average model-implied VSL among working-age individuals matches a UK policy benchmark, corresponding to a Value of a Prevented Fatality of approximately \pounds 900{,}000. Details on the definition, computation, and calibration of the VSL are provided in External Appendix~\ref{sec:svl}.

\paragraph{Initial conditions.}
\label{sec:initial-conditions}
The model is initialized at age 50. Initial health states are drawn from the empirical distribution of the residual \hindex at age 50, using observed health histories. Conditional on initial health, we initialize the remaining state variables using empirical conditional distributions estimated from the data.

Specifically, initial assets, pension wealth, and DI status are drawn from their distributions conditional on health at age 50. This procedure preserves the observed cross-sectional heterogeneity in economic resources and disability participation by health at model entry.

Initial earnings offers are constructed by combining the deterministic component of the earnings equation at age 50 with a draw of the persistent earnings component. This initialization ensures that the joint distribution of health, wealth, earnings capacity, and DI status at model entry closely matches that observed in the data.
\label{sec:initial-conditions-end}

\begin{table}[t]

    \centering
    \caption{Nonlinear model parameter estimates}
    \label{tab:estimates}
    \small
    
\begin{tabular}{llrllr}

  \multicolumn{3}{c}{Estimated} & \multicolumn{3}{c}{Calibrated} \\
  \cmidrule(lr){1-3} \cmidrule(lr){4-6}
  Cost of work & $\phi_w^0$ & 0.7 & Discount factor (biennial) & $\beta$ & 0.9 \\
   & $\phi_w^1$ & 1.38 & Risk aversion & $\nu$ & 3 \\
   & $\phi_w^2$ & 3.61 & Interest rate & $r$ & 0.029 \\
  Cost of health & $\phi_{h}$ & 0.25 & Consumption floor (biennial) & $\underbar{c}$ & 3320 \\
  Disability prob. & $\psi_d^1$ & 0.98 & Pension annuity rate & $p_r$ & 0.0378 \\
   & $\psi_d^2$ & 0.13 & Pension contribution rate & $c_w$ & 0.06 \\
   & $\psi_d^3$ & 0.03 &  &  &  \\
  Consumption weight & $\gamma$ & 0.47 &  &  &  \\
  Bequest motive & $\phi_B$ & 2098 &  &  &  \\
   & $K$ & 329350 &  &  &  \\

\end{tabular}

\end{table}

\paragraph{Estimated parameters and simulated method of moments.}
The remaining preference and cost parameters are estimated using a Simulated Method of Moments. For each candidate parameter vector, we solve the model and simulate 30{,}000 life-cycle histories. We minimize the distance between simulated and empirical moments using a simulated annealing routine (basinhopping). For the local minimizer, we employ BOBYQA, a derivative-free bound optimization algorithm designed for noisy objective functions. We also experimented with the Nelder-Mead simplex method as an alternative local minimizer.
Parameter estimates are reported in Table~\ref{tab:estimates}.

\paragraph{Identification.}
\label{sec:identification}
Labor force participation by age and health identifies the disutility of work and the health-related time cost, $\phi_w(\cdot)$ and $\phi_h(\cdot)$. Variation in participation across health states isolates the health-related time cost, while age profiles of labor supply identify the age-dependent component of work disutility. Participation behavior at older ages is particularly informative: because DI is no longer available after age 65, labor supply responses to health beyond this age identify the preference-based cost of working in poor health separately from DI incentives.

The consumption weight $\gamma$ is identified from the joint behavior of labor supply and asset accumulation over the life cycle. This parameter governs the trade-off between leisure and consumption in the presence of health and income uncertainty, and therefore affects how individuals adjust work and saving as health deteriorates with age.

Conditional on preferences, the health dependence of DI eligibility, $\psi_d(h_t)$, is jointly identified from the age profile of DI receipt and labor supply responses to health. Although DI participation by health is not directly targeted, the model must simultaneously match the sharp increase in DI receipt at older ages and the decline in labor supply with worsening health. This joint restriction affects the slope of $\psi_d(\cdot)$, as steeper health dependence would generate excessive DI participation and insufficient labor supply among individuals in poor health.

Average assets by age identify the strength of precautionary saving motives in the presence of health and earnings risk. These moments discipline the extent to which disability insurance substitutes for self-insurance through savings.

Bequest motives are identified using a combination of calibration and estimation. Following the literature, we fix the marginal propensity to bequeath out of an additional unit of wealth to 0.98, which corresponds to the marginal propensity to bequeath implied by the estimates in \citet{french2005effects}. Conditional on this calibration, the parameter $K$, which governs the level of wealth at which bequest motives become operative, is identified from the level and slope of asset decumulation at older ages. Because assets are targeted only up to age 75 and mortality is stochastic, fixing the marginal propensity to bequeath improves stability and interpretability of the estimates.

\paragraph{Model fit and validation.}
\label{sec:model-fit}
Fixing labor supply, the consumption weight $\gamma$'s estimate implies a coefficient of relative risk aversion for consumption of 1.93 ($\gamma(\nu-1)+1$), which falls within the range of previous estimates in the literature (e.g. \citet{french2005effects} and \citet{capatina2015life}). To compare the estimated values for the cost of bad health with previous estimates in the literature, we compute the average value of our \hindex when self-reported binary health is classified as fair/poor (-1.02). The corresponding time cost is 0.11. Being in bad health entails a 11\% reduction in the time endowment (increasing to 25\% for the lowest health realization). The same figure is 21\% in \citet{dalbianco2022}, 12\% in  \citet{french2011effects} and 14-21\% in \citet{capatina2015life}.

The estimated parameters imply that the bequest motive becomes operative when the consumption value of total wealth exceeds \pounds 6{,}532. This value is aligned with those commonly used in the literature and generate asset decumulation patterns consistent with the data. Conditional on the calibrated strength of bequest motives, the estimated threshold helps reproduce the observed slowdown in asset drawdown at older ages.

Finally, the disutility of work $\phi_w(t)$ at age 50 is 0.7, reaching 0.82 at age 68. The size of the effect is comparable with the estimates in \citet{hosseini2024important}, who estimate a baseline disutility of work of 0.62, increasing with frailty.

\begin{figure}[t]
    \centering
    \caption{Estimation fit: targeted moments} \label{fig:fit}
                \includegraphics[width=.85\linewidth]{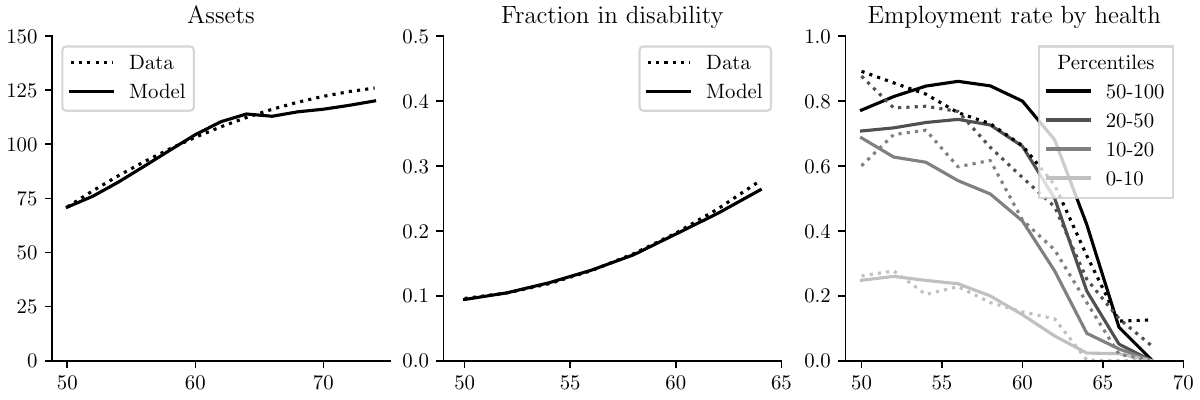}
    \caption*{\footnotesize\normalfont \textit{Note:} \targetedcaption}
\end{figure}

Figure~\ref{fig:fit} shows that the model reproduces well the targeted age profiles of average assets, labor force participation by health status, and DI receipt. Despite its parsimonious structure, the model also matches salient non-targeted patterns, including asset profiles by health and DI inflow and outflow rates (see Appendix Figure~\ref{fig:fit_nottarget}).

Importantly, the model captures a substantial share of cross-sectional dispersion in economic outcomes. In Appendix Figure \ref{fig:CVassetfit}, we show that the model reproduces the age profile of the coefficient of variation of assets reasonably well, although it somewhat understates dispersion at older ages. The model also generates a lower dispersion of earnings than observed in the data: the standard deviation of annual earnings is about \pounds 6{,}000 in the model, compared with roughly \pounds 11{,}000 in the data. This gap partly reflects the absence of part-time work in the model, as well as other intensive-margin adjustments. These moments are not directly targeted in the estimation but validate the model's ability to account for inequality over the life cycle.

We also estimate the model under a linear health process.\footnote{Parameter estimates are in Appendix Table \ref{tab:estimates-lll}.} While both specifications fit the targeted moments comparably well (see Figure \ref{fig:fit_lin} and \ref{fig:fit_lin_nottarget}), they generate different implications for health dynamics and economic behavior, reflecting differences in persistence patterns. We explore these differences in the next section.

\section{The effects of nonlinear health dynamics}\label{sec:results}

This section studies how the specification of health dynamics shapes economic outcomes over the later life cycle. We organize the analysis around a sequence of complementary exercises, each designed to isolate a different margin through which health affects behavior, inequality, and welfare. We simulate 30,000 life histories per experiment.

We begin by comparing life-cycle outcomes and their dispersion across alternative health process specifications. This comparison illustrates how simplifying assumptions about health dynamics affect economic predictions, and sets the stage for distinguishing between purely mechanical effects and the role of parameter re-estimation.

We then quantify the overall burden of realized bad health by comparing the baseline economy to a counterfactual in which all individuals experience persistently good health. This exercise removes realized health risk while keeping behavior, institutions, and preferences unchanged, providing an upper bound on the economic and welfare costs of adverse health realizations.

Next, we study impulse responses to one-time health shocks to isolate the marginal effects of innovations to the persistent component of health and to characterize state dependence and asymmetry in dynamic responses. Finally, we assess the insurance value of DI, both without fiscal adjustment and under revenue-neutral reforms, highlighting how nonlinear health dynamics interact with public insurance design.

Throughout the results section, we use two complementary welfare measures, depending on the experiment. When the policy environment and decision rules are held fixed and only realized health histories differ--such as in the analysis of realized health shocks--we measure welfare using a compensated-equivalent variation based on realized lifetime utility, following \citet{de2022lifetime}. In this setting, a standard ex-ante consumption-equivalent variation (CEV) is not meaningful, as expected lifetime utility is unchanged and welfare differences reflect only ex-post realizations of health risk. In contrast, when the experiment changes the economic environment or agents' optimal decisions--such as in the decomposition exercise or the removal of DI--we report the CEV computed from expected lifetime utility at age 50.\footnote{External Appendix \ref{sec:app-CEV} shows how we compute the CEV when survival uncertainty differs in the baseline and counterfactual scenarios, following \citet{dalbiancoetal2025}.}

\subsection{Comparing model specifications: aggregate outcomes and inequality}
\label{subsec:agg}

To assess the quantitative role of health dynamics for economic predictions, and to implement the comparison outlined above, we compare four simulated economies that differ only in how health evolves after age 50. The \textit{Nonlinear} model is our baseline and features state-dependent health dynamics with non-Gaussian innovations.

We then consider two alternative specifications that restrict the health process while keeping the rest of the  environment unchanged. In the \textit{Linear} model, we re-estimate the health process imposing state-invariant persistence, i.e. persistence varies with age but is independent of lagged health and the rank of the innovation. All other components of the model--including preferences, survival, and the mapping from health into labor supply, earnings capacity, and DI eligibility--are held fixed at their baseline values.
In the \textit{Normal} specification, we further restrict the health process by assuming that innovations to the persistent component are Gaussian, with variance matched to the nonlinear estimate. As in the linear case, all non-health parameters are kept fixed.

Because these alternative specifications differ from the baseline only in the restrictions imposed on health dynamics, the resulting counterfactuals isolate the mechanical implications of flattening persistence or imposing normality on health innovations. They are not intended to fit the data, but to assess how simplifying assumptions about health dynamics affect economic predictions.

Finally, \textit{Linear--Estimated} re-estimates the model under linear health dynamics and simulates the resulting economy using the same linear structure. This exercise captures the outcome of estimating and simulating a misspecified model, allowing parameters to partially compensate for the restricted health process. Comparing \textit{Linear--Estimated} to the baseline therefore quantifies the \emph{irreducible} implications of imposing linear health dynamics for inference and policy analysis.

In all simulations, individuals are initialized at age 50 from the same joint distribution of state variables as in the baseline model, including health, assets, and unobserved heterogeneity. Alternative health dynamics operate only after age 50 -- differences across specifications arise from the evolution of health, not from initial conditions.

\begin{table}[t!]
    \centering
    \caption{Life-Cycle Outcomes: nonlinear vs. linear health dynamics}
    \begin{tabular*}{0.8\textwidth}{@{\extracolsep{\fill}}l cccc}
        & Assets & Cumulated & Work & DI rate \\
        & at 70  & earnings  & & \\
        \textbf {All} &&&& \\ \midrule
        
Nonlinear   &      112618&      12.661&       0.518&       0.162\\
Linear      &       -2617&      -0.022&      -0.016&       0.014\\
Normal      &       -4415&      -0.043&      -0.030&       0.026\\
Linear--Estimated&        5417&      -0.018&      -0.013&       0.004\\
 \\
        \textbf {Low health} &&&& \\ \midrule
        
Nonlinear   &       94012&      12.636&       0.471&       0.319\\
Linear      &       -6014&      -0.052&      -0.032&       0.025\\
Normal      &       -7185&      -0.083&      -0.050&       0.053\\
Linear--Estimated&        -342&      -0.049&      -0.031&       0.012\\
 \\
        \textbf {Low health and wealth} &&&& \\ \midrule
        
Nonlinear   &       37870&       0.524&       0.509&       0.040\\
Linear      &       -2463&       0.016&       0.003&       0.000\\
Normal      &       -1732&       0.040&       0.006&       0.001\\
Linear--Estimated&       -1877&      -0.023&      -0.011&      -0.001\\
 \\
    \end{tabular*}
    \caption*{\footnotesize\normalfont \textit{Note:} The table reports baseline levels under the \textit{Nonlinear} model (first row in each panel) and deviations under alternative specifications: \textit{Linear} and \textit{Normal} modify the health transition in simulation holding all estimated parameters fixed; \textit{Linear--Estimated} reports a model re-estimated under linear health dynamics. Columns 2--5 report: assets at age 70 (in \pounds1000), log cumulated earnings at age 68, average DI receipt rate (ages 50--64), and average labor force participation rate (ages 50--68). Results are shown for all individuals (top panel), individuals with below-median initial health (middle panel), and individuals with below-median initial health and wealth (bottom panel).}
    \label{tab:Agg}
\end{table}

Table \ref{tab:Agg} summarizes aggregate life-cycle outcomes. Relative to the baseline \textit{Nonlinear} model, imposing linear health dynamics in simulation (\textit{Linear}) reduces asset accumulation at age 70 by about 2.3\%, lowers cumulative earnings modestly, decreases labor force participation by 1.6 percentage points, and increases DI enrollment by 1.4 percentage points. When innovations are further restricted to be Gaussian, these differences become more pronounced. These results show that the shape of health risk--particularly its state dependence and asymmetry--has first-order effects on savings and meaningful, but more limited, effects on labor supply and DI participation.

Misspecification matters most for economically vulnerable individuals, but in out\-come-specific ways. Among individuals entering the model with below-median health and wealth, linear and normal specifications generate substantially lower asset accumulation relative to the baseline, while effects on earnings and labor supply are small and sometimes slightly positive in this fixed-parameter simulations.
This divergerce in responses reflects the different sensitivity of stock and flow outcomes to misspecification of health dynamics. Asset accumulation is a forward-looking stock that responds to changes in the perceived distribution of future health risk; imposing linear or Gaussian dynamics relative to the nonlinear benchmark therefore leads to sizable reductions in assets, especially for individuals in poor health and with low wealth.
By contrast, cumulated earnings--an accumulated flow--move primarily through period-by-period labor supply, DI participation, and contemporaneous earnings capacity, which are less sensitive to misspecification of persistence concentrated in the lower tail.

Re-estimation under linear health dynamics produces heterogeneous effects across the population. Among economically vulnerable individuals--those with below-median health and wealth--the \textit{Linear--Estimated} model substantially mitigates the small positive responses of earnings and labor supply observed under fixed-parameter linear dynamics, with these effects either disappearing or turning negative. For this group, reduced asset accumulation also persists, though at attenuated levels relative to the fixed-parameter simulations. In contrast, for the population as a whole, re-estimation generates an increase in asset accumulation, reflecting compositional effects and preference adjustments that partially compensate for misspecified health dynamics. These patterns indicate that while preference re-estimation can mitigate short-run behavioral distortions--particularly for flow variables and among vulnerable groups--the aggregate implications for life-cycle saving depend on complex interactions between health dynamics misspecification and parameter adjustments with different effects across the initial conditions distribution.

\begin{table}[t]
    \centering
    \caption{Measures of inequality: Nonlinear vs. Linear Health Dynamics}
    \begin{tabular}{l*{4}{>{\centering\arraybackslash}m{1.5cm}}}

        & \multicolumn{2}{c}{Assets at 70} & \multicolumn{2}{c}{Cum. earnings} \\
        \textbf {All} & SD & CV& SD & CV \\ \midrule
Nonlinear   &       54212&       0.481&       0.436&       0.035\\
Linear      &         532&       0.016&       0.009&       0.001\\
Normal      &         411&       0.023&       0.018&       0.002\\
Linear--Estimated&        1525&      -0.009&       0.002&       0.000\\
 \\
        \textbf{Low health} \\ \midrule
        Nonlinear   &       49563&       0.527&       0.549&       0.044\\
Linear      &       -1151&       0.023&      -0.008&      -0.000\\
Normal      &       -1198&       0.030&       0.009&       0.001\\
Linear--Estimated&        -929&      -0.008&      -0.024&      -0.002\\
 \\
        \textbf{Low health and wealth} \\ \midrule
Nonlinear   &       37870&       0.524&       0.509&       0.040\\
Linear      &       -2463&       0.016&       0.003&       0.000\\
Normal      &       -1732&       0.040&       0.006&       0.001\\
Linear--Estimated&       -1877&      -0.023&      -0.011&      -0.001\\
 \\
    \end{tabular}
    \caption*{\footnotesize\normalfont \textit{Note:} The table reports levels for the \textit{Nonlinear} model and absolute changes from the \textit{Nonlinear} model for the other model specifications. Assets at age 70 and log cumulated earnings: standard deviation and coefficient of variation.}
    \label{tab:AggIn}
\end{table}

Table \ref{tab:AggIn} reports measures of inequality in assets and cumulated earnings. For assets, linear and normal health dynamics increase the coefficient of variation relative to the nonlinear baseline, while changes in the standard deviation are modest or negative. This pattern indicates that higher relative inequality is driven primarily by lower average asset holdings rather than by greater absolute dispersion. The effect is strongest in the fixed-parameter simulations and is attenuated after re-estimation.

For cumulated earnings, differences in inequality across specifications are small. Linear and normal dynamics slightly increase dispersion in the full population, but effects are muted among individuals with low initial health and wealth. Overall, misspecifying health dynamics mainly affects the level and relative dispersion of wealth, with more limited implications for earnings inequality.

The remainder of this section explains why these aggregate and distributional differences arise. We first quantify the overall burden of realized bad health by comparing the baseline economy to a counterfactual in which all individuals experience persistently good health. We then use impulse responses to isolate the marginal effects of innovations to the persistent component of health and to connect state-dependent shock responses to the aggregate patterns documented here.

For conciseness, in the rest of this section we focus on comparisons between the \textit{Nonlinear} baseline and the \textit{Linear} specification. The \textit{Normal} case delivers qualitatively similar but typically more pronounced effects and therefore does not add additional insights beyond those already captured by the linearization of health dynamics.

\subsection{The cost of realized bad health}
\label{subsec:burden}

We next quantify the economic and welfare costs of realized bad health by comparing the baseline economy to a counterfactual in which all individuals experience persistently good health throughout the life cycle, defined as health at the 99th percentile of the age-conditional distribution. This counterfactual removes all realized health risk-- initial heterogeneity, persistent shocks, and transitory innovations--while keeping preferences and decision rules unchanged. It therefore provides an upper bound on the cost of adverse health shocks, distinct from the channel-decomposition exercise discussed later.

\begin{table}[t!]
    \centering
    \caption{Welfare and income costs of realized bad health (health always at 99th percentile)}
    \begin{tabular*}{0.8\textwidth}{@{\extracolsep{\fill}} l*{3}{c}}
     Model & p25 & p50 & p75   \\
     \multicolumn{4}{c}{Welfare cost of bad health (\pounds)}\\ \hline
    Baseline            &        1417&        2346&        4107\\
Linear              &        1394&        2425&        4334\\
\linreest           &        1411&        2503&        4410\\
 \\
     \multicolumn{4}{c}{Income cost of bad health (\pounds)}\\ \hline  Baseline            &         218&        1226&        2454\\
Linear              &         113&        1212&        2545\\
\linreest           &         139&        1204&        2420\\
 \\
     \multicolumn{4}{c}{Earnings cost of bad health (\pounds)}\\\hline   Baseline            &         692&        1749&        3264\\
Linear              &         636&        1741&        3336\\
\linreest           &         623&        1699&        3146\\
 \\
    \end{tabular*}
    \caption*{\footnotesize\normalfont \textit{Note:} The counterfactual eliminates all realized health risk (initial heterogeneity, persistent and transitory components) while keeping preferences, policies, and decision rules unchanged.}
    \label{tab:welfare}
\end{table}

Table \ref{tab:welfare} reports welfare, income, and earnings costs of realized bad health in the population as a whole. Because the economic environment and decision rules are unchanged, welfare is evaluated using realized lifetime utility following \citet{de2022lifetime}.\footnote{
Our welfare measure follows \citet{de2022lifetime}. For each simulated life history, we compute realized lifetime utility under the baseline health process and under the counterfactual with health fixed at the 99th percentile. We then find the proportional change in consumption, $1-\lambda$, applied uniformly to consumption in all periods of the counterfactual, that equalizes realized lifetime utility across the two scenarios. To express welfare costs in monetary units, we multiply $\lambda$ by average consumption in the counterfactual economy.}
Income costs are defined as the discounted difference in lifetime disposable income between the baseline and the counterfactual economy, where disposable income includes labor earnings, disability benefits, pensions, and net transfers after taxes. Earnings costs isolate the labor market channel and are defined analogously as the discounted difference in lifetime labor earnings. Income and earnings are discounted to age 50 and summed up to age 70, while welfare costs reflect lifetime utility over the entire remaining life cycle.

Comparing welfare, income, and earnings costs highlights the economic channels through which adverse health shocks affect individuals. Earnings costs substantially exceed income costs across all specifications, reflecting the insurance provided by disability insurance, pensions, and the tax-transfer system, which partially offset lost labor income. While this insurance pattern is robust across models, linear health dynamics slightly alter its distribution across individuals. In contrast, welfare losses are considerably larger than income losses in all specifications, as they additionally capture the utility cost of reduced leisure, labor supply distortions, and survival risk, with differences across models primarily reflected in the dispersion of welfare losses rather than in their central tendency.

Welfare and income losses are sizable in all specifications, with median costs of similar magnitude across nonlinear and linear health dynamics. Linear specifications generate higher dispersion in welfare losses, reflecting a wider spread between the lower and upper tails. Differences in income and earnings costs are comparatively modest at the median, although linear dynamics compress losses at the bottom of the distribution.

\begin{table}[t]
    \centering
    \caption{Welfare and income costs of realized bad health: the role of heterogeneity}
    \begin{tabular}{lccccccccc}
        & \multicolumn{3}{c}{$\zeta$=1} & \multicolumn{3}{c}{$\zeta$=2} & \multicolumn{3}{c}{$\zeta$=3} \\
     Model & p25 & p50 & p75  & p25 & p50 & p75 & p25 & p50 & p75 \\
     \multicolumn{10}{c}{Welfare cost of bad health (\pounds)}\\ \hline
    Baseline            &        1808&        2902&        5115&        1643&        2446&        4078&        1233&        2004&        3434\\
Linear              &        1592&        2791&        5131&        1627&        2587&        4403&        1244&        2159&        3838\\
\linreest           &        1587&        2862&        5187&        1678&        2694&        4526&        1254&        2255&        3941\\
 \\
     \multicolumn{10}{c}{Income cost of bad health (\pounds)}\\ \hline  Baseline            &           0&        1536&        3247&         319&        1368&        2525&         265&        1045&        2047\\
Linear              &           0&        1362&        3240&         208&        1355&        2696&         229&        1092&        2219\\
\linreest           &           0&        1337&        2979&         243&        1338&        2538&         236&        1094&        2150\\
 \\
     \multicolumn{10}{c}{Earnings cost of bad health (\pounds)}\\\hline   Baseline            &         932&        2256&        4425&         865&        1897&        3338&         577&        1450&        2678\\
Linear              &         676&        2048&        4315&         827&        1914&        3491&         575&        1535&        2885\\
\linreest           &         654&        1958&        3856&         817&        1862&        3289&         554&        1513&        2774\\
 \\
    \end{tabular}
    \caption*{\footnotesize\normalfont \textit{Note:} The counterfactual eliminates all realized health risk (initial heterogeneity, persistent and transitory components) while keeping preferences, policies, and decision rules unchanged.}
    \label{tab:welfareZeta}
\end{table}

Table \ref{tab:welfareZeta} shows that heterogeneity in the cost of bad health is primarily driven by time-invariant heterogeneity between $\zeta$-types. Welfare, income, and earnings costs decline monotonically with $\zeta$, reflecting greater self-insurance capacity among higher types. Relative to the nonlinear baseline, linear specifications compress these differences, understating the burden of bad health for low-$\zeta$ individuals and overstating it for high-$\zeta$ individuals. As a result, while aggregate costs remain similar, misspecifying health dynamics distorts the distribution of health-related losses across individuals.

Beyond differences in levels, Table \ref{tab:welfareZeta} shows that misspecifying health dynamics also affects the composition of health-related losses across welfare, income, and earnings. In the nonlinear model, low-$\zeta$ individuals experience disproportionately large welfare losses relative to income and earnings losses, reflecting strong non-pecuniary costs of poor health and limited scope for self-insurance. Linear specifications flatten this gradient, reducing the wedge between welfare and income costs for low-$\zeta$ types and overstating it for high-$\zeta$ individuals. As a result, linear health dynamics not only compress heterogeneity in total costs, but also misrepresent how adverse health translates into welfare versus monetary losses across individuals.

Our analysis is complementary to \citet{de2022lifetime}. While they model health heterogeneity through discrete types and duration dependence, we work with a continuous health measure and a flexible nonlinear process that allows persistence and shock responses to vary smoothly across the health distribution. This makes it possible to unpack nonlinear health dynamics and to assess how misspecifying persistence primarily distorts the distribution of welfare and income losses rather than their aggregate magnitude.

In addition, focusing on a UK institutional setting without explicit medical expenditures shifts the role of insurance toward disability benefits, pensions, and the tax-transfer system. In this environment, nonlinear health dynamics amplify the insurance value of disability insurance for people facing persistent adverse health histories.

While this exercise removes all realized health risk, impulse responses to one-time health shocks studied in the next section isolate the marginal effects of innovations to the persistent component of health, allowing us to connect state-dependent shock responses to the aggregate and distributional patterns documented above.

\subsection{Economic and welfare effects of health shocks}
\label{subsec:shocks}

We study the economic and welfare consequences of a one-time adverse health realization at age 52, comparing the baseline nonlinear model to a linear-health specification using shocks of the same likelihood.\footnote{We do not report results for the re-estimated linear model, as the objective is to isolate how alternative health dynamics propagate a given realization.}

A key difference between nonlinear and linear health dynamics concerns not only persistence but also the magnitude of health shocks. As shown in Figure \ref{fig:health-count}, shocks generate substantially larger initial health changes in the nonlinear model, especially for individuals starting in poor health. Linear dynamics attenuate both the initial size of shocks and their subsequent persistence.
As a result, differences in economic and welfare outcomes reflect the interaction of two forces: larger initial health deviations and stronger state-dependent persistence under nonlinear dynamics. Linear health dynamics dampen both channels, producing smaller and more homogeneous responses even when shocks are equally likely.

\begin{table}[t]
    \centering
    \caption{Effects of bad shocks on average outcomes}
    \begin{tabular}{lcccccc}
    Wealth  & \multicolumn{3}{c}{Assets at 70} &  \multicolumn{3}{c}{Cumulated earnings} \\
    \multicolumn{1}{r}{$\tau_{init}=$} & 0.1 & 0.5 & 0.9 & 0.1 & 0.5 & 0.9  \\ \hline
     \\ \multicolumn{7}{l}{Nonlinear (baseline)} \\
10                  &      -17779&       -3510&       -1133&      -0.497&      -0.063&      -0.020\\
80                  &      -11499&       -4475&       -1605&      -0.569&      -0.109&      -0.032\\
 \\ \hline
 \multicolumn{7}{l}{Linear (same likelihood shocks)} \\
 10                  &       -2581&       -1621&       -1508&      -0.070&      -0.026&      -0.017\\
80                  &       -1979&       -2108&       -1969&      -0.086&      -0.041&      -0.028\\
 \\
    \end{tabular}
    \label{tab-shockBad}

    \caption*{\footnotesize\normalfont \textit{Note:} Effects of a shock that places individuals at the 10th percentile of the health distribution at age 52, relative to a median shock, by wealth (first column, in \pounds 1K), and by percentile (10th, 50th, 90th) of the persistent component of health at 50 (columns 2--7).}

\end{table}

Table \ref{tab-shockBad} reports the effects of a bad health realization relative to a median realization. Nonlinear dynamics strongly amplify long-run asset losses, particularly for individuals starting in poor health and low wealth ($\tau_{init}=0.1$ and initial wealth at \pounds10k). For these individuals, the nonlinear model predicts large reductions in assets at age 70, while the linear specification delivers much smaller losses. Differences in cumulated earnings between models are comparatively muted, consistent with earnings responding primarily through contemporaneous labor supply and DI participation rather than through long-run expectations.

This difference in sensitivity between wealth and labor-market outcomes helps helps reconcile the aggregate results in Section \ref{subsec:agg}. Asset accumulation reflects precautionary responses to the entire distribution of future health risk and is therefore highly sensitive to persistence and asymmetry in adverse shocks. Earnings, by contrast, adjust mainly in response to current health states and institutional margins, and are thus less affected by misspecification of persistence concentrated in the lower tail of the health distribution.

Figure \ref{fig-CVshock} shows that adverse shocks also have stronger distributional consequences for wealth under nonlinear dynamics. In particular, bad shocks hitting low-wealth individuals lead to a persistent increase in the dispersion of asset holdings, whereas the linear model generates much smaller changes. Changes in the dispersion of cumulated earnings are comparatively small (Appendix Table \ref{tab-shockBadCVcumearn}).

\newcommand{\twoslicegraphic}[5]{%
    \begin{subfigure}{\textwidth}
        \caption*{#2}%
        \includegraphics[
            width=\textwidth,
            trim=0in #3 0in #4,
            clip
        ]{#1}%
    \end{subfigure}
    \begin{subfigure}{\textwidth}
        \includegraphics[
            width=\textwidth,
            trim=0in 0.2in 0in #5,
            clip
        ]{#1}%
    \end{subfigure}%
}

\begin{figure}[t!]
\caption{Changes in the coefficient of variation of assets after shocks of different magnitudes}
\label{fig-CVshock}
\centering

\twoslicegraphic
  {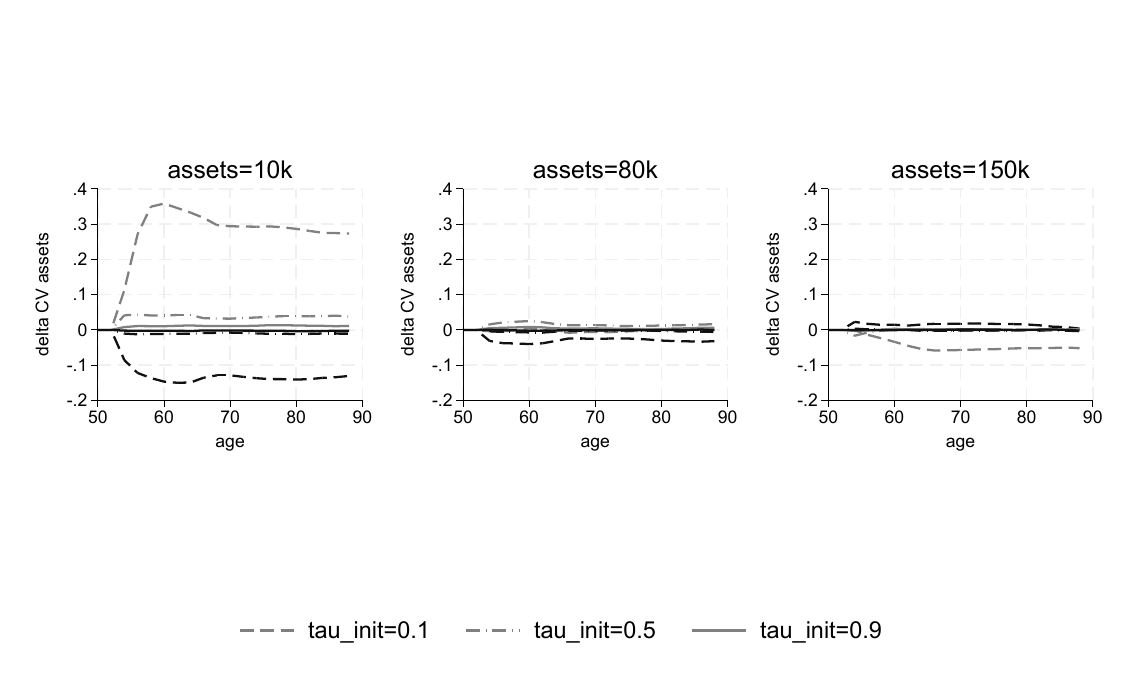}
  {Nonlinear} 
  {1.4in}     
  {1in}       
  {4.1in}     
\twoslicegraphic
  {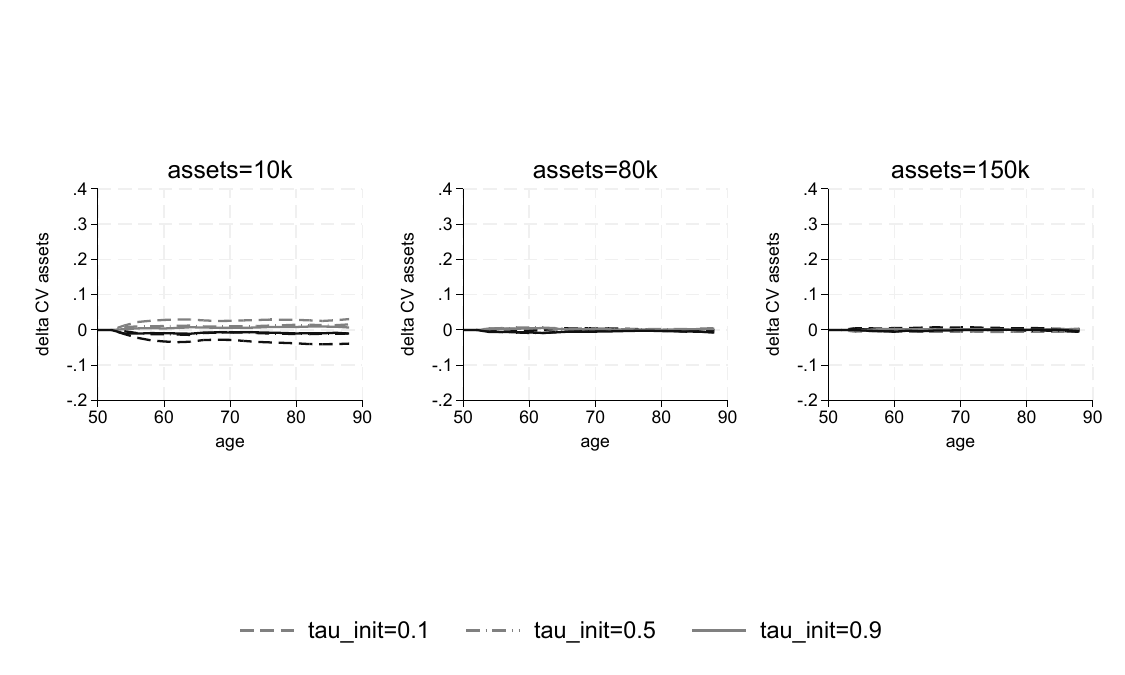}
  {Linear}
  {1.4in} 
  {1in}   
  {4.1in} 
\end{figure}

\begin{table}[t]
    \centering
    \caption{Welfare cost of a bad vs. median health realization: with and without DI (\pounds)}
    \begin{tabular}{lcccc}
          & \multicolumn{2}{c}{with DI} & \multicolumn{2}{c}{without DI} \\
    Model & p50 & p75 & p50 & p75   \\
    \multicolumn{5}{c}{$\tau_{init}=0.1$, initial wealth \pounds10k} \\ \hline
    Nonlinear(baseline) &         284&        1492&        1351&        3356\\
Linear (same likelihood shocks)&         107&         531&         206&         779\\
 \\
    \multicolumn{5}{c}{$\tau_{init}=0.5$, initial wealth \pounds10k} \\ \hline
    Nonlinear(baseline) &         421&        1047&         511&        1254\\
Linear (same likelihood shocks)&         178&         710&         209&         806\\
 \\
    \multicolumn{5}{c}{$\tau_{init}=0.1$, initial wealth \pounds80k} \\ \hline
    Nonlinear(baseline) &        -207&         872&         928&        2774\\
Linear (same likelihood shocks)&          86&         513&         163&         684\\
 \\
    \end{tabular}
    \caption*{\footnotesize\normalfont
\textit{Note:} The table reports median (p50) and upper-quartile (p75) welfare costs. Welfare is computed using realized lifetime utility, as the policy environment and decision rules are unchanged. Welfare effects compare a bad health realization (10th percentile at age 52) to a median realization.}

    \label{tab:welfareshock}
\end{table}

Table \ref{tab:welfareshock} reports median and upper-quartile welfare losses from a bad health shock relative to a median shock, with and without DI. Because welfare is evaluated from realized lifetime utility holding the policy environment and decision rules fixed, this exercise compares realized histories rather than ex-ante risk. As a result, the welfare effects distribution can have a negative median, even if adverse health shocks generate large welfare losses in the upper tail and DI has positive insurance value ex-ante.

Two patterns emerge. First, welfare losses are substantially larger under nonlinear dynamics, especially for individuals starting from low health and low wealth ($\tau_{init}=0.1$ and initial wealth at \pounds10k). Second, removing DI leads to a marked increase in welfare losses in all models, but the increase is considerably larger under nonlinear dynamics. Hence, DI provides more valuable insurance when adverse health histories are more persistent and concentrated in low-health states. Under linear dynamics, welfare losses rise by less when DI is removed, reflecting both weaker persistence and smaller initial deteriorations, and hence a smaller accumulation of losses over the life cycle.

These findings are consistent with the aggregate results in Section~\ref{subsec:agg} and Table~\ref{tab:Agg}, which show that misspecifying health dynamics primarily affects life-cycle saving rather than labor supply. The welfare costs documented here arise largely from the accumulation of asset losses following persistent adverse health realizations, a mechanism that is attenuated under linear health dynamics.


At first glance, these conditional shock responses may appear at odds with the aggregate inequality patterns documented in Table \ref{tab:AggIn}. In particular, while nonlinear dynamics generate larger increases in asset dispersion following adverse health shocks, aggregate asset inequality--as measured by the coefficient of variation--is higher under linear health dynamics for low-health, low-wealth individuals.

This difference reflects the distinction between conditional and unconditional objects. The shock exercise conditions on individuals being in vulnerable states and shows that, when adverse realizations occur, nonlinear dynamics amplify dispersion through larger and more persistent health deteriorations. Aggregate outcomes, instead, depend on how frequently individuals enter such vulnerable states and how long they remain there. Under linear dynamics, individuals accumulate less precautionary wealth on average, leading to lower mean assets and mechanically higher coefficients of variation, even though shocks are attenuated and absolute dispersion is smaller.

Thus, nonlinear dynamics amplify inequality conditional on adverse shocks, while linear dynamics generate higher relative inequality in the aggregate by compressing average wealth levels.

\subsection{Decomposition of health effects}
\label{subsec:decomposition}

We now decompose the economic and welfare effects of health to clarify the mechanisms underlying the aggregate and dynamic results documented above.
Specifically, we consider four channels through which health affects economic outcomes: mortality, the time cost of bad health, earnings capacity, and DI eligibility. We shut down each channel by setting the relevant health-dependent function to its value at the 99th percentile of the age-conditional health distribution, effectively assigning all individuals excellent health for that margin only, while leaving all other channels unchanged. Each counterfactual is solved as a new optimization problem, so that individuals re-optimize their behavior given the modified environment.

\begin{table}[t!]
    \caption{Decomposition of the effects of health}
    \centering
    \begin{tabular}{lcccccccc}\\
        & \multicolumn{3}{c}{Assets at 70} & \multicolumn{2}{c}{Work} & DI & \multirow{2}{*}{CEV} \\
        Effects removed  & Mean & SD & CV & 50--59 & 60--69 & 50--65 &  \\ \toprule
        \\ \multicolumn{7}{l}{\textbf{Nonlinear}}\\ \midrule
            None (baseline)      &      112618&       54212&        0.48&        0.73&        0.29&        0.16&           .\\
All (relative to baseline)&       28205&       -3603&       -0.12&        0.14&        0.25&       -0.14&       15.05\\
Mortality           &       -4779&         -30&        0.02&       -0.01&       -0.02&        0.00&        8.98\\
Time cost           &       34037&       -1763&       -0.12&        0.12&        0.22&       -0.07&        6.96\\
Earnings            &         905&         505&        0.00&        0.00&        0.00&        0.00&        0.17\\
DI                  &         759&        -632&       -0.01&        0.05&        0.04&       -0.11&       -0.95\\
 \\
        \multicolumn{7}{l}{\textbf{Linear}}\\ \midrule
            None (baseline)      &      110001&       54743&        0.50&        0.72&        0.27&        0.18&           .\\
All (relative to baseline)&       30837&       -4108&       -0.14&        0.15&        0.27&       -0.15&       15.35\\
 \\
        \multicolumn{7}{l}{\textbf{Linear -- Estimated}}\\ \midrule
            None (baseline)      &      118034&       55737&        0.47&        0.72&        0.27&        0.17&           .\\
All (relative to baseline)&       32623&       -5748&       -0.14&        0.15&        0.26&       -0.14&       15.70\\

    \end{tabular}
    \label{tab:dec_abb}
    \caption*{\footnotesize\normalfont \textit{Note:} All entries report changes relative to the \emph{None (baseline)} row (where no health channels are removed). The table reports outcomes after removing health's effects on the channels reported in the first column. Results are shown for three model specifications (Nonlinear, Linear, Linear--Estimated). Assets at 70, Work 50--59 and 60--69 report employment rates for those ages; DI 50--65 reports DI receipt rates; CEV denotes standard ex-ante consumption-equivalent variation at age 50
    (welfare effect).}
\end{table}

Table \ref{tab:dec_abb} reports the resulting changes in asset accumulation, labor supply, DI participation, and welfare across model specifications. For the nonlinear model, we present a full channel-by-channel decomposition. For the other specifications, we report only the aggregate effects of health, as the detailed decomposition yields a similar qualitative ranking of channels and therefore adds limited additional insight.

Three results stand out. First, in the nonlinear specification, the time cost of bad health is the dominant channel for most economic outcomes. Removing the time cost leads to the largest increases in asset accumulation and labor force participation, as well as sizable welfare gains. This reflects the central role of health-related time losses in shaping lifetime economic outcomes.

Second, welfare effects in the nonlinear model are primarily driven by mortality risk and the time cost of bad health. Shutting down the earnings and DI channels has noticeable effects on labor supply and income, but their contribution to welfare, measured by consumption-equivalent variation, is comparatively smaller.

Third, while the qualitative ranking of channels is similar across model specifications, their quantitative importance differs. Relative to the linear specifications, the nonlinear model implies smaller average effects of health on assets, labor supply, and welfare. This is because under nonlinear dynamics, large health shocks and high persistence are concentrated in relatively rare low-health states. As a result, aggregate effects are muted despite large conditional impacts following severe adverse health shocks.

The re-estimated linear model yields the largest aggregate effects of health, as reflected in the ``All'' rows of Table~\ref{tab:dec_abb}. In this case, preference and technology parameters adjust to compensate for the misspecified health process, amplifying the role of health in determining economic outcomes even though the linear dynamics fail to capture the state dependence and asymmetry observed in the data.

Appendix Table \ref{tab:dec_sdearn_abb}  focuses on the decomposition of health effects on cumulated earnings and their dispersion. Consistent with the evidence for the U.S. in \citet{hosseini2024important}, earnings dispersion increases with age in all model specifications. In our estimates, however, the time cost of bad health--rather than DI--emerges as the main contributor to earnings inequality. This difference is consistent with institutional features of the U.K. disability system, which is less generous and largely flat-rate, and may also reflect differences in labor supply margins and sample composition.

A final caveat is worth noting when comparing our decomposition results to those in \citet{hosseini2024important}. Unlike the PSID, which provides detailed longitudinal information on earnings histories over the full working life, the ELSA data contain more limited information on individual labor market careers. As a result, the earnings process in our model is identified primarily from later-life outcomes rather than from rich career-long variation. While this is appropriate for our focus on older individuals in the UK, it may attenuate the estimated role of the productivity channel and, more generally, affect the relative importance of earnings-related channels.

\label{subsec:decomposition-capatina}
Our decomposition is closely related to the accounting exercise of \citet{capatina2015life}. Consistent with her findings, we show that the time cost of bad health is a central channel through which health affects labor supply, asset accumulation, and welfare. This result emerges robustly in our setting despite differences in institutions and model structure. In addition, our framework allows us to explicitly quantify the welfare cost of mortality risk, using a consumption-equivalent measure that is internally consistent with changes in survival probabilities. While our analysis abstracts from medical expenditures--a reasonable simplification in the UK context where most health spending is publicly provided--and focuses on later-life outcomes rather than the full working age, it incorporates DI as a key insurance margin. Within this setting, we find that mortality and time costs jointly account for the bulk of welfare losses.

The decomposition clarifies that health dynamics affect economic outcomes primarily through health-related time costs and survival, and that their quantitative importance depends critically on how shock magnitudes and persistence are distributed across health states. These insights provide a natural foundation for the policy experiments on DI presented in the next section.

\subsection{The insurance value of Disability Insurance}
\label{subsec:DI}

We conclude by assessing the value of disability insurance. Unlike the decomposition exercise, which isolates DI as one among several channels, this experiment treats DI as a policy instrument. We compare the baseline economy to a counterfactual in which DI is removed and individuals optimally adjust labor supply, saving, and claiming decisions. Welfare is measured using standard ex-ante CEV at age 50.

\begin{table}[t]
    \centering
    \caption{CEV of removing DI}
    \begin{tabular}{lccc}
    Model & p25 & p50 & p75 \\ \hline
    Baseline            &       -1.17&       -0.65&       -0.37\\
Linear              &       -1.63&       -0.69&       -0.35\\
Linear - reest      &       -1.69&       -0.70&       -0.35\\
 \\
    \multicolumn{4}{c}{Revenue neutral} \\ \hline
    Baseline            &        2.65&        4.18&        5.45\\
Linear              &        2.64&        4.43&        5.78\\
Linear - reest      &        2.56&        4.44&        5.78\\
 \\
    \end{tabular}
    \label{tab:nodi_welfare}
\end{table}

Table \ref{tab:nodi_welfare} reports the welfare effects of removing DI. In all model specifications, eliminating DI generates welfare losses, confirming its role as insurance against future health and earnings risk. These losses are concentrated in the lower part of the welfare distribution (p25) and decline toward the upper tail (p75), consistent with heterogeneity in self-insurance capacity through assets and labor supply.
Table \ref{tab:nodi_welfareZ} shows that welfare losses are highly uneven across unobserved types. The cost of removing DI is largest for low-$\zeta$ individuals and substantially smaller for high-$\zeta$ types, reflecting their higher earnings capacity and greater scope for self-insurance.
Welfare losses from removing DI are sizable across health dynamics specifications, with larger losses under linear dynamics for low-$\zeta$ individuals.

At first glance, these results may appear at odds with Table~\ref{tab:welfare}, which shows that nonlinear dynamics generate more severe welfare losses conditional on realized bad health. The two findings are not contradictory, as they refer to fundamentally different objects.
Table~\ref{tab:welfare} evaluates the ex-post cost of realized adverse health histories, holding behavior fixed and focusing on the severity and persistence of health shocks once they occur. In contrast, Table~\ref{tab:nodi_welfareZ} evaluates the ex-ante insurance value of DI, which depends on how health risk is distributed across individuals and over the life cycle, as well as on the scope for self-insurance through the choice of savings and labor supply.

\begin{table}[t]
    \centering
    \caption{CEV of removing DI by heterogeneity type}
    \begin{tabular}{lccccccccc}
    & \multicolumn{3}{c}{$\zeta=1$} & \multicolumn{3}{c}{$\zeta=2$} & \multicolumn{3}{c}{$\zeta=3$} \\
    Model & p25 & p50 & p75 & p25 & p50 & p75 & p25 & p50 & p75 \\ \hline
    Baseline            &       -1.45&       -0.93&       -0.72&       -1.01&       -0.58&       -0.43&       -0.91&       -0.41&       -0.28\\
Linear              &       -2.50&       -1.63&       -1.18&       -0.87&       -0.53&       -0.38&       -0.77&       -0.39&       -0.27\\
\linreest           &       -2.60&       -1.67&       -1.21&       -0.89&       -0.52&       -0.38&       -0.78&       -0.39&       -0.27\\
 \\
    \multicolumn{4}{l}{Model with revenue neutrality} \\ \hline
    Baseline            &        2.26&        3.94&        5.35&        2.70&        4.15&        5.42&        2.88&        4.30&        5.54\\
Linear              &        1.40&        3.53&        5.21&        3.17&        4.64&        5.97&        3.26&        4.74&        6.06\\
\linreest           &        1.22&        3.48&        5.14&        3.08&        4.70&        5.99&        3.22&        4.79&        6.09\\
 \\
    \end{tabular}
    \label{tab:nodi_welfareZ}
\end{table}

When DI removal is implemented in a revenue-neutral way by reducing labor and pension tax rates,\footnote{See External Appendix \ref{sec:rev_neu_DI} for details on revenue neutrality.} welfare effects turn positive for all individuals in our sample (Table \ref{tab:nodi_welfare}). Table \ref{tab:nodi_welfareZ} shows that these gains are increasing in $\zeta$: individuals with better underlying health experience larger welfare improvements. This pattern reflects the redistributive role of DI, which compresses welfare differences across health types rather than generating net welfare losses for any group under revenue neutrality.

The positive welfare effects under revenue-neutral DI removal reflect the limited tax base of the population considered--older, low-educated individuals with weak labor market attachment--rather than a general inefficiency of DI. Consistent with this interpretation, the similarity of welfare gains across alternative health dynamics indicates that these results are driven primarily by the fiscal adjustment implicit in revenue neutrality, rather than by differences in the propagation of health risk.

Taken together, these results show that the welfare consequences of health risk and social insurance depend critically on the interaction between health dynamics, labor supply margins, and policy design. Accounting for nonlinear health dynamics is therefore essential for evaluating both the distributional and welfare effects of DI.

\section{Conclusion}\label{sec:conclusion}

This paper studies how rich health dynamics shape economic outcomes, inequality, and welfare over the later life cycle. We construct a continuous health index from objective health indicators, estimate its evolution using a flexible quantile-based framework, and embed the estimated process into a life-cycle model of consumption, saving, labor supply, and DI participation.

A key finding of the paper is that the health process displays pronounced nonlinear features--such as state-dependent dispersion, asymmetry, and persistence--that can be directly observed in the data and are well captured by our quantile-based specification. While these features are consistent with insights from recent work emphasizing the importance of state-dependent health risk, our contribution is to provide a unified and flexible representation of health dynamics that reproduces this richness without imposing parametric restrictions. We show that capturing these aspects of health dynamics is quantitatively important: models that abstract from them substantially understate the cumulative effects of adverse health shocks on asset accumulation and welfare, particularly for older individuals in poor health and with limited wealth.

Decomposing the effects of health also highlights how model specification shapes the interpretation of underlying mechanisms. In the nonlinear model, the decomposition reveals a clear distinction between the channels driving economic outcomes and those driving welfare losses: health-related time costs primarily affect assets, labor supply, and program participation, while mortality risk is the main determinant of welfare.
Although the qualitative ranking of channels is broadly similar across specifications, linear health dynamics smooth health risk across individuals and states, reducing the concentration of large and persistent losses in poor health. As a result, linear models attenuate the welfare impact of adverse health shocks for individuals with low underlying health, even when average outcomes and aggregate decompositions appear similar.
This attenuation reflects differences in the distribution of health risk rather than in average effects: under linear dynamics, adverse realizations are less persistent and less concentrated in low-health states, leading to smaller conditional welfare losses despite comparable aggregate magnitudes.

Comparing disability insurance counterfactuals across model specifications further highlights the distinction between ex-post and ex-ante evaluations of health risk. Removing DI without fiscal compensation generates welfare losses in all models, confirming its role as insurance against health and earnings risk. However, the magnitude and distribution of these losses differ across specifications. While nonlinear health dynamics amplify welfare losses following severe and persistent adverse health realizations, linear dynamics can imply larger ex-ante welfare losses from removing DI for individuals with low permanent health, reflecting lower average precautionary saving and greater reliance on public insurance. As a result, the insurance value of DI is shaped not only by the persistence of adverse health histories, but also by how health dynamics affect saving behavior and exposure to risk ex ante.

Overall, our results show that evaluating health risk and social insurance requires accounting for both the magnitude and persistence of health shocks and for how these features interact with saving behavior over the life cycle. Models that impose linear health dynamics may appear to fit average outcomes reasonably well, yet do not fully account for the effect of heterogeneous persistence on economic outcomes.

\label{sec:conclusion-limitations}
Our analysis abstracts from several potentially important dimensions. For computational reasons, we do not model joint nonlinearities in health and income processes, endogenous health dynamics, spousal labor supply and income, or private health expenditures. Extending the framework along these dimensions would allow for a richer analysis of household-level insurance and labor supply decisions.

The model also abstracts from health investments and medical spending decisions. While this is appropriate for the UK institutional environment, where healthcare is largely publicly provided, incorporating endogenous prevention and treatment choices would be important for studying settings with substantial out-of-pocket medical spending. Exploring the interaction between health dynamics, medical spending risk, and economic behavior represents a natural direction for future research.

Finally, while education plays an important role in shaping both health dynamics and economic outcomes, our structural analysis focuses on a single education group to maintain tractability and to study a population most exposed to disability risk. An important extension would be to allow health dynamics to differ by education and to incorporate educational heterogeneity directly into the life-cycle model. This would allow for a richer analysis of inequality and the interaction between health, human capital, and insurance over the life cycle.

Despite these limitations, our results highlight the importance of modeling health as a continuous process with nonlinear dynamics when assessing economic outcomes and their distribution over the life cycle.

\appendix
\section{Appendix}
\renewcommand{\thetable}{\thesection.\arabic{table}}
\renewcommand{\thefigure}{\thesection.\arabic{figure}}
\setcounter{table}{0}
\setcounter{figure}{0}

\subsection{Model solution and estimation details} \label{sec:app-bellman}

\paragraph{State space and choices}

Let $X_t=\{a_t,h_t,\vartheta_t,p_t,di_{t-1}\}$ denote the state vector at the beginning of period $t$, where $a_t$ are assets, $h_t$ health, $\vartheta_t$ the persistent component of earnings, $p_t$ accumulated pension wealth, and $di_{t-1}$ indicates whether the individual was receiving DI in the previous period. Given $X_t$, individuals choose consumption $c_t$ and next-period assets $a_{t+1}$. They also make discrete choices regarding labor supply $w_t\in\{0,1\}$ and, when eligible, DI participation.

The model is solved by discretizing the continuous state variables on finite grids. Assets $a_t$ are discretized using a grid of 30 points. Health $h_t$ is discretized using a total of 26 points, combining 19 grid points for the persistent component and 7 points for the transitory component. The persistent earnings component $\vartheta_t$ is discretized on a grid of 5 points, while accumulated pension wealth $p_t$ is discretized on 5 points. Disability insurance status $di_{t-1}$ is binary. Details on the construction of the health grids and the discretization of the health process are provided in External Appendix~\ref{sec:discretization}.

\paragraph{Timing within the period}

At the beginning of period $t$, individuals observe the state vector $X_t$. If not already enrolled in DI, they decide whether to apply for DI and choose labor supply in the event of rejection. Labor income, disability benefits, pensions, taxes, and transfers are realized within the period. Consumption takes place, assets evolve according to the budget constraint, and survival to period $t+1$ is realized at the end of the period.

\paragraph{Value functions}

Let $V(X_t)$ denote the value function at time $t$. The individual maximizes expected lifetime utility by choosing among a finite set of discrete options. If $di_{t-1}=0$, the individual can choose to work or not work and may apply for DI. If $di_{t-1}=1$, the individual can either continue receiving disability benefits or exit the program and return to work.

When applying for DI, the individual also chooses whether to work in the event of rejection. The value of applying therefore takes the form:
\[
V^{app}(X_t)=\psi_d(h_t)V^{DI}(X_t)+(1-\psi_d(h_t))V^{j}(X_t),
\]
where $j \in \{\text{work},\text{inactive}\}$ denotes the labor supply choice following rejection, and $\psi_d(h_t)$ denotes the probability of being granted DI.

As an example, when $w_t=1$, $di\_app_t=0$ the value function is the following:
\begin{eqnarray}
V^{work}(X_{t}) & = & \max_{a_{t+1}} \Bigg\{ U(c_t,l_t) + \beta \pi^{t+1}\iint_{\substack{h_{t+1},\\e_{t+1}}} V(X_{t+1}|X_{t})dF(X_{t+1}|X_t) \nonumber\\
	& & +  \beta(1-\pi^{t+1})b(a_{t+1})\Bigg\} \label{vf} \\
	& & \text{s.t.:} \nonumber \\
    a_{t+1} & = &  (1+r)\cdot a_t + e_t\cdot(1-c_p\mathbbm{1}(age<65))  - tax_t + tr_t - c_t  \label{atplus1young} \\
    l_t  & = &  1-\phi_w(t)-\phi_{h}(h_t) \nonumber	\\
    tax_t & = & f( e_t\cdot(1-c_p\mathbbm{1}(age<65)) + r_p\cdot p_t\mathbbm{1}(age\geq65), r\cdot a_t) \nonumber	\\
    tr_t & = & \max
      \left(0, \underbar{c}-
        \left( a_t + r_p\cdot p_t\mathbbm{1}(age\geq65) + e_t\cdot(1-c_p\mathbbm{1}(age<65))
      \right)              \right.  \\
          & & \hspace{3em} \left. + r\cdot a_t-tax_t
      \right) \nonumber \\
    a_t & \geq 0 \nonumber
\end{eqnarray}
We rule out borrowing by imposing a non-negativity constraint on assets.

\paragraph{Bequest motives and marginal propensity to bequeath.}

To provide an economic interpretation of the bequest function parameters $(\phi_B,K)$, we follow \citet{de2010elderly} and characterize bequest motives in terms of the marginal propensity to bequeath (\emph{mpb}) and the level of resources at which the bequest motive becomes operative.
Parameter $K$ governs the curvature of the bequest function and determines the level of resources at which bequests become operative, while $\phi_B$ controls the strength of bequest motives.
This characterization allows us to interpret the estimated values of $(\phi_B,K)$ in terms of intuitive economic objects and to compare them with values reported in the literature, such as \citet{french2005effects} and \citet{de2010elderly}. The marginal propensity to bequeath is the fraction of an additional unit of wealth left as a bequest (rather than consumed) for an individual who enters
period $t$ with cash-on-hand, is in the best possible health state, and dies with probability one at the end of the period.
\paragraph{Disability insurance acceptance probability}

We model the acceptance probability as a flexible function of health, decreasing in $h_t$. Specifically, $\psi_d(h_t)$ is specified as a linear spline over the health distribution, with knots defined at selected percentiles of health. The spline is defined over the interval $[\underline{h}, \hat{h}]$, where $\underline{h}$ denotes the minimum value of health in the data and $\hat{h}$ the median. We impose the boundary conditions $\psi_d(\underline{h})=1$ and $\psi_d(\hat{h})=0$, reflecting that individuals in very poor health are always granted benefits, while individuals in good health are never accepted.
The spline coefficients are estimated jointly with the remaining structural parameters of the life-cycle model.
Once enrolled in DI, individuals continue to receive benefits in subsequent periods without re-evaluation, independently of their current health status.

\subsection{Additional Tables and Figures} \label{sec:app-figtab}

\begin{table}[H]
    \centering
    \caption{Descriptive statistics by age}
    \label{tab:descriptives}
 \begin{tabular}{
                  >{\raggedright}m{0.3\linewidth}
                  >{\centering}m{0.12\linewidth}
                  >{\centering}m{0.12\linewidth}
                  >{\centering \arraybackslash}m{0.12\linewidth}
 }
                     & All   & 50--64   & 65--75    \\ \midrule
         \% couple   & 81.7  & -       & - \\
         \% low-educated & 44.9 & - & - \\ \midrule
         \% working  & 44.8 & 68.9 & 15.2  \\
         \% in DI    & -  & 13.8 & - \\
         annual earnings (\pounds) & 13,166 & 14,345 & 6,577 \\
         wealth (1000 \pounds) & 108 & 98 & 121 \\
 \end{tabular}
 \caption*{\footnotesize\normalfont \textit{Note:} ELSA data, waves 1--7. Sample of low-educated males living with a partner.}
 \end{table}

\begin{figure}[H]
\centering
\caption{Moments of health shocks by age and previous health deciles. Data (top panel) and Simulations from the linear model (bottom panel).}
\includegraphics[width=.8\textwidth]{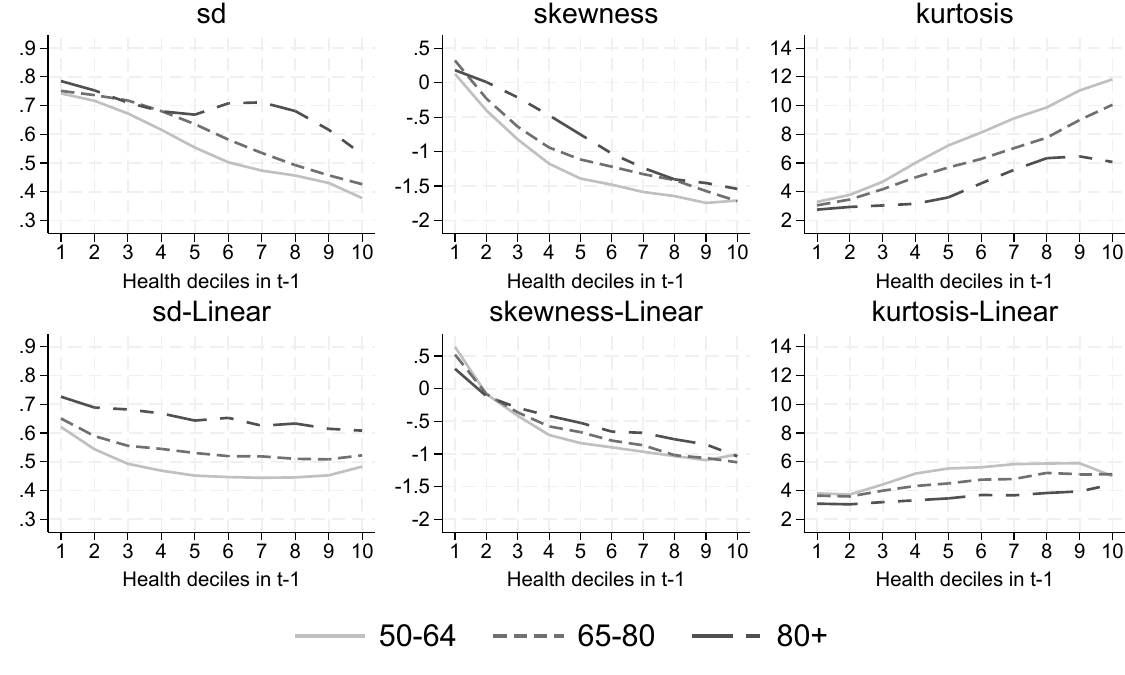}
\label{fig-1linear}
\end{figure}

\begin{figure}[H]
\centering
\caption{Variances and covariances of health shocks, conditional on health in $t-1$ and age. Data (top panel) and Simulations from the linear model (bottom panel).}
\label{fig-cond2L}
\begin{subfigure}{0.25\textwidth}
    \centering
    \includegraphics[width=\textwidth,
    trim=0.15in 0 0.15in 0, clip]{images-IERrev/DataCMoments_tertile_1_alone.png}
    \label{fig-cond2aa}
\end{subfigure}%
\begin{subfigure}{0.25\textwidth}
    \centering
    \includegraphics[width=\textwidth,
    trim=0.15in 0 0.15in 0, clip]{images-IERrev/DataCMoments_tertile_2_alone.png}
    \label{fig-cond2bb}
\end{subfigure}%
\begin{subfigure}{0.25\textwidth}
    \centering
    \includegraphics[width=\textwidth,
    trim=0.15in 0 0.15in 0, clip]{images-IERrev/DataCMoments_tertile_3_alone.png}
    \label{fig-cond2cc}
\end{subfigure}

\begin{subfigure}{0.25\textwidth}
    \centering
    \includegraphics[width=\textwidth,
    trim=0.15in 0 0.15in 0, clip]{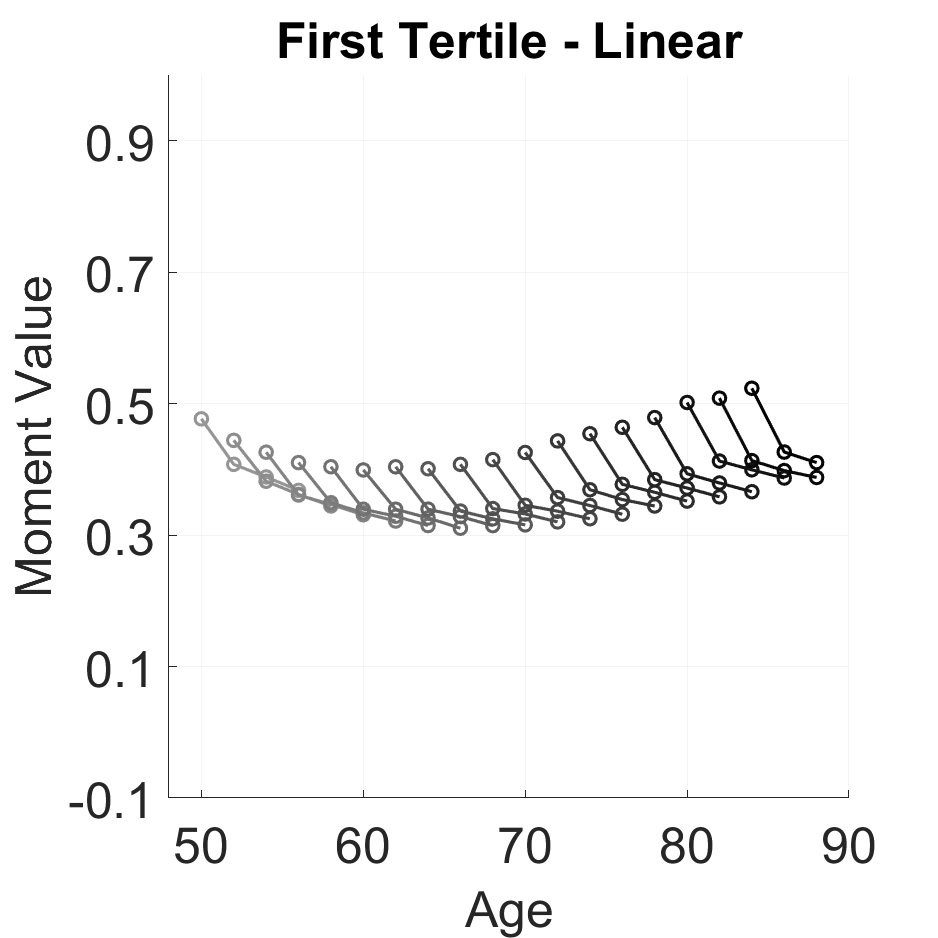}
    \label{fig-cond2aSC}
\end{subfigure}%
\begin{subfigure}{0.25\textwidth}
    \centering
    \includegraphics[width=\textwidth,
    trim=0.15in 0 0.15in 0, clip]{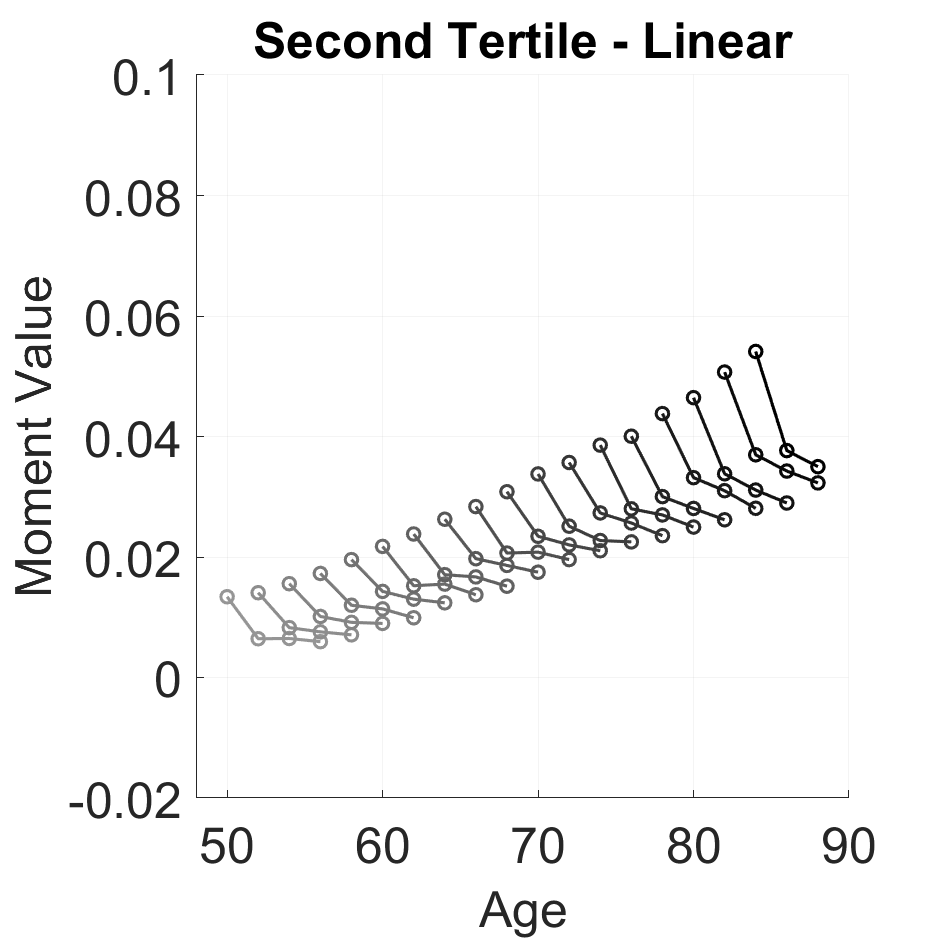}
    \label{fig-cond2bSC}
\end{subfigure}%
\begin{subfigure}{0.25\textwidth}
    \centering
    \includegraphics[width=\textwidth,
    trim=0.15in 0 0.15in 0, clip]{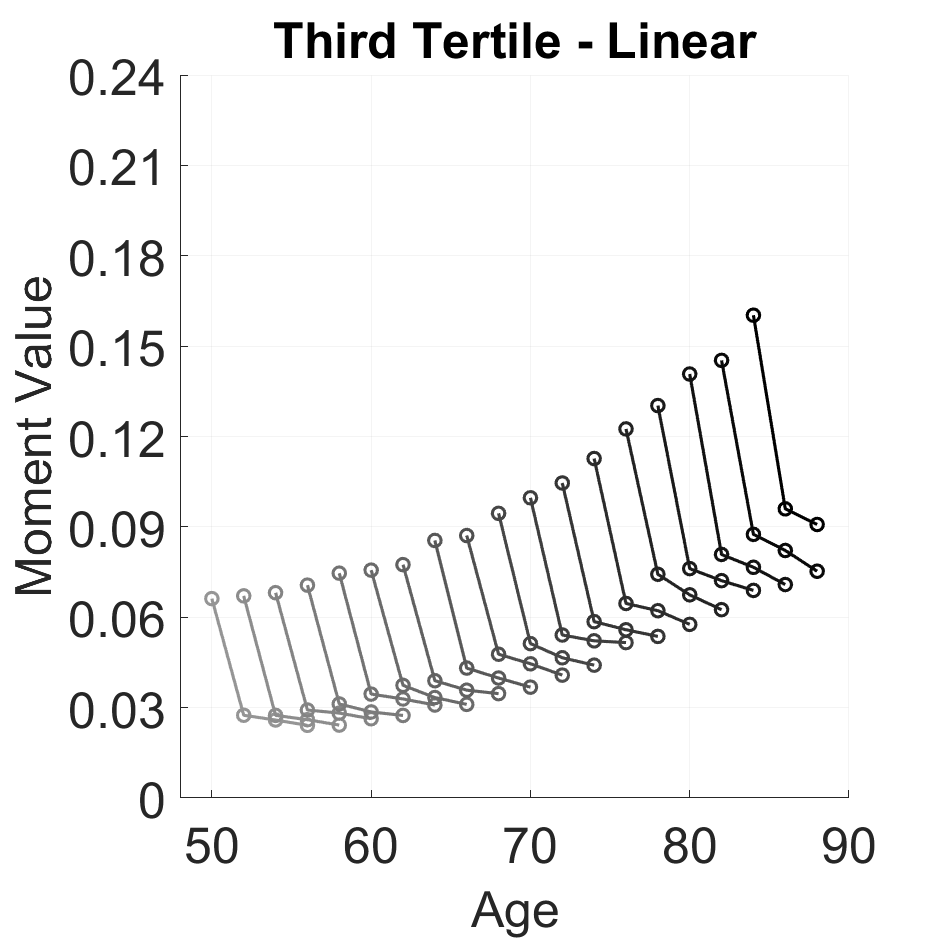}
    \label{fig-cond2cSC}
\end{subfigure}
\end{figure}

\begin{figure}[H]
    \centering
    \caption{Health care expenditure as a share of GDP in the UK and the US}
    \begin{tabular}{rl}
    \includegraphics[
            width=.42\linewidth,
            trim=0in 1.1in 0.2in 0in,
            clip
        ]{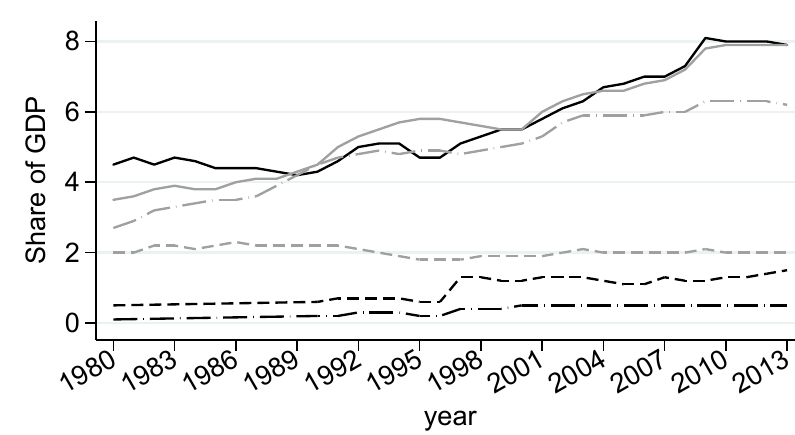}
         &
    \raisebox{0.32in}{%
      \includegraphics[
            width=.5\linewidth,
            trim=.8in 0in 0in 2.9in,
            clip
        ]{images/Healthcareexp.pdf}
        }
    \end{tabular}
    \label{fig:medicalexp}
\caption*{\footnotesize\normalfont \textit{Note:} Health care expenditure as share of GDP by source. Black lines: UK. Grey lines: US. Source: OECD statistics.}
\end{figure}
\FloatBarrier

\begin{figure}[H]
    \centering
    \caption{Estimation fit: non-targeted moments, nonlinear model} \label{fig:fit_nottarget}
                \includegraphics[width=.85\linewidth]{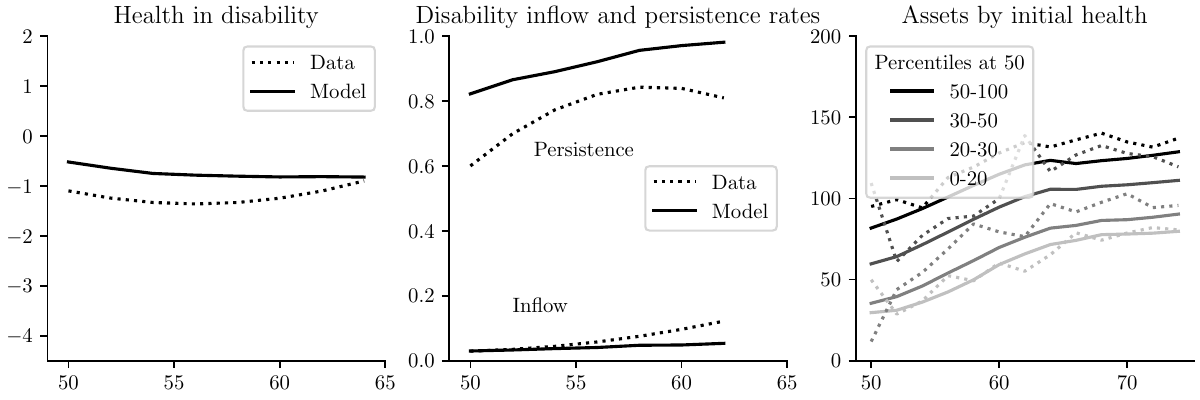}
    \caption*{\footnotesize\normalfont \textit{Note:} \nontargetcaption}
\end{figure}

\begin{figure}[H]
    \centering
    \caption{Estimation fit: CV of assets (non-targeted moment), nonlinear model} \label{fig:CVassetfit}
                \includegraphics[width=.5\linewidth]{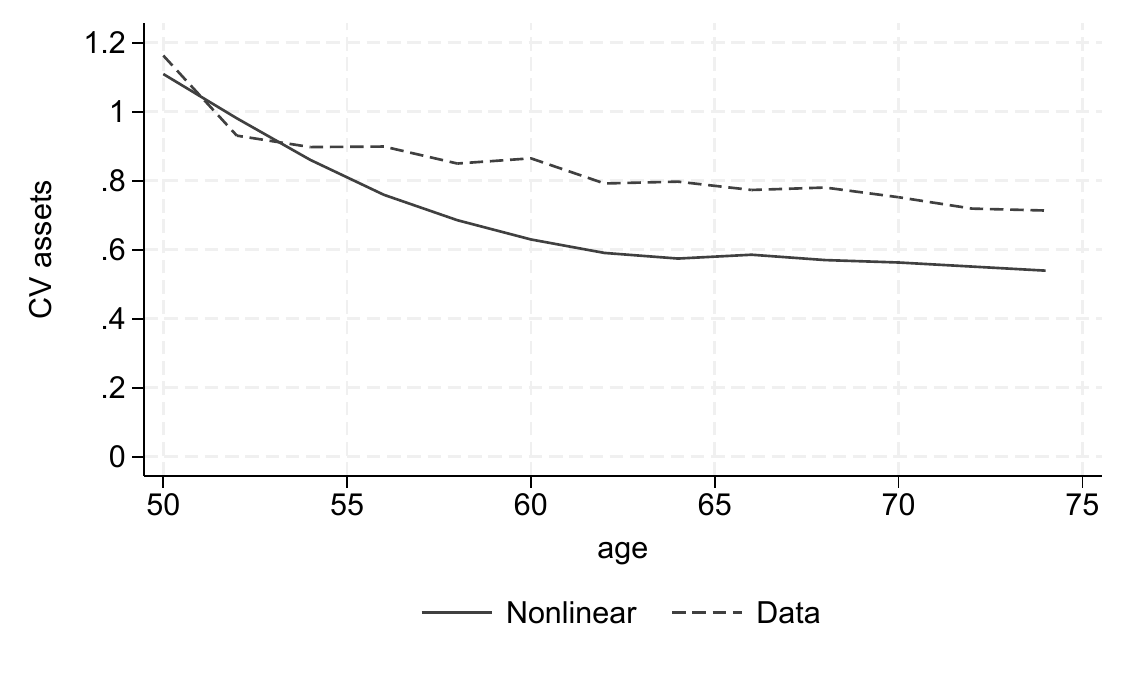}
    \caption*{\footnotesize\normalfont \textit{Note:} \nontargetcaption}
\end{figure}

\begin{figure}[H]
    \centering
    \caption{Estimation fit: targeted moments, linear model} \label{fig:fit_lin}
                \includegraphics[width=.8\linewidth]{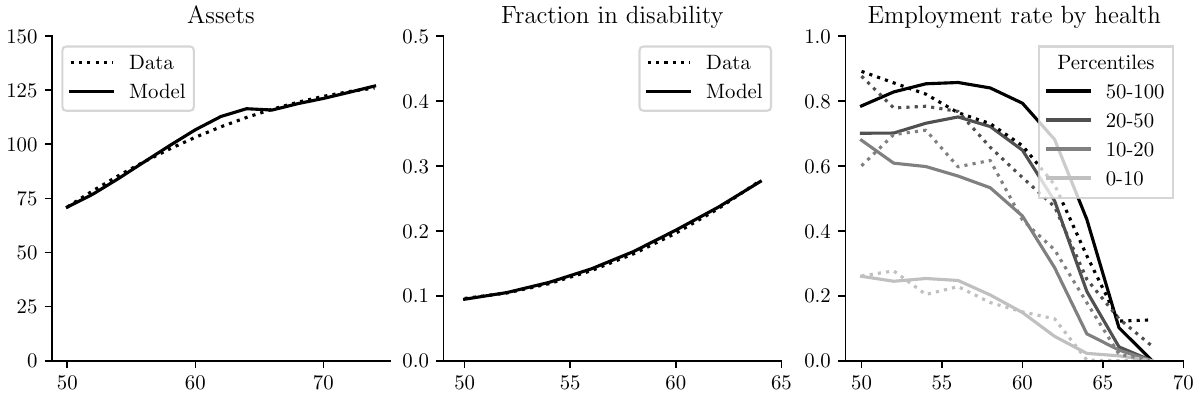}
    \caption*{\footnotesize\normalfont \textit{Note:} \targetedcaption}
\end{figure}

\begin{figure}[H]
    \centering
    \caption{Estimation fit: non targeted moments, linear model} \label{fig:fit_lin_nottarget}
                \includegraphics[width=.9\linewidth]{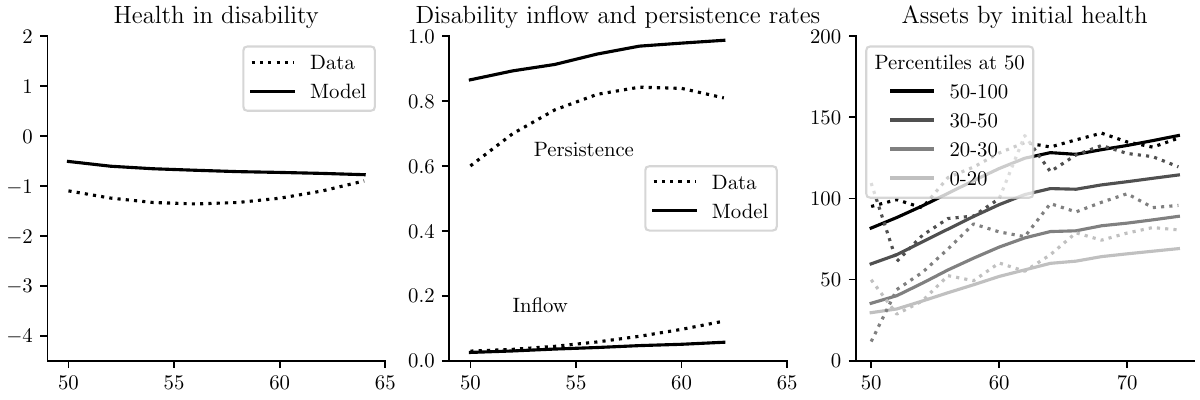}
    \caption*{\footnotesize\normalfont \textit{Note:} \nontargetcaption}
\end{figure}

\begin{table}[H]

    \centering
    \caption{Linear model parameter estimates}
    \label{tab:estimates-lll}
    \small
    
\begin{tabular}{llclll}
  \toprule
  \multicolumn{3}{c}{Estimated} & \multicolumn{3}{c}{Calibrated} \\
  \cmidrule(lr){1-3} \cmidrule(lr){4-6}
  Cost of work & $\phi_w^0$ & 0.74 & Discount factor (biennial) & $\beta$ & 0.9 \\
   & $\phi_w^1$ & 1.21 & Risk aversion & $\nu$ & 3 \\
   & $\phi_w^2$ & 3.57 & Interest rate & $r$ & 0.029 \\
  Cost of health & $\phi_{h}$ & 0.24 & Consumption floor (biennial) & $\underbar{c}$ & 3320 \\
  Disability prob. & $\psi_d^1$ & 1.0 & Pension annuity rate & $p_r$ & 0.0378 \\
   & $\psi_d^2$ & 0.05 & Pension contribution rate & $c_w$ & 0.06 \\
   & $\psi_d^3$ & 0.03 &  &  &  \\
  Consumption weight & $\gamma$ & 0.48 &  &  &  \\
  Bequest motive & $\phi_B$ & 2334 &  &  &  \\
   & $K$ & 268744 &  &  &  \\
  \bottomrule
\end{tabular}

\end{table}

\begin{table}[H]
    \centering
    \caption{Effects of bad shocks on coefficient of variation of cumulated earnings}
    \begin{tabular}{lcccc}
Wealth & \multicolumn{4}{c}{CV of Cumulated earnings} \\
 & \multicolumn{2}{c}{Nonlinear} &  \multicolumn{2}{c}{Linear}\\
    $\tau_{init}$ & 0.1 & 0.5 &  0.1 & 0.5  \\ \hline
10                  &      -0.014&      -0.002&      -0.004&      -0.003\\
80                  &      -0.017&      -0.003&      -0.003&      -0.003\\
 \\

    \end{tabular}
    \label{tab-shockBadCVcumearn}
\end{table}


\begin{table}[h!]
    \caption{Decomposition of the effects of health on the dispersion of cumulated earnings}
    \centering
    \begin{tabular}{lcccccc}\\
        & \multicolumn{5}{c}{SD} & Mean \\
        Effects removed  & 52 & 56 & 60 & 64 & 68 & 68\\ \toprule
        \\ \multicolumn{6}{l}{\textbf{Nonlinear}}\\ \midrule
            None(baseline)      &        0.26&        0.40&        0.48&        0.54&        0.55&       12.47\\
All (relative to baseline)&       -0.06&       -0.10&       -0.18&       -0.18&       -0.15&        0.31\\
Mortality           &        0.00&        0.01&        0.01&        0.00&        0.00&       -0.02\\
Time cost           &       -0.05&       -0.08&       -0.14&       -0.13&       -0.11&        0.26\\
Earnings            &       -0.00&        0.00&        0.01&        0.01&        0.01&        0.01\\
DI                  &       -0.01&       -0.04&       -0.07&       -0.07&       -0.06&        0.09\\
 \\
        \multicolumn{6}{l}{\textbf{Linear}}\\ \midrule
            None(baseline)      &        0.26&        0.41&        0.49&        0.55&        0.55&       12.44\\
All (relative to baseline)&       -0.06&       -0.10&       -0.19&       -0.19&       -0.16&        0.33\\
 \\
        \multicolumn{6}{l}{\textbf{Linear -- Estimated}}\\ \midrule
            None(baseline)      &        0.26&        0.40&        0.48&        0.53&        0.55&       12.45\\
All (relative to baseline)&       -0.08&       -0.13&       -0.20&       -0.19&       -0.15&        0.33\\

    \end{tabular}
    \label{tab:dec_sdearn_abb}
    \caption*{\footnotesize\normalfont \textit{Note:} All entries report changes relative to the baseline row (where no health channels are removed). The table decomposes the effects of health on cumulated earnings dispersion by removing health's effects on different channels: time cost of work, DI, mortality, and wages. Results are shown for three model specifications (Nonlinear, Linear, Linear--Estimated) from ages 52--68. SD: standard deviation of log cumulated earnings; Mean: average log cumulated earnings at age 68. }
\end{table}

\setlength{\bibsep}{2pt}
\begin{spacing}{1.04}
    \bibliographystyle{econ}
    \bibliography{references}
\end{spacing}

\clearpage

\begin{center}
\Large{\textbf{External appendices to: ``The economic effects of nonlinear health dynamics: estimates from a dynamic life-cycle model,'' \\ by Chiara dal Bianco and Andrea Moro }}
\end{center}
\setcounter{page}{1}

\section{Health: measurement and dynamics} \label{sec:app-health}
\subsection{Data and health measurement}\label{sec:app-health1}

We use data from the English Longitudinal Study of Ageing (ELSA), one of the Heath and Retirement Study (HRS) sister surveys. As the HRS, ELSA targets individuals aged 50 and above residing in England and Wales. It starded in 2002 and collects biennial data on several domains relevant to the study of ageing: health, wealth, labour supply and earnings, family networks, among the others.

Table \ref{tab:healthlist} lists the set of variables used to construct the health index. The first three variables are used to run a principal component analysis and contruct $h^s_{it}$, the outcome in equation \ref{eq:Hhat1}.

\textit{Self-reported health} is the answer to the following question: ``Would you say your health is 1. excellent, 2. very good, 3. good, 4. fair, 5. poor?''. \textit{Health limits activities}, refers to any activity-related limitations reported by the ELSA respondents: ``My health stops me from doing things I want to do: 1. Often, 2. Sometimes, 3. Not often, 4. Never.''. Finally, \textit{health limits work} focuses on work-related health limitations: ``Do you have any health problem or disability that limits the kind or amount of paid work you could do, should you want to? Yes/No.''

\footnotesize
{
\def\sym#1{\ifmmode^{#1}\else\(^{#1}\)\fi}
\begin{longtable}{l*{4}{c}}
\caption{Health measures descriptive statistics}
\label{tab:healthlist} \\\hline\hline
            &\multicolumn{3}{c}{Age}                               \\
            &       50-59&       60-69&         70+&       Total\\ \hline
&&&& \\
\emph{Self-reported health measures}: &&&& \\
self-reported health (1 for excellent , 5 for poor health)         &        2.47&        2.63&        2.84&        2.66\\
health limits activities (1 for never, 4 for often)     &        0.84&        1.03&        1.39&        1.10\\
health limits work (Yes/No)       &        0.21&        0.27&        0.38&        0.29\\
&&&& \\
eyesight (from 1=excellent to 6=blind)       &        2.37&        2.40&        2.59&        2.46\\
hearing (from 1=excellent to 5=poor)      &        2.53&        2.75&        3.04&        2.79\\
&&&& \\
\emph{ADL/IADL}: difficulty &&&& \\
walking 100 yards     &       0.061&       0.085&        0.14&       0.096\\
sitting 2 hours     &        0.11&        0.12&        0.11&        0.11\\
getting up from chair after sitting long periods     &        0.15&        0.19&        0.25&        0.20\\
climbing several flights stairs without resting     &        0.16&        0.23&        0.37&        0.26\\
climbing one flight stairs without resting     &       0.060&       0.082&        0.14&       0.096\\
stopping, kneeling or crouching     &        0.22&        0.27&        0.39&        0.30\\
reaching or extending arms above shoulder level     &       0.069&       0.080&       0.094&       0.082\\
pulling or pushing large objects     &       0.074&       0.095&        0.13&        0.10\\
lifting or carrying weights over 10 pounds     &       0.084&        0.12&        0.17&        0.13\\
picking up 5p coin from table      &       0.027&       0.037&       0.055&       0.040\\
dressing, including putting on shoes and socks     &       0.087&        0.11&        0.16&        0.12\\
walking across a room     &       0.014&       0.015&       0.023&       0.017\\
bathing or showering     &       0.044&       0.056&        0.10&       0.069\\
eating, such as cutting up food     &      0.0081&       0.010&       0.017&       0.012\\
getting in and out of bed     &       0.043&       0.042&       0.045&       0.043\\
using the toilet, including getting up or down     &       0.017&       0.019&       0.027&       0.021\\
using map to figure out how to get around strange place     &       0.013&       0.017&       0.027&       0.019\\
preparing a hot meal     &       0.019&       0.018&       0.032&       0.023\\
shopping for groceries     &       0.038&       0.040&       0.060&       0.046\\
making telephone calls     &      0.0097&       0.014&       0.035&       0.020\\
managing money, eg paying bills,keeping track expenses     &       0.018&       0.015&       0.024&       0.019\\
&&&& \\
depression: CES-D questions answered yes (from 1 to 8)      &        1.19&        1.04&        1.13&        1.11\\
&&&& \\
\emph{disgnosed conditions}: &&&& \\ 
angina	    &       0.040&       0.074&        0.12&       0.082\\
heart attack     &       0.031&       0.058&       0.090&       0.062\\
congestive heart failure     &      0.0031&      0.0063&       0.010&      0.0067\\
heart murmur     &       0.025&       0.026&       0.043&       0.032\\
abnormal heart rhythm     &       0.049&       0.069&        0.10&       0.074\\
stroke     &       0.013&       0.030&       0.058&       0.035\\
high blood pressure or hypertension     &        0.33&        0.41&        0.50&        0.42\\
diabetes or high blood sugar     &       0.069&       0.075&        0.10&       0.082\\
chronic lung disease such as chronic bronchitis or emphysema     &       0.037&       0.063&       0.088&       0.064\\
asthma     &       0.088&       0.086&       0.080&       0.084\\
arthritis (including osteoarthritis , or rheumatism)     &        0.19&        0.22&        0.26&        0.23\\
osteoporosis     &       0.010&       0.016&       0.025&       0.018\\
cancer or a malignant tumor     &       0.024&       0.052&       0.097&       0.060\\
Parkinson's disease     &      0.0015&      0.0040&       0.010&      0.0054\\
any emotional, nervous or psychiatric problems     &       0.092&       0.065&       0.032&       0.061\\
Alzheimer's disease     &           0&     0.00022&      0.0038&      0.0014\\
dementia (serious memory impairment)     &      0.0048&      0.0048&      0.0098&      0.0065\\
&&&& \\
\emph{eye problems}: &&&& \\
glaucoma or suspected glaucoma     &       0.023&       0.046&       0.094&       0.056\\
diabetic eye disease     &       0.013&       0.023&       0.026&       0.021\\
macular degeneration     &      0.0088&       0.017&       0.049&       0.026\\
cataracts     &       0.036&        0.11&        0.33&        0.17\\
&&&& \\
incontinence      &       0.038&       0.076&        0.13&       0.083\\
BMI      &        28.5&        28.3&        27.6&        28.1\\
grip stength (measure of sarcopenia)  &        45.7&        41.9&        35.3&        40.7\\
\hline
Number of observations       &       24014&            &            &            \\
\hline\hline
\end{longtable}
}

\normalsize

\begin{table}[H]
    \centering
    \caption{Percentiles of the health distribution in the data}
    \begin{tabular}{cccc}
    p10 & p20 & p30 & p50 \\
    \hline
    -1.56 & -0.73 & -0.21 & 0.33 \\

    \end{tabular}
    \label{tab:healthperc}
\end{table}

\subsection{Comparing alternative health measures and dynamics} \label{app:comparinghealth}
\subsection*{Health measures}
\begin{figure}[H]
\centering
    \caption{Frailty index distribution}\label{fig-frailty}
    \includegraphics[width=0.7\textwidth]{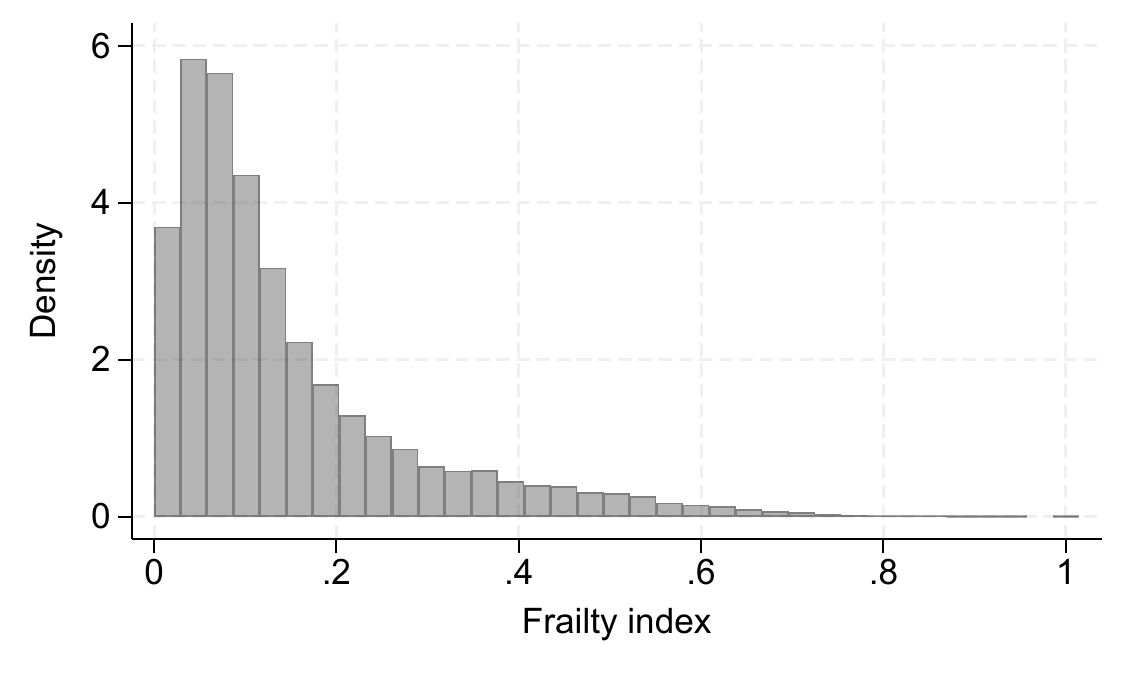}
\caption*{\footnotesize\normalfont \textit{Note:} Cross-sectional distribution of the frailty index in the data (ELSA, waves 1-7).
}
\end{figure}

\begin{figure}[h!]
\centering
\caption{Correlation of the Health index with SRH (a) and Frailty (b)}
\begin{subfigure}{0.8\textwidth}
    \centering
    \includegraphics[width=\textwidth]{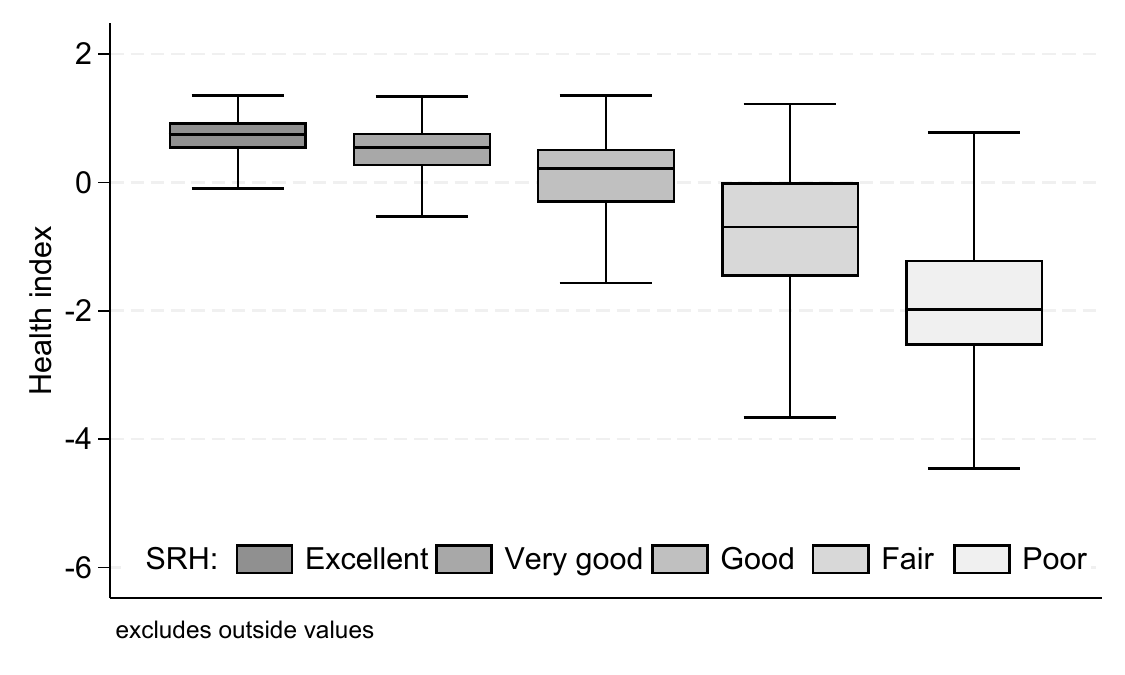}
    \caption{SRH}
    \label{fig-corrSRH}
\end{subfigure}
\hfill
\begin{subfigure}{0.8\textwidth}
    \centering
    \includegraphics[width=\textwidth]{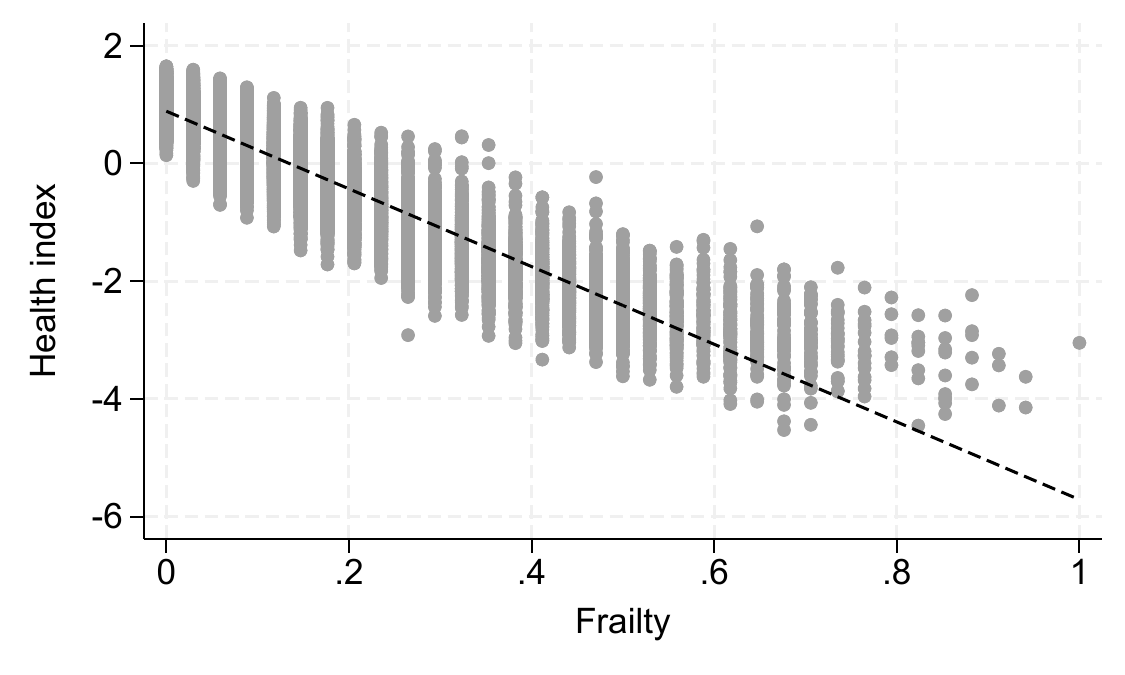}
    \caption{Frailty}
    \label{fig-corrFrailty}
\end{subfigure}
\label{fig-corr}
\end{figure}

Figure \ref{fig-frailty} shows the distribution of the frailty index and Figure \ref{fig-corr} the correlation of the \hindex with SRH (top panel) and frailty (bottom panel).

Tables \ref{tab:predictionM}, \ref{tab:predictionW_DI}, \ref{tab:predictionCARE} report the estimated predictive power of three alternative health measures--\hindex, frailty and SRH--on DI receipt, labour supply, recipiency of formal or informal support, and mortality in period $t+1$. We regress the event dummy in $t+1$ on a second order polynomial in health (for the health index and frailty) or a set of dummies (for SRH) measured at time $t$, a second order polynomial in age, a dummy for being in a couple (at time $t$), school leaving age (below compulsory, between compulsory and 18, at age 19 or above), and a set of wave dummies. Overall, \hindex and frailty outperform SRH. \hindex performs slightly better in predicting mortality, DI and labour supply, whereas frailty in predicting (in)formal help receipt. \\

\begin{table}[htbp]\centering
\def\sym#1{\ifmmode^{#1}\else\(^{#1}\)\fi}
\caption{Outcome: 1 if died in t+1}
\label{tab:predictionM}
\begin{tabular}{l*{3}{c}}
\toprule
                    &\multicolumn{1}{c}{(1)}&\multicolumn{1}{c}{(2)}&\multicolumn{1}{c}{(3)}\\
                    &\multicolumn{1}{c}{\hindex}&\multicolumn{1}{c}{Frailty}&\multicolumn{1}{c}{SRH}\\
\midrule
\hindex              &      -0.016\sym{***}&                     &                     \\
                    &     (0.001)         &                     &                     \\
\addlinespace
Frailty             &                     &       0.109\sym{***}&                     \\
                    &                     &     (0.009)         &                     \\
\addlinespace
SRH=2               &                     &                     &       0.004         \\
                    &                     &                     &     (0.003)         \\
\addlinespace
SRH=3               &                     &                     &       0.017\sym{***}\\
                    &                     &                     &     (0.003)         \\
\addlinespace
SRH=4               &                     &                     &       0.037\sym{***}\\
                    &                     &                     &     (0.004)         \\
\addlinespace
SRH=5               &                     &                     &       0.095\sym{***}\\
                    &                     &                     &     (0.009)         \\
\midrule
Observations        &       21827         &       21827         &       21827         \\
Pseudo R-squared    &       0.221         &       0.217         &       0.227         \\
\bottomrule
\multicolumn{4}{l}{\footnotesize Standard errors in parentheses}\\
\multicolumn{4}{l}{\footnotesize Note: AME (average marginal effect). SE in parenthesis.}\\
\multicolumn{4}{l}{\footnotesize * p<0.10, ** p<0.05, *** p<0.01}\\
\multicolumn{4}{l}{\footnotesize \sym{*} \(p<0.10\), \sym{**} \(p<0.05\), \sym{***} \(p<0.01\)}\\
\end{tabular}
\end{table}

\begin{table}[htbp]\centering
\def\sym#1{\ifmmode^{#1}\else\(^{#1}\)\fi}
\caption{Outcome: 1 if working in t+1}
\label{tab:predictionW_DI}
\begin{tabular}{l*{6}{c}}
\toprule
                    &\multicolumn{1}{c}{(1)}&\multicolumn{1}{c}{(2)}&\multicolumn{1}{c}{(3)}&\multicolumn{1}{c}{(4)}&\multicolumn{1}{c}{(5)}&\multicolumn{1}{c}{(6)}\\
                    &\multicolumn{2}{c}{\hindex}&\multicolumn{2}{c}{Frailty}&\multicolumn{2}{c}{SRH}\\
                    &\multicolumn{1}{c}{DI}&\multicolumn{1}{c}{work}&\multicolumn{1}{c}{DI}&\multicolumn{1}{c}{work}&\multicolumn{1}{c}{DI}&\multicolumn{1}{c}{work}\\
\midrule
\hindex              &      -0.069\sym{***}&       0.113\sym{***}&                     &                     &                     &                     \\
                    &     (0.002)         &     (0.008)         &                     &                     &                     &                     \\
\addlinespace
Frailty             &                     &                     &       0.509\sym{***}&      -1.051\sym{***}&                     &                     \\
                    &                     &                     &     (0.016)         &     (0.060)         &                     &                     \\
\addlinespace
SRH=2               &                     &                     &                     &                     &       0.012\sym{***}&      -0.009         \\
                    &                     &                     &                     &                     &     (0.003)         &     (0.012)         \\
\addlinespace
SRH=3               &                     &                     &                     &                     &       0.054\sym{***}&      -0.060\sym{***}\\
                    &                     &                     &                     &                     &     (0.005)         &     (0.012)         \\
\addlinespace
SRH=4               &                     &                     &                     &                     &       0.179\sym{***}&      -0.234\sym{***}\\
                    &                     &                     &                     &                     &     (0.010)         &     (0.016)         \\
\addlinespace
SRH=5               &                     &                     &                     &                     &       0.374\sym{***}&      -0.507\sym{***}\\
                    &                     &                     &                     &                     &     (0.020)         &     (0.019)         \\
\midrule
Observations        &       10939         &       10939         &       10939         &       10939         &       10939         &       10939         \\
Pseudo R-squared    &       0.363         &       0.218         &       0.359         &       0.214         &       0.285         &       0.183         \\
\bottomrule
\multicolumn{7}{l}{\footnotesize Standard errors in parentheses}\\
\multicolumn{7}{l}{\footnotesize Note: AME (average marginal effect) - Individuals aged 50-64. SE in parenthesis.}\\
\multicolumn{7}{l}{\footnotesize * p<0.10, ** p<0.05, *** p<0.01}\\
\multicolumn{7}{l}{\footnotesize \sym{*} \(p<0.10\), \sym{**} \(p<0.05\), \sym{***} \(p<0.01\)}\\
\end{tabular}
\end{table}

\begin{table}[htbp]\centering
\def\sym#1{\ifmmode^{#1}\else\(^{#1}\)\fi}
\caption{Outcome: 1 if receives (in)formal help in t+1}
\label{tab:predictionCARE}
\begin{tabular}{l*{6}{c}}
\toprule
                    &\multicolumn{1}{c}{(1)}&\multicolumn{1}{c}{(2)}&\multicolumn{1}{c}{(3)}&\multicolumn{1}{c}{(4)}&\multicolumn{1}{c}{(5)}&\multicolumn{1}{c}{(6)}\\
                    &\multicolumn{2}{c}{\hindex}&\multicolumn{2}{c}{Frailty}&\multicolumn{2}{c}{SRH}\\
                    &\multicolumn{1}{c}{formal}&\multicolumn{1}{c}{informal}&\multicolumn{1}{c}{formal}&\multicolumn{1}{c}{informal}&\multicolumn{1}{c}{formal}&\multicolumn{1}{c}{informal}\\
\midrule
\hindex              &      -0.022\sym{***}&      -0.120\sym{***}&                     &                     &                     &                     \\
                    &     (0.001)         &     (0.002)         &                     &                     &                     &                     \\
\addlinespace
Frailty             &                     &                     &       0.160\sym{***}&       0.899\sym{***}&                     &                     \\
                    &                     &                     &     (0.007)         &     (0.014)         &                     &                     \\
\addlinespace
SRH=2               &                     &                     &                     &                     &       0.003\sym{*}  &       0.030\sym{***}\\
                    &                     &                     &                     &                     &     (0.002)         &     (0.004)         \\
\addlinespace
SRH=3               &                     &                     &                     &                     &       0.019\sym{***}&       0.099\sym{***}\\
                    &                     &                     &                     &                     &     (0.002)         &     (0.005)         \\
\addlinespace
SRH=4               &                     &                     &                     &                     &       0.051\sym{***}&       0.270\sym{***}\\
                    &                     &                     &                     &                     &     (0.004)         &     (0.008)         \\
\addlinespace
SRH=5               &                     &                     &                     &                     &       0.101\sym{***}&       0.520\sym{***}\\
                    &                     &                     &                     &                     &     (0.009)         &     (0.014)         \\
\midrule
Observations        &       22392         &       22392         &       22392         &       22392         &       22392         &       22392         \\
Pseudo R-squared    &       0.255         &       0.314         &       0.263         &       0.327         &       0.211         &       0.214         \\
\bottomrule
\multicolumn{7}{l}{\footnotesize Standard errors in parentheses}\\
\multicolumn{7}{l}{\footnotesize Note: AME (average marginal effect). SE in parenthesis.}\\
\multicolumn{7}{l}{\footnotesize * p<0.10, ** p<0.05, *** p<0.01}\\
\multicolumn{7}{l}{\footnotesize \sym{*} \(p<0.10\), \sym{**} \(p<0.05\), \sym{***} \(p<0.01\)}\\
\end{tabular}
\end{table}

\subsection*{Health dynamics}

\begin{figure}[t]
\centering
\caption{Variances and covariances of health shocks by age}
\includegraphics[width=0.5\textwidth]{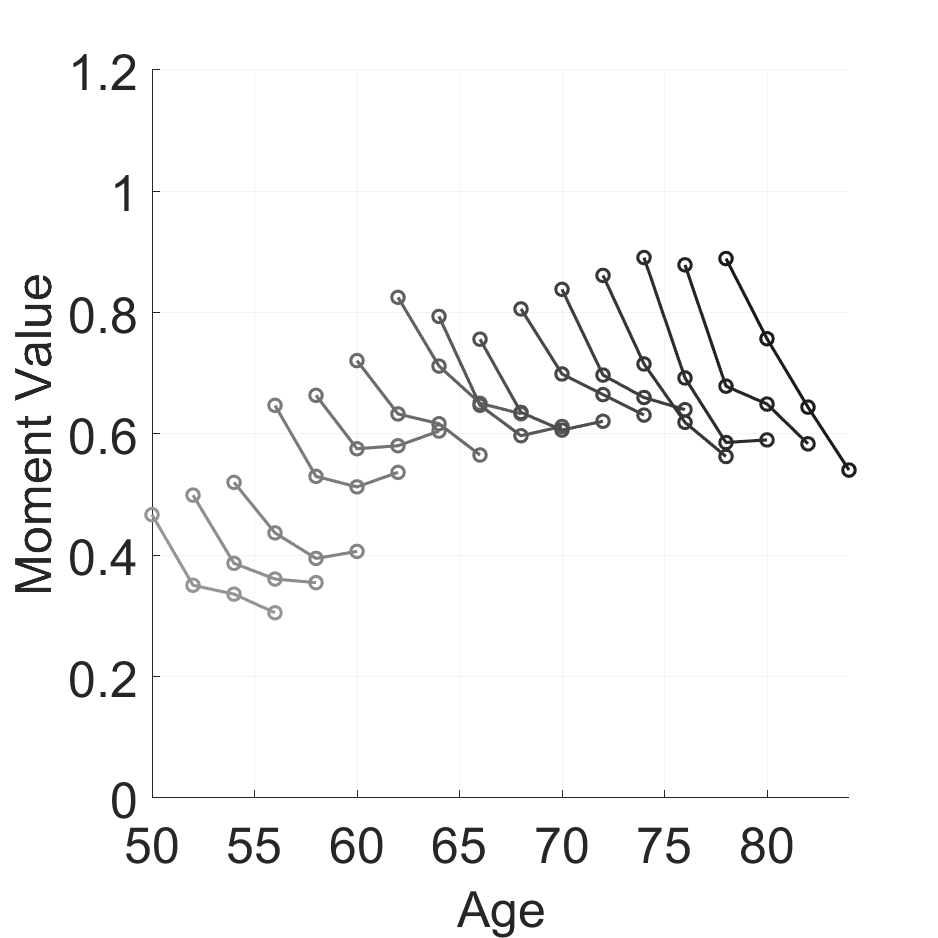}
\label{fig-uncond2}
\caption*{\footnotesize\normalfont \textit{Note:} ELSA, waves 1-7. Men observed in at least two waves.}
\end{figure}

Figure \ref{fig-uncond2} reports the variances and covariances of health shocks by age for the \hindex, unconditional of previous health realizations (reported in Figure \ref{fig-1nonlinear} in the main text).

\begin{figure}[t]
\centering
\caption{Moments of frailty shocks by age and previous frailty quantiles}
    \begin{subfigure}{\textwidth}
        \includegraphics[
            width=.9\textwidth,
            trim=0in 1.2in 0in 0.5in,
            clip
        ]{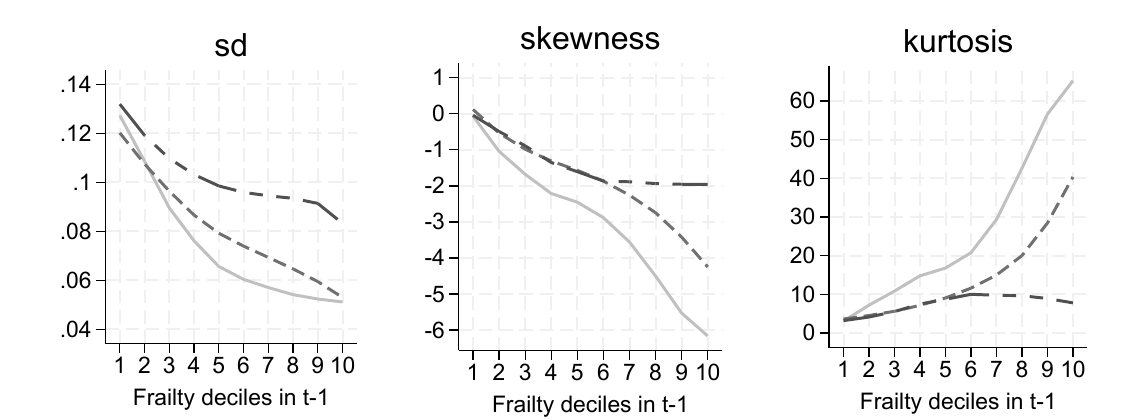}
    \end{subfigure}
    \begin{subfigure}{\textwidth}
        \includegraphics[
            width=.9\textwidth,
            trim=0in 0.2in 0in 4.0in,
            clip
        ]{images/hshockDATA_Frailty.pdf}%
    \end{subfigure}

\label{fig-1nonlinearF}
\caption*{\footnotesize\normalfont \textit{Note:} ELSA, waves 1-7. Men observed in at least two waves.}
\end{figure}

Figure \ref{fig-1nonlinearF} shows the same moments of Figure \ref{fig-1nonlinear} in the main text computed for the frailty index. The main facts are confirmed. Noticed that left skewness is more pronounced for frailty as it is excess kurtosis.

\subsection{Discretization of the health process}\label{sec:discretization}

Following \citet{de2020nonlinear}, after estimating the health process using \citeauthor*{ABB2017}'s procedure, we simulate 100{,}000 health histories and discretize the persistent $(\eta)$, transitory $(\varepsilon)$ and fixed-effect $(\zeta)$ components of health at each age using non–equally spaced grids.
The grid size is set to $N_\eta=19$ for the persistent component, to $N_\varepsilon=7$ for the transitory component and to $N_\zeta=3$ for the time invariant  heterogeneity.

For the persistent component, we first select a set of percentile cutoffs,
\[
\begin{aligned}
\{&0.025,\, 0.05,\, 0.075,\, 0.10,\, 0.125,\, 0.15,\, 0.20,\, 0.25,\, 0.30,\, 0.35,\\
  &0.40,\, 0.45,\, 0.50,\, 0.55,\, 0.65,\, 0.75,\, 0.85,\, 0.95\}
\end{aligned}
\]
and construct the grid points as the midpoint of each interval defined by these percentiles.
Specifically, the first grid point is the midpoint between the minimum of the simulated distribution and the 0.025 percentile, the next is the midpoint between the 0.025 and 0.05 percentiles, and so on, with the last grid point given by the midpoint between the 0.95 percentile and the maximum.
We then compute the transition matrices for the persistent component from period $t$ to period $t+1$ (recall that the health data are biennial).
The resulting age-specific transition matrices for the persistent component of health are used as inputs in the life-cycle model described in Section~\ref{sec:model}.

For the transitory component, we follow the same discretization strategy adopted for the persistent component. Specifically, we select a set of percentile cutoffs
\[
\{0.05,\,0.15,\,0.30,\,0.45,\,0.60,\,0.80\},
\]
and construct a grid with $N_\varepsilon=7$ points defined as the midpoints of the intervals determined by these percentiles. As before, the first grid point is given by the midpoint between the minimum of the simulated distribution and the 0.05 percentile, while the last grid point corresponds to the midpoint between the 0.80 percentile and the maximum. This non–equally spaced grid allows for a finer representation of the central mass of the transitory shock distribution while retaining sufficient support in the tails.

Given the assumption that the transitory component is i.i.d. over time, discretization only requires computing the unconditional probabilities associated with each grid point, which are obtained from the empirical frequencies in the simulated data.

Finally, we discretize the individual fixed effect component of health using a coarse grid with $N_\zeta=3$ points. The grid is constructed using the cutoffs $\{0.3,\,0.5\}$ of the simulated fixed-effect distribution, yielding three intervals. As for the other components, grid points are defined as midpoints of these intervals.

The discretized persistent, transitory, and fixed-effect components jointly define the health state space used in the structural model. This approach preserves the key distributional features and dynamics of the estimated continuous health process while ensuring computational tractability.


\section{Earnings process} \label{sec:app-wage}

This appendix describes the estimation of the earnings process used as an input in the life-cycle model. All parameters of the earnings process are estimated outside the structural model and are taken as given in the solution and simulation of the life-cycle problem.

To estimate the earnings process and control for selection into participation, we closely follow \citet{low2015disability}. We report below the main steps of the procedure. We write the earnings equation and the labor supply participation equation as
\begin{align*}
	\log e_{it} & = f_i + \omega_e(h_{it},age_{it}) + \vartheta_{it}+\upsilon_{it}  \\
	&\upsilon_{it} \sim N(0,\sigma^2_{\upsilon^e}) \\
    &\vartheta_{it} = \vartheta_{i,t-1}+\nu^e_{it}, \quad \nu^e_{it} \sim N(0,\sigma^2_{\nu^e}).\\
	P^*_{it} & =\omega_P(h_{it},age_{it}) + \psi G_{it} +\phi_{it} \\
					 & = p_{it} + \phi_{it}
\end{align*}
where $G_{it}$ is the vector of the exclusion restrictions and $P_{it}=1$ if $P^*_{it}>0$.
To get rid of the time-invariant eterogeneity $f_i$, we rewrite the wage equation using data in differences, with $s$ denoting a generic lag with $s\geq1$.
\begin{align*}
	\Delta^s\log e_{it} &=\omega_e(\Delta^sh_{it},\Delta^sage_{it}) + \Delta^s\vartheta_{it} + \Delta^s\upsilon_{it} \\
	&=\omega_e(\Delta^sh_{it},\Delta^sage_{it}) + \sum_{j=0}^{s-1}\nu^e_{it-j} + \Delta^s\upsilon_{it}
\end{align*}
We observe earnings growth only for individuals working in both $t$ and $t-s$, therefore the conditional expectation takes the following form:
\footnotesize
\begin{align*}
	E(\Delta^s\log e_{it}| P_{it}=P_{it-s}=1) &=\omega_e(\Delta^s h_{it},\Delta^sage_{it}) + E\left(\sum_{j=0}^{s-1}\nu^e_{it-j}| P_{it}=P_{it-s}=1\right) \\
	 &=\omega_e(\Delta^s h_{it},\Delta^sage_{it}) + E\left(\sum_{j=0}^{s-1}\nu^e_{it-j}| \phi_{it}>-p_{it}, \phi_{it-s}>-p_{it-s} \right)
\end{align*}
\normalsize
where $P_{it}=1$ if $s_t={1}$, the individual is working, and zero otherwise.
Assuming $(\phi_{it} \phi_{it-s})' \sim N(0,I)$, the conditional expectation can be written as:
\begin{align*}
	E(\Delta^s\log e_{it}| P_{it}=P_{it-s}=1) &=\omega_e(\Delta^sH_{it},\Delta^sage_{it}) + \\
	& \left( \sigma_{\nu^e}\sum_{j=0}^{s-1}\rho_{\nu^e_{t-j}\phi_t} \right) \lambda_{it}+\left(\sigma_{\nu^e}\sum_{j=0}^{s-1}\rho_{\nu^e_{t-j}\phi_{t-s}}\right)\lambda_{it-s}
\end{align*}
where $\lambda_{it}$ is the inverse Mills' ratio, $\sigma^2_{\nu^e}$ is the variance of $\nu^e_{it}$, $\rho_{\nu_{k}\phi_{\ell}}$ is the correlation between $\nu_{ik}$ and $\phi_{i\ell}$.
The regression of the earnings growth on the controls in differences and the inverse Mills' ratios for each lag $s$ allows to consistently estimate the parameters of the earnings process.

In particular, we assume that the variables in $G$ serve as exclusion restrictions. They include institutional characteristics and family characteristics that affect the labor supply decision: (i) whether the individual is above state pension age, (ii) having children aged below 25, (iii) partner's health, and (iv) whether the partner is above state pension age.

\begin{table}[H]
\centering
\caption{Estimates of the deterministic and stochastic components of the earnings process.}
\begin{tabular}{l c  c c} 		\hline\hline
\multicolumn{4}{l}{Deterministic component} \\ \hline
            &\multicolumn{1}{c}{(1)}&\multicolumn{1}{c}{(2)}&\multicolumn{1}{c}{(3)}\\
            &\multicolumn{1}{c}{Employment}&\multicolumn{1}{c}{Earnings growth}&\multicolumn{1}{c}{Earnings levels}\\
\hline
age         &       0.101         &     0.059           &       0.465\sym{***}\\
            &     (0.084)         &    (0.144)          &     (0.059)         \\
age2        &      -0.002\sym{*}  &    -0.001           &      -0.004\sym{***}\\
            &     (0.001)         &    (0.001)          &     (0.001)         \\
Hindex      &       0.304\sym{***}&     0.007           &       0.097\sym{**} \\
            &     (0.048)         &    (0.049)          &     (0.034)         \\
Hindex2     &      -0.138\sym{***}&     0.092           &       0.046\sym{*}  \\
            &     (0.035)         &    (0.048)          &     (0.023)         \\
Hindex3     &       0.007         &    -0.042           &      -0.007         \\
            &     (0.029)         &    (0.025)          &     (0.021)         \\
Hindex4     &       0.001         &    -0.019           &      -0.008         \\
            &     (0.009)         &    (0.013)          &     (0.005)         \\
\hline
N &       14171         &        4238         &        6505         \\
p-value excl. restr.&       0.000         &                     &                     \\
p-value sel. corr.&                     &       0.001         &                     \\
p-value health&                     &       0.246         &       0.000         \\
 \\ \hline
\multicolumn{4}{l}{Stochastic component}  \\ \hline
$\sigma^2_{\upsilon^e}$ &   & 0.291\sym{***} & \\
                    &       & (0.025) & \\
$\sigma^2_{\nu^e}$ &        & 0.054\sym{***}     & \\
                &           & (0.017) &   \\
\hline\hline
\end{tabular}
\label{tab:wage_proc}
\end{table}

The first column of Table \ref{tab:wage_proc} reports probit parameter estimates for the selection equation. Additional controls included are having a partner and time fixed effects. The exclusion restrictions are jointly significant (p-value 0.000). The probit selection equation allows to construct the inverse Mills's ratio to be included in the earnings growth equation to account for selection bias. Estimates are shown in the second column of Table \ref{tab:wage_proc}. For comparability, column (3) of Table \ref{tab:wage_proc} reports the earnings equation estimated in levels, without controlling for selection. Given  the low number of observations, estimates in column (2) are not very precise. The earnings offer decreases with age and increases with health status up to the first decile of the unconditional health distribution, beyond which it levels off.

Finally, the following set of moment conditions on the adjusted error term are used to identify the parameters of the random component of the wage process.
\footnotesize
\begin{align*}
E(\Delta^s(\vartheta_{it}+\upsilon_{it})|\phi_{it}>-p_{it}, \phi_{it-s}>-p_{it-s}) & =\sigma_{\nu^e}\lambda_{it}\sum_{j=0}^{s-1}\rho_{\nu^e_{t-j}\phi_t}  +  \sigma_{\nu^e}\lambda_{it-s}\sum_{j=0}^{s-1}\rho_{\nu^e_{t-j}\phi_{t-s}} \\
E(\Delta^s(\vartheta_{it}+\upsilon_{it})^2|\phi_{it}>-p_{it}, \phi_{it-s}>-p_{it-s}) & =\sigma^2_{\nu^e}\left(s - p_{it}\lambda_{it}\sum_{j=0}^{s-1}\rho_{\nu^e_{t-j}\phi_t} - p_{it-s}\lambda_{it-s}\sum_{j=0}^{s-1}\rho_{\nu^e_{t-j}\phi_{t-s}}\right) + 2\sigma^2_\upsilon \\
E(\Delta^s(\vartheta_{it}+\upsilon_{it})\Delta^\ell(\vartheta_{i,t-s}+\upsilon_{i,t-s})) & = -\sigma^2_\upsilon
\end{align*}
\normalsize

Estimates are reported in the bottom panel of Table \ref{tab:wage_proc}. Note that, given the biennial nature of ELSA data, the typical lag (difference in the interview years) between consecutive waves is 2 (70\%). However, due to several reasons such as fieldwork length, interview schedule, possible work interruptions and no participation of the respondent in a particular wave, $s$ can take value 1 (13\%) or values greater than 2 (18\%), most commonly 3 (15\%). To avoid noise in the parameter estimates, we set the wave distance to be two years for consecutive waves, therefore our lags are multiple of two.

Given that the fixed effect canceled out when computing the adjusted residuals $g_{it}$, we recover the fixed effect $\hat{f}_i$ for those observed working and set the fixed effect equal to minimum of the $\hat{f}_i$'s distribution for those not working.

In the life-cycle simulations, we fix the individual fixed effect to the average value estimated for low-educated men born between 1948 and 1952, ensuring consistency with the estimation sample used for the structural model.

\section{Survival process and correction for mortality selection}

\subsection{Survival process}\label{sec:app-mortality}
This appendix describes the construction of age- and health-specific survival probabilities used in the life-cycle model.

To compute mortality rates we discretize health in four quantiles defined by the 20, 30, and 50th percentile cutoffs. We assume that mortality risks perceived by the individuals are consistent with the life tables, and rescale estimated mortality in each health-age group in order to match the life tables' mortality rates. In this way, we use the hetogeneity by health in mortality obtained from the data, but anchor the aggregate mortality to match the cross-sectional life tables.

ELSA data are linked to administrative death records which allow to know the exact year of death of any individual (including attriters) up until February 2013. Therefore, we can estimate biennial death probabilities. The steps followed are reported below.

\begin{enumerate}
\item We estimate the probability of being of health level \emph{i} ($\hat{Pr}(H_t=i)$) and of dying by $t+1$ conditional on health level $i$ ($\hat{Pr}(death^D_{t+1}|H_{t}=i)$) using all observations for male respondents. To control for cohort effects, we estimate these probabilities using fixed-effect regressions. When we predict from the estimated regressions, we set the fixed effect equal to the average fixed effect for those born between 1948 and 1952;
\item the probability of dying by $t+1$ at each age $t$ is given by:
\begin{equation}
    \hat{Pr}(death^D_{t+1})=\sum_{i=1}^4 \hat{Pr}(H_t=i)*\hat{Pr}(death^D_{t+1}|H_{t}=i);
    \label{eq:mortalityagg}
\end{equation}
\item we compare the estimated probability with the life tables for each age $t$:
$$\frac{\hat{Pr}(death^{LT}_{t+1})}{\hat{Pr}(death^D_{t+1})}=\alpha_t$$
\item we rescale each conditional probability in such a way that the unconditional probability matches the life tables:
$$ \hat{Pr}(death^{LT}_{t+1})=\sum_{i=1}^4 \hat{Pr}(H_t=i)*\hat{Pr}(death^C_{t+1}|H_{t}=i) $$
with $\hat{Pr}(death^C_{t+1}|H_{t}=i)=\alpha_t*\hat{Pr}(death^D_{t+1}|H_{t}=i)$.
\end{enumerate}

The procedure adopted is the same proposed by \citet{dalbianco2022}, see External Appendix B.3 of that paper for a detailed description of the derivation.

\subsection{Mortality selection} \label{sec:app-mortality_sel}
Ideally, health and mortality would be estimated jointly in a unified framework. However, the complexity of the estimation procedure required to recover the nonlinear health process prevents us from doing so in a computationally feasible way. To account for endogenous selection into survival, we therefore adopt an iterative procedure that ensures consistency between the estimated health dynamics and the observed age profile of health among survivors.

As described in the main text (Section \ref{sec:hmeasure}), we first estimate the deterministic component of health directly from the data. In particular, we recover an age profile using a polynomial in age, while conditioning on a fixed set of observed characteristics--cohort, education, and partnership--which are held constant throughout the procedure.

In a second step, we use the residuals from this regression to estimate the stochastic component of health $h_{it}$, which is modeled as the sum of a persistent component $\eta$, a transitory shock $\varepsilon$, and an individual fixed effect $\zeta$, following \citet{ABB2017}.

We then simulate health histories from the estimated process and apply the mortality process described in this section, thereby generating a simulated health distribution that accounts for selective survival.

Next, we compare the simulated mean health-by-age profile among survivors with the deterministic age profile estimated in the first step. Any discrepancy between the two profiles is used to update the parameters of the deterministic age polynomial. This corrected profile is then used to recompute health residuals and re-estimate the stochastic health process.

We iterate on these steps until convergence, defined as the point at which the simulated mean health-by-age profile--after accounting for mortality--matches the deterministic age profile estimated from the data. This procedure allows us to control for selection induced by mortality while preserving the flexibility of the health process estimation.

\section{Statistical Value of Life} \label{sec:svl}
The statistical value of life (SVL) represents the willingness to pay for a marginal increase in the probability of survival, and is defined as the marginal rate of substitution between wealth and survival probability. In the model, the SVL at age $t$ is given by
\[
SVL_t=\frac{dV(X_t)/d\pi^{t+1}}{dV(X_t)/da_t},
\]
where $\pi^{t+1}$ denotes the probability of surviving to period $t+1$ conditional on being alive at $t$.

The derivative of the value function with respect to survival probability is
\[
\frac{dV(X_t)}{d\pi^{t+1}}
=\beta\left( E_t V(X_{t+1})
-\phi_B\frac{(a_{t+1}+K)^{(1-\nu)\gamma}}{1-\nu} \right),
\]
while the derivative with respect to assets is
\[
\frac{dV(X_t)}{da_t}
=(1+r)\,\gamma\,
c_t^{\gamma(1-\nu)-1}\,
\bar{\ell}_t^{(1-\gamma)(1-\nu)},
\]
where $\bar{\ell}_t$ denotes leisure when not working,
\[
\bar{\ell}_t
=1-\phi_h\frac{h_{\max}-h_t}{h_{\max}-h_{\min}}.
\]

Combining these expressions yields
\[
SVL_t=
\frac{\beta}{1+r}\,
\frac{E_t V(X_{t+1})
-\phi_B\frac{(a_{t+1}+K)^{(1-\nu)\gamma}}{1-\nu}}
{\gamma\;
c_t^{\gamma(1-\nu)-1}\;
\bar{\ell}_t^{(1-\gamma)(1-\nu)}}.
\]

For a given value of the utility shifter $\bar b$, we solve and simulate the model and compute the average SVL across working-age individuals. We then calibrate $\bar b$ so that the model-implied average SVL matches a UK policy benchmark of approximately \pounds 900{,}000, corresponding to the value of a prevented fatality used in early-2000s UK regulatory analysis. This value lies at the lower end of the range typically reported in the European literature and reflects the more conservative valuation framework adopted in the UK relative to the United States.

The calibration target is chosen to be consistent with policy-relevant valuations of mortality risk reductions used in the United Kingdom in the early 2000s. In particular, we target a Value of a Prevented Fatality (VPF) of approximately \pounds 900{,}000, which corresponds to the central value adopted in UK cost--benefit analyses during the period covered by our data. This benchmark is substantially lower than typical values used in US regulatory analyses, but lies well within the range of estimates reported for European countries.

The difference reflects both institutional and methodological factors. Relative to the US literature, European studies tend to rely on more conservative valuation frameworks and on stated-preference or revealed-preference evidence derived from labor market and transportation contexts with lower estimated willingness to pay for risk reductions. As discussed in \citet{johansson2002definition}, cross-country differences in the value of a statistical life are expected to arise from differences in income levels, risk perceptions, labor market institutions, and policy environments. Meta-analytic evidence for Europe reported in \citet{doucouliagos2012estimates} places typical VSL estimates below those commonly used in the US, with values around \pounds 1 million being standard in UK regulatory practice at the time.

By calibrating $\bar b$ to this benchmark, we ensure that the utility value of survival embedded in the model is consistent with contemporaneous UK policy evaluations of mortality risk, without imposing US-based valuations that would be less appropriate for the institutional context under study.


\section{Consumption Equivalent Variation and revenue neutrality} \label{sec:app-CEV}

We follow \citet{dalbiancoetal2025} and derive consumption equivalent variation as follows. Let $\pi_t^i$ denote the conditional survival probability from $t$ to $t+1$ in scenario $i\in\{1,2\}$ and
$\Pi_t^i \equiv \prod_{s=0}^{t-1}\pi_s^i$ the unconditional probability of being alive at $t$. Moreover, define the utility net of the constant term as $\widetilde U(c,l)\equiv U(c,l)-\bar b$.

The generalized CEV $\mu$ is defined as the uniform proportional change in consumption and in bequeathed
resources $(a_{t+1}+K)$ along the counterfactual allocation (scenario 2) that satisfies the indifference condition
\begin{align}
E_0\sum_{t\ge 0}\beta^t \Pi_t^2
\left[
U\big((1-\mu)c_t^2,l_t^2\big)
+
(1-\pi_t^2)\, b\big((1-\mu)(a_{t+1}^2+K)\big)
\right]
= \nonumber \\
E_0\sum_{t\ge 0}\beta^t \Pi_t^1
\left[
U\big(c_t^1,l_t^1\big)
+
(1-\pi_t^1)\, b\big(a_{t+1}^1\big)
\right].
\label{eq:cev_def}
\end{align}

Let $\alpha\equiv \gamma(1-\nu)$. Using the scaling properties implied by the functional forms in the main text,
\begin{equation}
\widetilde U\big((1-\mu)c,l\big)=(1-\mu)^{\alpha}\widetilde U(c,l),
\qquad
b\big((1-\mu)(a+K)\big)=(1-\mu)^{\alpha}b(a),
\label{eq:scaling}
\end{equation}
we can rewrite the indifference condition \eqref{eq:cev_def} as
\begin{align}
&(1-\mu)^{\alpha}
E_0\sum_{t\ge 0}\beta^t \Pi_t^2
\Big[
\widetilde U(c_t^2,l_t^2)
+
(1-\pi_t^2)\, b(a_{t+1}^2)
\Big]
+
\bar b\,E_0\sum_{t\ge 0}\beta^t \Pi_t^2
\nonumber\\
&\qquad\qquad=
E_0\sum_{t\ge 0}\beta^t \Pi_t^1
\Big[
\widetilde U(c_t^1,l_t^1)
+
(1-\pi_t^1)\, b(a_{t+1}^1)
\Big]
+
\bar b\,E_0\sum_{t\ge 0}\beta^t \Pi_t^1 .
\label{eq:cev_collect}
\end{align}
Define
\begin{equation}
E_0\widetilde W^{\,i}
\equiv
E_0\sum_{t\ge 0}\beta^t \Pi_t^i
\Big[
\widetilde U(c_t^i,l_t^i)
+
(1-\pi_t^i)\, b(a_{t+1}^i)
\Big],
\qquad
\widetilde \Pi^{\,i}\equiv E_0\sum_{t\ge 0}\beta^t \Pi_t^i .
\label{eq:defs}
\end{equation}
Then \eqref{eq:cev_collect} implies
\begin{equation}
(1-\mu)^{\alpha}E_0\widetilde W^{\,2}
=
E_0\widetilde W^{\,1}
+
\bar b\left(\widetilde \Pi^{\,1}-\widetilde \Pi^{\,2}\right),
\label{eq:cev_step}
\end{equation}
and therefore the generalized CEV is
\begin{equation}
\mu
=
1-
\left[
\frac{
E_0\widetilde W^{\,1}
+
\bar b\left(\widetilde \Pi^{\,1}-\widetilde \Pi^{\,2}\right)
}{
E_0\widetilde W^{\,2}
}
\right]^{\frac{1}{\alpha}}
,
\qquad \alpha=\gamma(1-\nu).
\label{eq:cev_closed}
\end{equation}

When survival probabilities coincide across scenarios, $\pi_t^1=\pi_t^2$ for all $t$, then $\Pi_t^1=\Pi_t^2$
and $\widetilde \Pi^{\,1}=\widetilde \Pi^{\,2}$, so the $\bar b$ term cancels out and \eqref{eq:cev_closed}
reduces to the standard CEV expression.

\paragraph{Revenue neutrality under DI removal.}
\label{sec:rev_neu_DI}

This subsection describes how revenue neutrality is implemented when the counterfactual scenario $i=2$ corresponds to the complete removal of DI, as discussed in Section~\ref{subsec:DI}.

Let $GR^i$ and $GC^i$ denote, respectively, the present discounted value of government revenues and expenditures in scenario $i\in\{1,2\}$. Government revenues are given by
\begin{equation}
GR^i
=
\sum_{n=1}^{N}\sum_{t\ge 0} \beta^t \pi_t^i \, \left[ tax_{nt}^i , nic_{nt}^i\right]
\label{eq:gr_app}
\end{equation}
where $tax_{nt}^i$ denotes total taxes paid by individual $n$ in period $t$, including taxes on labor income, pension income, and asset income, as defined in the budget constraint in Section~\ref{sec:model}, and $nic_{nt}^i$ denotes national insurance contributions.

Government expenditures are
\begin{equation}
GC^i
=
\sum_{n=1}^{N}\sum_{t\ge 0} \beta^t \Pi_t^i \, tr_{nt}^i,
\label{eq:gc_app}
\end{equation}
where $tr_{nt}^i$ includes DI benefit, state pension benefit and the tranfers needed to reach the consumption floor.
In the baseline economy ($i=1$), individuals may apply for and receive DI benefits. In the counterfactual economy ($i=2$), DI is entirely removed: individuals cannot apply for DI and never receive DI benefits.

Removing DI generates mechanical government savings through lower program expenditures, as well as behavioral responses affecting labor supply, earnings, transfers, and tax revenues. To ensure comparability across scenarios, we impose that the present discounted value of net government revenues is held constant:
\begin{equation}
GR^1 - GC^1 = GR^2 - GC^2 .
\label{eq:rev_neutrality}
\end{equation}

Revenue neutrality is achieved by proportionally adjusting the income tax function in the counterfactual economy. Specifically, all marginal tax rates in the baseline tax schedule are multiplied by a common factor $(1+\Delta)$ in scenario $2$, where $\Delta$ is chosen such that condition~\eqref{eq:rev_neutrality} holds. The adjustment is applied uniformly across all tax brackets, excluding capital income.

The value of $\Delta$ is computed iteratively: starting from the baseline tax schedule, we simulate the counterfactual economy without DI, evaluate the implied government budget imbalance, and update $\Delta$ until convergence to revenue neutrality is achieved.

\section{Moment profiles} \label{sec:app-momest}

This appendix describes how empirical moment profiles are constructed for the sample used in the estimation, consisting of low-educated men living with a partner.\\

\paragraph{Wealth profile} Our measure of wealth includes both housing and non-housing wealth.
\citet{blundell2016dynamic} report real house prices in England from 2002 to 2013 and document a 40\% increase between 2002 and 2004, the first two waves of ELSA. We assume that the house price increase and the resulting wealth increase for homeowners do not affect individual decisions in terms of consumption, retirement, and labor market participation. Therefore, we strip out house price changes by dividing net primary housing wealth by the house price index, using as reference year 2004, and we assume a price increase equal to the real rate of return on other financial assets. The corrected net primary housing wealth is added up to net non-housing wealth and used to estimate the wealth profile.

To correct for cohort effects, we regress wealth $a_{it}$, on an individual specific effect $f_i$, a polynomial in age and unemployment rate $U_t$, proxying for aggregate time effects.
\begin{equation}
a_{it} = f_i + \sum_{n=1}^S{\pi_n age_{it}^n} + \pi_U U_t + u_{it}
\label{eq:momest}
\end{equation}
This specification allows the estimation of age parameters accounting for individual fixed effects and time effects.

The estimated fixed effects $\hat{f}_i$ are regressed on a set of ten-year cohort dummies, this allows to compute the conditional expectation of $\hat{f}_i$ for a specific cohort of individuals, $E[\hat{f}_i|cohort=c]$. We then simulate from the estimated model fixing unemployment rate at $4.9\%$ and the individual fixed effect with the average fixed effect for the cohort of interest. Specifically, we replace $f_i$ with $\tilde{f}_i=\hat{f}_i-E[\hat{f}_i|cohort_i]+E[\hat{f}_i|cohort=c]$. The reference cohort $c$ includes individuals born between 1948 and 1952, which is the same cohort targeted in the structural estimation.

\paragraph{Labor force participation and disability insurance profiles}
Labor force participation and DI receipt profiles are constructed using regression-adjusted age profiles estimated on the full sample, including individuals with different education levels and partnership status. Participation and DI receipt are regressed on a polynomial in age, health indicators--for participation only, cohort controls, education, and partnership status.

The estimated coefficients are used to construct smooth age profiles that are representative of the 1948--1952 cohort. These profiles are then evaluated at the characteristics of low-educated men living with a partner, which is the group targeted in the structural estimation. This approach reduces sampling noise while preserving systematic variation by age and health.

To ensure consistency between the institutional environment in the data and the DI program modeled in the paper, labor force participation and DI receipt profiles are constructed using observations up to wave 4 of ELSA only. This restriction corresponds to the period during which the UK disability system was based on the Incapacity Benefit program. After wave 4, the structure of DI in the UK changed substantially, making later waves less comparable to the institutional setting captured by the model.

\section{Tax function}\label{sec:tax}

The tax function reproduces the one that applied in 2003/04 in the UK. The tax unit in the UK system is the individual.  The income tax schedule is based on three bands. The tax base includes earnings, pensions and interest income net of personal tax-free allowances. The main tax allowances are listed in Table \ref{tab:taxall}.

For those aged less than SPA, National Insurance payments are levied on earnings between a lower limit ($\pounds$4,628) and the upper earnings limit (UEL  $\pounds$30,940) at a rate of 11\%. Those having gross earning below the lower limit do not pay social insurance contributions, whereas those with earnings above UEL are subject to a rate of 1\%.

\footnotesize{\begin{table}[ht]
\centering
\caption{Income tax schedule}
\begin{tabular}{l c c} 		\hline\hline
Band				    & Rate on  		   & Rate on  	  \\
						& earned income    & investment income \\	\hline
0-1960			        & 0.1			   & 0.2					\\
1961-30500	            & 0.22			   & 0.2					\\
30501-			        & 0.4			   & 0.4						\\ \hline\hline
\end{tabular}
\label{tab:taxsch}
\end{table}}
\normalsize

\footnotesize{\begin{table}[ht]
\centering
\caption{Personal tax allowances and credits}
\begin{tabular}{p{8cm} p{7cm} } 		\hline\hline
Allowance/credit		&		Amount per year ( $\pounds$) 	\\	\hline
Single personal allowance: all individuals		&  $\pounds$4,615	\\
Age allowance: Age 65-74											&  $\pounds$6,610 reduced to  $\pounds$4,615 (50\% of income over  $\pounds$18,300) \\
Age allowance: Age 75+												&  $\pounds$6,720 reduced to  $\pounds$4,615 (50\% of income over  $\pounds$18,300) \\
Married Couples age allowance: Age 65-74			&  $\pounds$5,565 reduced to  $\pounds$0 (50\% of income over  $\pounds$18,300, less any reduction to personal age allowance) \\
Married Couples age allowance: Age 75+				&  $\pounds$5,635 reduced to  $\pounds$0 (50\% of income over  $\pounds$18,300, less any reduction to personal age allowance) \\ \hline\hline
\end{tabular}
\label{tab:taxall}
\end{table}}
\normalsize

\end{document}